\documentclass[graybox]{svmult}

\usepackage{mathptmx}       
\usepackage{helvet}         
\usepackage{courier}        
\usepackage{type1cm}        
\usepackage{makeidx}         
\usepackage{graphicx}        
\usepackage{multicol}        
\usepackage[bottom]{footmisc}

\usepackage{amsmath}
\usepackage{amssymb}

\newcommand{\nc}{\newcommand}		
\newcommand{\renc}{\renewcommand}	
\nc{\nuc}[2]    {$^{#1}${#2}}           
\nc{\vc}[1]	{\mbox{\boldmath $#1$}}	
\nc{\wtil}      {\widetilde}            
\nc{\del}       {\partial}              
\nc{\al}        {\alpha}                
\nc{\beq}     {\begin{eqnarray}}
\nc{\eeq}    {\end{eqnarray}}
\nc{\bra}       {\langle}               
\nc{\ket}       {\rangle}               
\nc{\bras}[1]   {\langle #1|}           
\nc{\kets}[1]   {|#1\rangle}            
\nc{\nn}      {\\ \nonumber}
\nc{\mapleft}[1]{			
 \smash{\mathop{\,			%
  \hbox to 1.5cm{\rightarrowfill}\, }\limits_{#1}}}
  
\makeindex             

\begin{document}

\title*{Di-neutron clustering and deuteron-like tensor correlation in nuclear structure focusing on $^{11}$Li}
\titlerunning{Di-neutron clustering and deuteron-like tensor correlation in $^{11}$Li} 
\author{Kiyomi Ikeda, Takayuki Myo, Kiyoshi Kato and Hiroshi Toki}
\institute{
     Kiyomi Ikeda \at RIKEN Nishina Center, Wako, Saitama 351-0198, Japan, \email{k-ikeda@postman.riken.go.jp}
\and Takayuki Myo \at General Education, Faculty of Engineering, Osaka Institute of Technology, Osaka 535-8585, Japan, \email{myo@ge.oit.ac.jp}
\and Kiyoshi Kato \at Division of Physics, Graduate School of Science, Hokkaido University, Sapporo 060-0810, Japan, \email{kato@nucl.sci.hokudai.ac.jp}
\and Hiroshi Toki \at Research Center for Nuclear Physics (RCNP), Osaka University, Ibaraki, Osaka 567-0047, Japan, \email{toki@rcnp.osaka-u.ac.jp}
}
%
%
\maketitle

\abstract{
$^{11}$Li is a Borromean nucleus, where two out of three objects as $^9$Li + $n$ 
and two neutrons independently do not form bound systems. Two neutrons 
should form a di-neutron cluster in the nuclear field generated by the 
$^9$Li core nucleus. We treat di-neutron clustering by solving the two 
neutron relative wave function precisely by using the bare 
nucleon-nucleon interaction so that the spatial clustering structure 
is obtained quantitatively within the whole $^{11}$Li nucleus. 
This di-neutron clustering is an essential dynamics to form the halo 
structure by making a compact di-neutron cluster, which distributes 
loosely around the $^9$Li core. This concept of di-neutron clustering 
should be clearly distinguished from the BCS pairing correlation, 
where no consideration of spatial clustering is made. The di-neutron 
clustering is a new concept and is a general phenomenon in neutron 
skin and neutron halo nuclei.
\newline\indent
This quantitative description of di-neutron clustering has made it 
necessary to consider another important deuteron-like tensor 
correlation, which is caused by strong tensor interaction in the 
nucleon-nucleon interaction. The tensor interaction originates from 
pion exchange and known to provide large attraction to form the $^4$He 
nucleus. The unique feature of the tensor correlation is to make 
highly correlated deuteron-like excitation, which interferes with 
shell model like structure in a unique way. This dynamical effect 
removes the magic number effect and makes easy the participation of 
the s-wave neutrons.  Hence, there are pairing and deuteron-like tensor correlations in addition to the mean field structure in $^9$Li.  The combined system of two additional neutrons with the correlated $^9$Li provides the halo phenomenon, in which the di-neutron clustering  develops with the help of large $s$-wave component caused by the deuteron-like tensor correlation.
\newline\indent
These two effects, the di-neutron clustering and the deuteron-like 
tensor correlation, are quite new and essential to provide the halo 
structure of $^{11}$Li. In this lecture note, we would like to introduce 
these two new concepts in a systematic manner and fill a gap between 
the halo phenomenon and the microscopic reason for this interesting 
phenomenon.
}


\section{Unstable nuclei and the halo structure of $^{11}$Li}
\label{theory:1}

We are in the era of being able to study experimentally unstable nuclei up to the drip lines and even up to super-heavy nuclei.  We are able now to provide precious informations for astrophysics and cosmo-physics directly from experiment.  All these activities started with the discovery of the halo structure of $^{11}$Li \cite{Ta88b}.  Hence, there are many experimental data on $^{11}$Li and surrounding nuclei.  First of all, we would like to discuss the characteristics of $^{11}$Li found by various experiments.  We should present all the existing experimental facts about the Li-isotopes, in particular $^9$Li and $^{10}$Li.

The formation of the halo structure in $^{11}$Li is a difficult subject to understand theoretically.  The fact that the binding energy is extremely small urges us to develop a theory to handle continuum states as precisely as bound states.  This fact of small binding energy of the last two neutrons forces us to consider the pairing interaction at low density, which leads to the concept of the di-neutron clustering phenomenon.  Most fascinating physics necessary for the quantitative understanding of the halo structure turned out to be the discovery of the deuteron-like correlation caused by the strong tensor interaction.  Theoretically we have experienced these conceptual developments on the theoretical framework to treat the Li isotopes.  We discuss first the experimental facts on the Li isotopes and then discuss the theoretical tools for the understanding of $^{11}$Li.  All these theoretical tools developed for the Li isotopes are to be used for many new phenomena found in unstable nuclei.

\subsection{Experimental facts on $^{9,10}$Li and $^{11}$Li}

\begin{figure}[th]
\sidecaption[t]
\includegraphics[width=4.0cm]{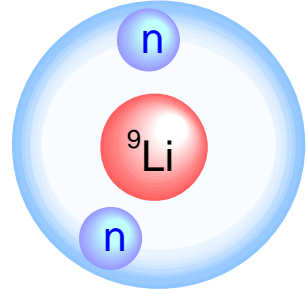}
\caption{A schematic image of halo structure in $^{11}$Li.}
\label{fig:momentum}
\end{figure}

\begin{figure}[th]
\sidecaption[t]
\includegraphics[width=7.0cm]{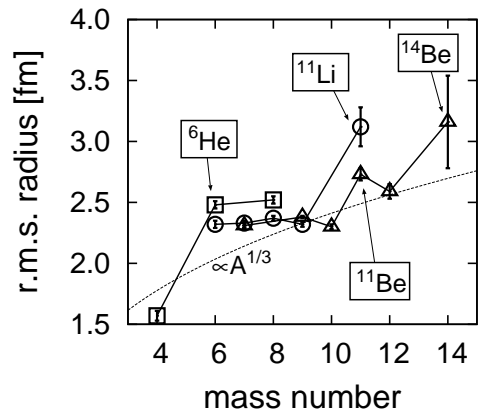}
\caption{\small Observed rms matter radii of He, Li and Be isotopes.  Shown by the dashed curve is the one proportional to $A^{1/3}$.  Those points marked by arrows are the ones with extraordinary large radii.
Data are taken from Ref.~\cite{Ta88b}.}
\label{fig:rms}
\end{figure}

There are many experimental data indicating that the $^{11}$Li nucleus has a halo structure.  A schematic picture of the halo structure is shown in Fig.\ref{fig:momentum}.  The halo structure of $^{11}$Li was discovered as an anomalously large reaction cross section of this nucleus with target nuclei.  The matter radii extracted from the reaction cross sections and other standard methods are shown in Fig.\ref{fig:rms}.  We can see from this figure that the matter radius of $^{11}$Li is much larger than that of the neighboring nucleus $^{9}$Li.  The matter radius of $^{11}$Li corresponds to those of medium mass nuclei.  If we were to pick up nuclei, whose radii are suddenly increased from the neighboring nuclei, they are $^6$He, $^{11}$Be and $^{14}$Be in addition to $^{11}$Li.  Detailed studies on these sudden jumps of the matter radii made these phenomena as caused by the halo formation.

There are many experiments performed on the halo structure of unstable nuclei.  Out of all these experiments, those of $^{11}$Li are very interesting due to important nuclear many body physics behind the experimental data.  In order to find the reason of the sudden increase, experiments were performed for the neutron separation momentum distributions.  The momentum distributions of neutron separation were measured experimentally by Kobayashi et al.  \cite{Ko92} by bombarding unstable nuclei on some stable target nuclei.  The momentum distribution for the case of $^{11}$Li is much narrower than other cases.  The narrowness of the momentum distribution is related with large extension of the neutron distributions due to the uncertainty principle.  Hence, this is a data showing directly the halo structure or at least a large spatial extension of the neutron distribution in $^{11}$Li.  As for the structure of $^{11}$Li, there was an experiment performed by Simon et al. \cite{simon99}, who bombarded $^{11}$Li on a carbon target and measure the momentum distribution of the $^{10}$Li fragments.  From the shape of the distribution, the $(1s_{1/2})^2$ contribution to the mixture of $(1s_{1/2})^2$ and $(0p_{1/2})^2$ components was determined to be $(45 \pm 10)\%$.

\begin{figure}[th]
\sidecaption[t]
\includegraphics[width=6.0cm]{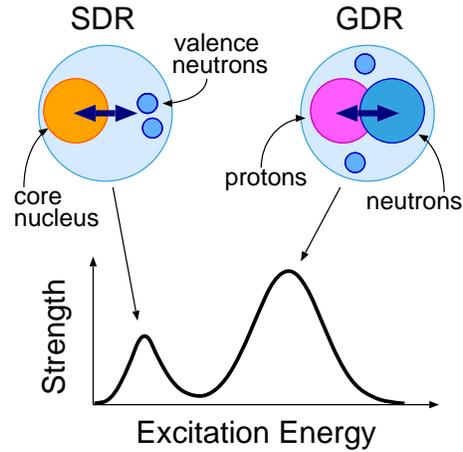}
\caption{A schematic spectrum of E1 excitation of soft dipole resonance and giant dipole resonance modes of halo nuclei.  Shown above in the left is a schematic view of the soft dipole resonance (SDR) and the one in the right is that of the giant dipole resonance (GDR).}
\label{fig:soft}
\end{figure}

\begin{figure}[th]
\sidecaption[t]
\includegraphics[width=6.0cm,clip]{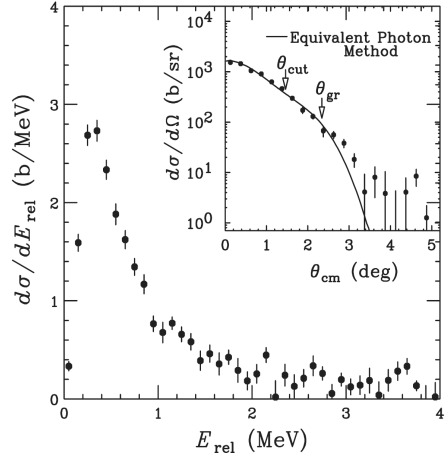}
\caption{Coulomb excitation of $^{11}$Li in $^{11}$Li+Pb at 70 MeV/nucleon as a function of the $^9$Li+$n$+$n$ relative energy.
Details are explained in the original papar \cite{Na06}.}
\label{fig:nakamura}
\end{figure}

If there is a halo structure, we may expect an interesting excitation mode of $^{11}$Li.  If the core nucleus $^9$Li is surrounded by neutrons in the nuclear halo region, there is a possibility of making oscillation of the core nucleus in the neutron sea.  This is called a soft dipole resonance to be excited by photo-disintegration as shown in Fig. \ref{fig:soft} \cite{Ha87,Ik88}.  The excitation energy is expected around a few MeV as compared to the standard giant $E1$ resonance of a few tens MeV.  
There are several experiments of Coulomb excitation of $^{11}$Li.  
We show the most recent experimental data on Coulomb excitation taken by Nakamura et al. in Fig.\ref{fig:nakamura} \cite{Na06}. 
There is a bump structure just above the threshold energy.  
However, it is still debated if the bump structure is caused by the soft dipole resonance or not due to the complicated nuclear structure of $^{11}$Li.  There are many other experimental data on $^{11}$Li as the magnetic moment and quadrupole moment.  These experimental data will be presented together with theoretical results later.

We should also see experimental  data for $^{10}$Li, which is the neighboring nucleus, although this nucleus is unbound.  We list the neutron single and double separation energies in $^{9,10,11}$Li in Table 1. The two neutron separation energy in $^{11}$Li is very small as 0.32MeV.  The separation energies in $^9$Li are large and this nucleus should be considered as a standard shell model type nucleus.  In $^{10}$Li, the single neutron separation energy is $-0.3$ MeV, indicating this nucleus is not stable.  This resonance structure seems to have positive parity and is assigned to have the $(p_{3/2})_\pi(p_{1/2})_\nu$ structure.  At the same time, there are few informations on virtual states.  There is an experimental data on neutron scattering with $^9$Li at the threshold energy \cite{Je06}.  The scattering length comes out to be $a \sim -20$ fm, which is comparable with the one of neutron-neutron scattering  $a=-18.5\pm 0.4 $ fm \cite{Te87}. Hence, the large scattering length indicates
  the existence of strong attraction in the s-wave channel, which is closed to the condition of forming a bound state in the free space.  This means that the s-state structure appears close to the threshold energy of $^9$Li + $n$.
\begin{table}[htdp]
\caption{\small Single and double neutron separation energies in unit of MeV in $^{9,10,11}$Li.}
\begin{center}
\begin{tabular}{c | c c}
~~Nucleus             &~~    $S_{2n}$[MeV]    &~~ $S_n$[MeV] \\
\hline
~~$^{11}$Li & ~~ 0.32 & ~~0.62  \\
~~$^{10}$Li & ~~ -      & ~~ -0.3  \\
~~$^{9}$Li   & ~~ 6.10 & ~~ 4.07 
\end{tabular}
\end{center}
\label{default}
\end{table}

\subsection{Theoretical studies on the halo structure in $^{11}$Li}

The halo structure was completely new in nuclear physics community.  Hence, there were many theoretical studies to describe this interesting phenomenon.  We have recognized immediately that the standard shell model approach badly fails due to the fact that the two additional neutrons in $^{11}$Li ought to enter in the $p_{1/2}$ neutron orbit but not in the $s_{1/2}$ orbit due to the N=8 magic structure.  Hence, most of theoretical studies introduce some phenomenology to bring down the $s_{1/2}$ orbit.  For example, in the work of Thompson and Zhukov \cite{To94}, they treat $^9$Li as a core and add two neutrons by taking state dependent neutron-core interactions.  The additional attraction for s-wave component makes the $(s_{1/2})^2$ state energetically close to the $(p_{1/2})^2$ state.  In this case, the $(s_{1/2})^2$ state has a large component in the ground state, which provides the halo structure for $^{11}$Li.  

There is a theoretical study on the pairing property and the E1 excitation in $^{11}$Li by Esbensen and Bertsch \cite{Es92}.  In their study, it is essential to bring down the $s_{1/2}$ orbit to reproduce the experimental E1 excitation spectrum.  As for the pairing correlation, there are many studies to describe $^{11}$Li as the BCS state.  In the study of Meng and Ring \cite{meng}, they describe $^{11}$Li in terms of a relativistic Hartree-Bogoliubov model.  In this study, they can include the continuum effect in their pairing correlations.  In the relativistic Hartree-Bogoliubov model, the s-wave contribution comes out to be about a quarter of the p-wave contribution for the paired two neutrons.  We need more participation of the s-wave component as compared to the finding of the experimental data of Simon et al. \cite{simon99}.

There is another interpretation on the halo structure as due to deformation.  In the work of Varga, Suzuki and Lovas \cite{Va02}, they try to break the $^9$Li core and introduce the cluster structure.  The wave function of $^{11}$Li is written as $^4$He$+t+4n$ and take the interaction among them by a phenomenological central interaction.  In this way, they can introduce the effect of the deformation and pairing correlations among the nucleons.  The deformation effect provides a large matter radius and some s-wave component in the wave function.

The theoretical challenge on the halo structure is therefore summarized as follows.  There are many indications that the s-wave component is very large in the ground state wave function.  Hence, we have to find a mechanism to bring down the $s_{1/2}$ orbit with the amount to wash out the N=8 magic structure.  The pairing properties are also very important to cause admixture of $(p_{1/2})^2$ and $(s_{1/2})^2$ states.  In the halo nucleus, we ought to consider the di-neutron pairing correlation in a small nuclear matter density.  All these new phenomena should be understood in terms of the many body framework with the nucleon-nucleon interaction.

\subsection{Nucleon-nucleon interaction and the deuteron and the di-neutron system}

\begin{figure}[bht]
\begin{center}
\includegraphics[width=10.0cm,clip]{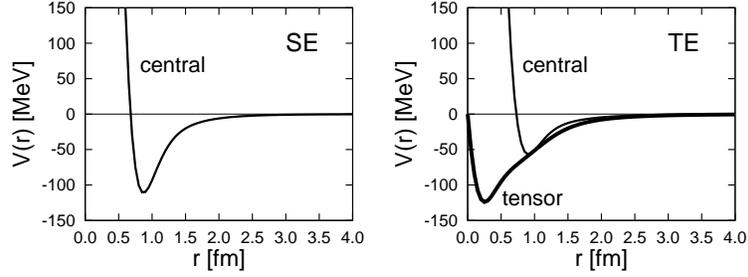}
\caption{Central and tensor interactions of the AV8' potential \cite{pudliner97} in singlet even (SE) and triplet even (TE) channels.}
\label{fig:av8}
\end{center}
\end{figure}

We should learn the properties of the nucleon-nucleon interaction in order to understand the halo structure in $^{11}$Li.  To this end, we would like to show the central and the tensor interactions in the $^3 S_1$ channel of the AV8' potential \cite{pudliner97}, which are shown in Fig.~\ref{fig:av8}.  In the central interaction, there are strong hard core (short range repulsive interaction) and intermediate range attraction of moderate strength.   As for the tensor interaction, the long range part drops with the pion range, while the short range part increases until $0.2$ fm and goes to zero at the origin due to the form-factor coming from the nucleon finite size.  On the other hand, we have the similar structure for the central interaction in the $^1 S_0$ channel, where there is a strong hard core due to the short range quark dynamics.  In this channel, there is no tensor contribution due to zero total spin.  The deuteron-like tensor correlation is produced by the $NN$ interaction in the $^3 S_1$ 
 channel, while the di-neutron clustering is produced by the $NN$ interaction in the $^1 S _0$ channel.

\begin{figure}[b]
\begin{minipage}[l]{5.5cm}
\includegraphics[width=5.5cm,clip]{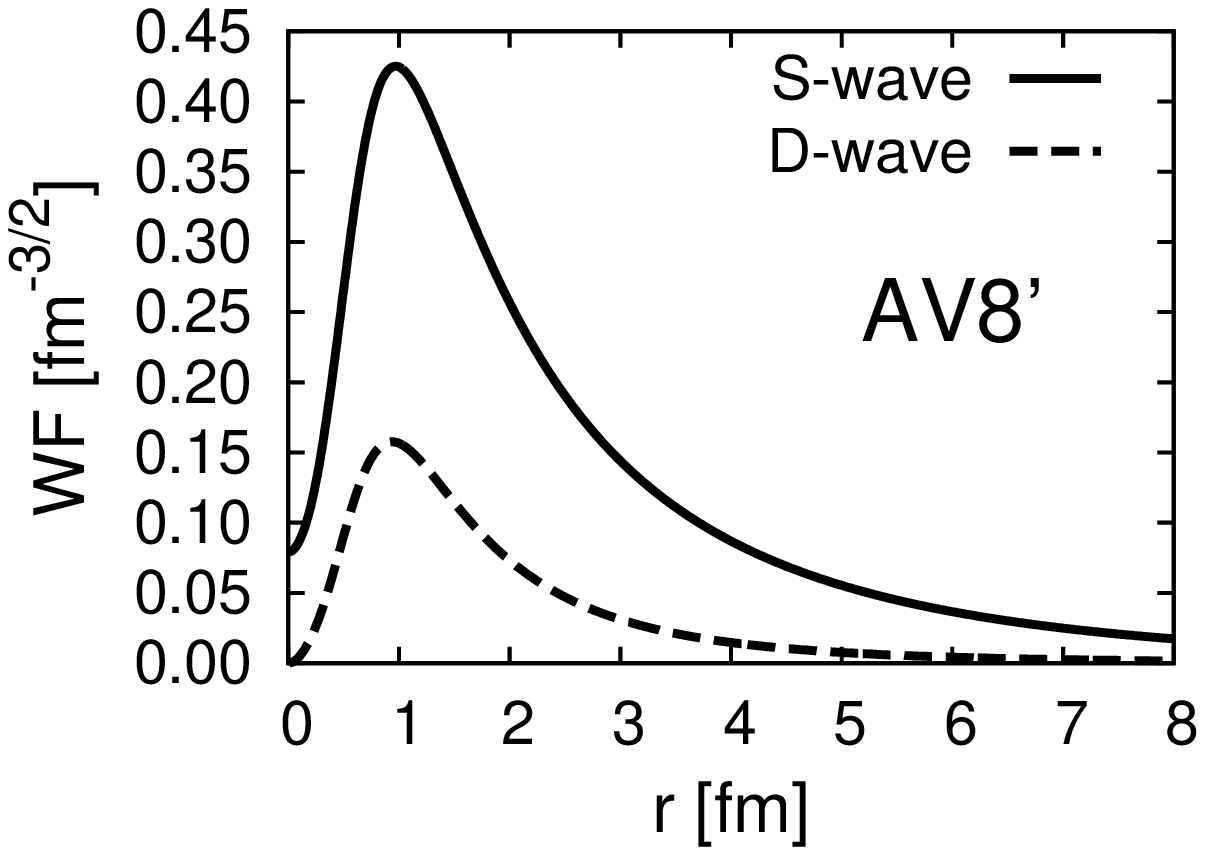}
\caption{Deuteron s-wave and d-wave functions obtained using the AV8' nucleon-nucleon potential.}
\label{fig:deuteron}
\end{minipage}
\hspace*{0.8cm}
\begin{minipage}[r]{4.5cm}
{\makeatletter\def\@captype{table}
\caption{Deuteron properties using the AV8' nucleon-nucleon potential.}
\begin{tabular}{p{2cm}p{2.4cm}p{2cm}p{4.9cm}}
\noalign{\smallskip}\svhline\noalign{\smallskip}
Energy  & $-2.24$ [MeV] \\
\noalign{\smallskip}\hline\noalign{\smallskip}
Kinetic & $19.88$  \\
~~~(SS)  & $11.31$  \\
~~~(DD)  & $ 8.57$  \\
\noalign{\smallskip}\hline\noalign{\smallskip}
Central & $-4.46$  \\
~~~(SS)  & $-3.96$  \\
~~~(DD)  & $-0.50$  \\
\noalign{\smallskip}\hline\noalign{\smallskip}
Tensor & $-16.64$ \\
~~~(SD)  & $-18.93$ \\
~~~(DD)  & $2.29$   \\
\noalign{\smallskip}\hline\noalign{\smallskip}
LS      & $-1.02$  \\
\noalign{\smallskip}\svhline\noalign{\smallskip}
P($D$)  & 5.78 [\%] \\
\noalign{\smallskip}\svhline\noalign{\smallskip}
Radius  & $1.96$ [fm] \\
~~~(SS)    & $2.00$ [fm] \\
~~~(DD)    & $1.22$ [fm] \\
\hline\noalign{\smallskip}
\end{tabular}
\label{tab:deuteron}       
\makeatother}
\end{minipage}
\end{figure}
In order to understand the role of the hard core and the tensor interaction, let us solve the Schr\"odinger equation for the deuteron by using the AV8' nucleon-nucleon interaction.   The wave function of the deuteron is written as
\begin{eqnarray}
\Psi_d &=& u(r) [Y_0(\hat r) \otimes \chi_1(\sigma_1 \sigma_2)]_{1M}+w(r)[Y_2(\hat r) \otimes \chi_1(\sigma_1 \sigma_2)]_{1M}~.
\end{eqnarray}
The deuteron wave function is written by the $s$-wave and $d$-wave components.  The tensor interaction mixes these two components.  In Fig.~\ref{fig:deuteron} and Table \ref{tab:deuteron}, we show the deuteron properties for the wave functions and the various energy contributions and radii.  In the wave function, the $s$-wave component is dominant and shows the long-tail due to the weak binding of 2 MeV.  In the short-range part less than the 0.5 fm region, the $s$-wave function is largely reduced due to the short-range repulsion in the central interaction.  Looking at the $d$-wave component, its amplitude starts from the origin, because of the centrifugal barrier in the $L=2$ partial wave, which is absent in the $s$-wave case.  These features of the $s$-wave and $d$-wave components have to be expressed using the shell model framework in finite nuclei as the deuteron-like tensor correlation.  These features will be treated for finite nuclei by using the tensor optimized shell
  model (TOSM) and the unitary correlation operator method (UCOM) to be discussed later.

Among the expectation values of all the two-body interactions in the deuteron, the tensor interaction has the largest contribution of about $-17$ MeV.   This expectation value is four times of that of the central interaction.  This tensor interaction is the origin of the $d$-wave mixing in the wave function.  From these results, it is found that the dominant energy contribution comes from the coupling of the $s$- and $d$-wave components by changing the relative orbital angular momentum by 2, $\Delta L=2$.  As for the radius of the deuteron, we decompose the radius into the $s$-wave and $d$-wave components and normalize them by using the corresponding amplitudes.  It is interesting to see the size difference between two angular components, where the $d$-wave size is much smaller than the $s$-wave size.  This compact $d$-wave structure produces high momentum component caused by the tensor interaction, namely the pion exchange effect.  Hence, we can learn the role and the properties of the tensor interaction in finite nuclei from the deuteron. That is, the tensor interaction creates the relative $d$-wave component in the wave function, which is spatially compact and involves high momentum components.   These features should appear in finite nuclei as the deuteron-like tensor correlation.  We shall treat this correlation in terms of the tensor optimized shell model (TOSM) in finite nuclei. 

As for the di-neutron correlation, we have a moderate intermediate attraction with a short range repulsion as shown in Fig.~\ref{fig:av8}.  There is no tensor interaction and the relative motion is completely described by the central interaction.  We are aware that there is no bound state in the $^1 S_0$ channel, but that the scattering length is  negatively very large $a= -18.5 \pm 0.4$ fm \cite{Te87}.  This negatively large scattering length indicates that the di-neutron system is close to develop a bound state.  Hence, for a system like $^{11}$Li, we expect a strong di-neutron clustering phenomenon in the halo region.  For the quantitative account we ought to use the $NN$ interaction for this phenomenon.

\subsection{Wave functions for $^{9,10}$Li and $^{11}$Li}

We write the wave functions of the Li isotopes in order to understand the standard shell model state, the di-neutron clustering and the deuteron-like tensor correlation.  It is illustrative to start writing the $^9$Li wave function.
\begin{eqnarray}
|^9 Li \ket &=&C_1 |(s_{1/2})_\pi^2(s_{1/2})_\nu^2(p_{3/2})_\pi(p_{3/2})_\nu^4 \ket_{J=3/2}
\label{eq:9Li}
\nn
&+&C_2 |(s_{1/2})_\pi^2(s_{1/2})_\nu^2(p_{3/2})_\pi(p_{3/2})_{\nu J=0}^2(p_{1/2})^2_{\nu J=0}\ket_{J=3/2} \nn
&+&C_3  |[(s_{1/2})_\pi(s_{1/2})_\nu]_{J=1}(p_{3/2})_\pi(p_{3/2})_\nu^4[(p_{1/2})_\pi(p_{1/2})_\nu]_{J=1}\ket_{J=3/2} \nn
&+&...
\end{eqnarray}
We have written here only the dominant components explicitly where $\pi$ and $\nu$ for each configuration denote proton and neutron, respectively.  The term with the amplitude $C_1$ corresponds to the standard shell model state.  The term with the amplitude $C_2$ corresponds to the main component of the two neutron pairing states, where a two-neutron pair couples to $J^\pi=0^+$.  The term with the amplitude $C_3$ corresponds to the main component of the deuteron-like tensor correlation states, where a proton-neutron pair couples to $J^\pi=1^+$.

The di-neutron clustering correlation, which is associated with the $C_2$ amplitude component, should involve further particle states in $sd$ and higher shells.  As for $^9$Li, the di-neutron clustering correlation provides a similar structure as the BCS state due to the fact that the nuclear density of the surface neutrons is ordinary as expected from the standard size neutron separation energies listed in Table.1.  With the increase of the neutron number, the nuclear density of the surface neutrons becomes very small and hence the di-neutron clustering correlation should show up.   This change of the di-neutron clustering correlation due to the nuclear density is related with the BCS-BEC crossover.  On the other hand, the deuteron-like tensor correlation, which is associated with the $C_3$ amplitude component, needs excitation of a proton-neutron pair with $J^\pi=1^+$ from occupied states to unoccupied states.  We have to include each particle state of the proton-neutron pair up to very high angular momentum state.

We write the $^{10}$Li wave function in terms of the $^9$Li wave function.
\begin{eqnarray}
|^{10}Li\ket &=& {\cal A}[|^9 Li\ket \times |\chi_n\ket]
\end{eqnarray}
The additional neutron may enter the $p_{1/2}$ orbit in the shell model state.  The addition of one more neutron to the pair correlated state has an effect to weaken the pairing correlation due to the blocking effect of the neutron.  The addition of one more neutron to the deuteron-like configuration is very interesting, since the additional neutron may go into the $p_{1/2}$ orbit or into the $s_{1/2}$ orbit in the shell model state.  If the last neutron goes into the $p_{1/2}$ orbit, the deuteron-like correlation is weakened by the additional neutron.  Instead, if the last neutron goes into the $s_{1/2}$ orbit, the deuteron-like correlation is not weakened, because 2p-2h states with the use of $s_{1/2}$ orbit are not important for the deuteron-like correlation.  Hence, there should appear the competition of the neutron $s_{1/2}$ and $p_{1/2}$ configurations in $^{10}$Li.

We write the $^{11}$Li wave function in terms of $^9$Li wave function.
\begin{eqnarray}
|^{11}Li\ket &=& {\cal A}[|^9 Li\ket \times |\chi_{nn}\ket]
\end{eqnarray}
In this case, there is an important physics to be added in addition to all the interesting phenomena in $^{10}$Li.  As for the pair correlated state, two neutrons block the pair correlated state by entering in the $p_{1/2}$ orbit.  More interesting is the case of the two neutrons going into the higher shell orbits.  In this case, the two neutrons stay in a low density region far from the $^9$Li core and hence the di-neutron clustering phenomenon is expected.  How large is the attraction due to the di-neutron clustering effect for the $^{11}$Li binding energy needs full account of all the effects.  As for the $C_3$ component, two neutrons going into the $p_{1/2}$ orbit generate a strong blocking effect of the deuteron-like correlation and this configuration is disfavored by the tensor interaction.  On the other hand, when the two neutrons go into the $s_{1/2}$ orbit, the deuteron-like correlation is not disturbed and therefore this configuration is favored.  Hence, as the consequence of these di-neutron clustering correlation and the deuteron-like tensor correlation, these two correlations cooperate to wash out the N=8 magic structure and provide the interesting halo phenomenon in $^{11}$Li.

\section{Di-neutron clustering and the hybrid-$VT$ model}
\label{sec:2}

\begin{figure}[t]
\centering
\includegraphics[width=10.0cm]{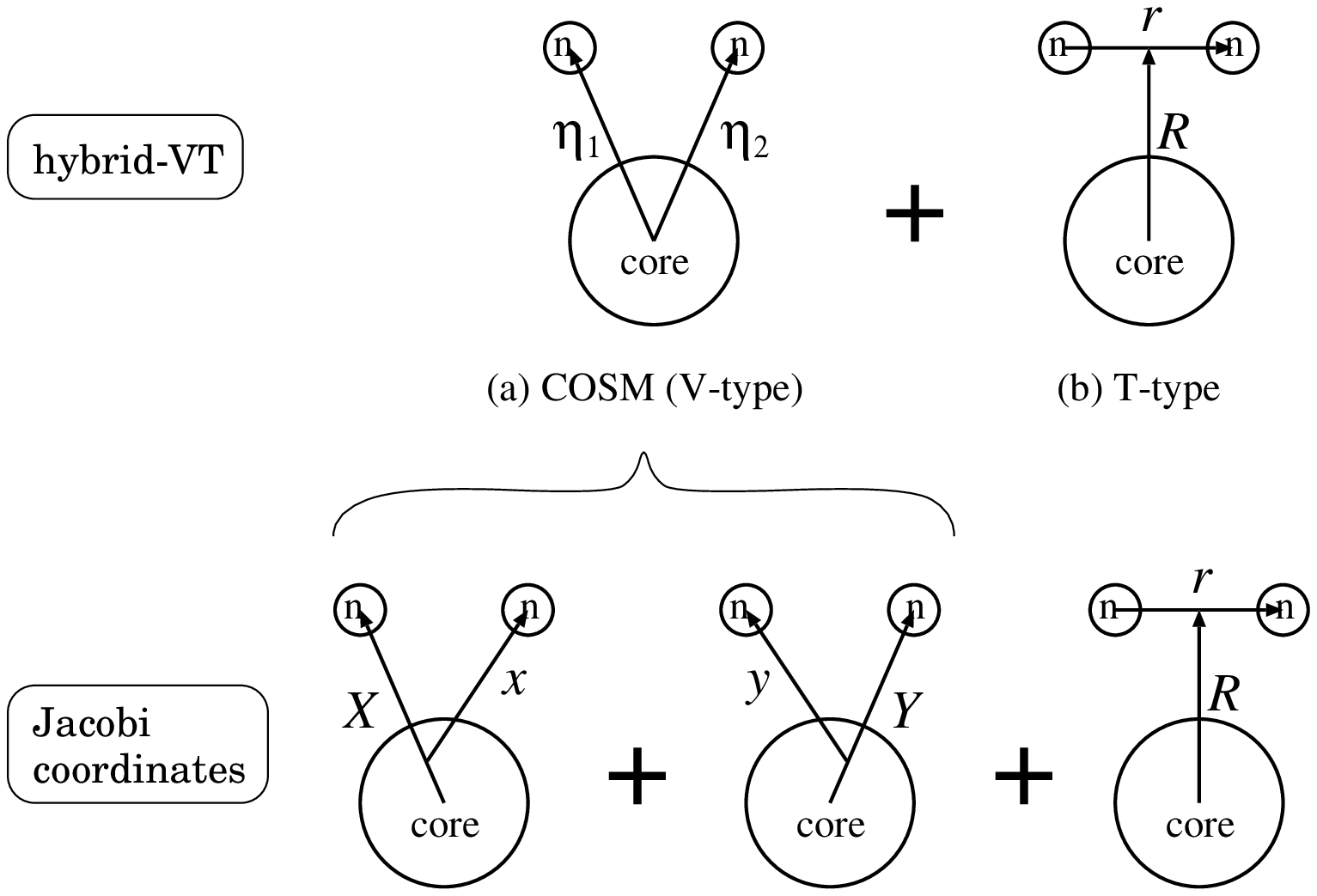}
\caption{The coordinates of the hybrid-$VT$ model and their relation with those of the Jacobi coordinates.  The $V$-type coordinates in (a) and the $T$-type coordinates in (b) are shown in the upper figures and the corresponding ones in the Jacobi coordinates are shown in the lower figures.} 
\label{fig:VT2}
\end{figure}

In $^{11}$Li \cite{Ta85a} and $^6$He \cite{Ta85b}, abnormally large matter radii were observed experimentally.   This phenomenon was interpreted as a result of the halo structure, where two valence neutrons are spatially extended around the core nucleus due to their weak binding.  It is important to investigate the dynamics of the motion of valence neutrons for the understanding of the halo structure.  In such a situation, it is necessary to develop a theoretical method to handle spatially extended structure of the Borromean system consisting of core nucleus and two neutrons.
For this purpose, we have developed the hybrid-$VT$ model as the most suitable model to describe the halo structure.  In the hybrid-$VT$ model for two-neutron halo nuclei, shown in Fig.~\ref{fig:VT2}, the mean field nature of each valence neutron can be described in the $V$-type basis states (the cluster orbital shell model (COSM), Fig.~\ref{fig:VT2} (a)). Further, the explicit neutron-neutron correlation is treated in the $T$-type (Fig. \ref{fig:VT2} (b)) basis states.  Here, the basic idea of this hybrid-$VT$ model is presented, and in the next sub-section we explain the formulation of the hybrid-$VT$ model in detail. 
 
In the weak binding system it is necessary to consider the large spatial extension of single particle wave functions of valence neutrons.  This situation corresponds to the coupling of the valence neutrons to continuum states.  Suzuki and Ikeda proposed the cluster orbital shell model (COSM, $V$-type) \cite{Su88}, and applied to $^6$He and $^{11}$Li \cite{Su90,To90a,To90b,Su90b,Su91,Ik92}.  The COSM is one extension of the shell model, in which the spatially extended character of valence neutrons can be treated.   The $V$-type coordinates in the COSM are the suitable coordinates to express the mean field property of the valence neutrons, and therefore can express the shell model properties of $^{5,6}$He and $^{10,11}$Li most effectively.   However, from the analysis of neutron-rich nuclei with the COSM, it was shown that the binding energies of the Borromean nuclei can not be described quantitatively.  
Furthermore, it was found that in a weakly bound system, the $n$-$n$ clustering correlation, namely the di-neutron clustering correlation, becomes important to provide an extra binding energy.  
This is characterized by the participation of many $J^\pi=0^+$ pair configurations with large single particle orbital angular momenta \cite{Ao95a,Myo02}.

Hence, it is important to include the physical effect of di-neutron clustering correlation explicitly in two-neutron halo nuclei, and the $T$-type wave function is suitable for this purpose \cite{To90a,Ik92,Ao95a,Mu98}. 
We combine the $T$-type basis functions with the $V$-type ones as shown in Fig.~\ref{fig:VT2}.  It has been shown that this hybrid-$VT$ model describes the di-neutron clustering correlation with full convergence of the binding energy and radius as will be discussed in the following sub-sections.  Hence, the hybrid-$VT$ model wave function involves two-kinds of physical correlations as the mean-field and di-neutron clustering correlations, in the two-neutron halo nuclei.  The hybrid-$VT$ model is also a special case of few-body systems; a three-body system is generally described by using the Jacobi-coordinates \cite{Fun94} as shown in Fig.~\ref{fig:VT2}.   In the case of core+$n$+$n$, the core nucleus has a large mass in comparison with that of valence neutrons.   Therefore, as shown in Fig.~\ref{fig:VT2}, the $V$-type coordinates (\vec{\eta}$_1$,\vec{\eta}$_2$)  correspond to the symmetric $Y$-type Jacobi coordinates, 
(\mbox{\boldmath$X$},\mbox{\boldmath$x$}) and (\mbox{\boldmath$Y$},\mbox{\boldmath$y$}).  

In this section, we explain the construction of the COSM and the hybrid-$VT$ model \cite{Ao95a,Myo02,Mu98,Myo01,Myo03,Ao95b,Ka95,Ao97,Ao97a,Ao98a,Ao99a,Ao01,Ao02} and its application to the three-body systems: $^6$He and $^{11}$Li.  The subject of the di-neutron clustering is discussed in terms of BCS-BEC cross over in recent literature \cite{hagino,matsuo,sun}.

\subsection{Formulation of hybrid-$VT$ model}
\label{subsec:2-1}

We consider the hybrid-$VT$ model for a spatially extended core+$n$+$n$ system.   Here, we derive the three-body Hamiltonian from the $A$-nucleon system.  This consideration is useful when we extend this model to include the core excitation in the halo nuclear system.   The $A$-body Hamiltonian is given as 
\begin{eqnarray}
H &=& \sum_{i=1}^{A}t_i - T_{cm} + \sum_{i>j}^A v_{ij}=T+V,
\end{eqnarray}
where $T =\sum t_i - T_{cm}$ is the kinetic energy operator of the system after removing the center-of-mass motion ($T_{cm}$) and $V=\sum v_{ij}$ is the two-body potential energy.  We decompose the Hamiltonian of an $A$-nucleon system into a core part with $A_c$ nucleons and $N(=A-A_c)$ valence neutrons.  The relative coordinates between the core and the valence nucleons are given as $\vec{\eta}_i=\vec{r}_i-\frac{1}{A_c}\sum_{i=1}^{A_c}\vec{r}_i$, as  shown in Fig. $\ref{fig:VT2}$.  
The kinetic energy term $T$ is rewritten as
\begin{eqnarray}
T &=& \sum_{i=1}^{A}t_i - T_{cm}\\
&=& T_c
+ \sum_{i=1}^N \frac{\vec{p}_i^2}{2\mu}
+ \sum_{i<j}^N \frac{\vec{p}_i\cdot\vec{p}_j}{(A_c+1)\mu},
\end{eqnarray}
Here, $T_c=\sum_{i=1}^{A_c}t_i - T^{c}_{cm}$ is the kinetic energy with $T_{cm}^{c}$ being the center of mass motion of the core nucleus.  The operator $\vec{p}$=$-i\hbar\nabla_{\vec{\eta}}$ is the momentum conjugate to $\vc{\eta}$ and $\mu=A_c/(A_c+1)m$ is the reduced mass between the core and a single neutron.  The term of $\vec{p}_i\cdot\vec{p}_j$ is the recoil motion from the center of mass system.   The potential term is similarly decomposed as
\begin{eqnarray}
    V 
&=& \sum_{i<j}^{A_c} v_{ij} + \sum_{i=1}^{N}\sum_{j=1}^{A_c} v_{ij} 
  + \sum_{i<j}^N v_{ij},\\
&=& V_c + \sum_{i=1}^N V_i + \sum_{i<j}^N v_{ij},
\end{eqnarray}
where the mean field potential $V_i$ for each valence neutron is given as
\begin{eqnarray}
    V_i 
&=& \sum_{j=1}^{A_c} v_{ij}.
\end{eqnarray}
Here, $V_c$ are the potential term of the core nucleus, and the Hamiltonian is rewritten as
\begin{eqnarray}
  H 
&=&  \Biggl[T_c + \sum_{i=1}^N \frac{\vec{p}_i^2}{2\mu}+
     \sum_{i<j}^N \frac{\vec{p}_i\cdot\vec{p}_j}{(A_c+1)\mu}\Biggr]+
     \Biggl[V_c + \sum_{i=1}^N V_i + \sum_{i<j}^N v_{ij}\Bigr],
     \\
&=&  H_c + \sum_{i=1}^N \Bigl[\frac{\vec{p}_i^2}{2\mu}+V_i\Bigr]
      + \sum_{i<j}^N \Bigl[ v_{ij}
+\frac{\vec{p}_i\cdot\vec{p}_j}{(A_c+1)\mu}\Bigr],
\label{eq:COSM}
\end{eqnarray}
where the first term $H_c=T_c+V_c$ is the Hamiltonian of the core and the second and third terms are for valence neutrons.
The second term is the single particle Hamiltonian for the relative motion between the single neutron and the core.  
This defines the orbitals for the valence nucleons.  
The third term is the two-body operator between the valence neutrons, which produces the coupling between valence neutrons, such as the di-neutron correlation.

We start with the core part $\Phi(A_c)$ and write the Schr\"odinger equation as
\begin{eqnarray}
\Phi(A_c)&=&\sum_{\alpha} C_\alpha\phi_\alpha(A_c),
\label{eq:corewf}
\\
H_c\Phi(A_c)&=&E_c\Phi(A_c)~,
\label{eq:core}
\end{eqnarray}
where the index $\alpha$ is the label to distinguish various configurations of the core nucleus and the amplitudes $C_\alpha$ are those used in Eq.~(\ref{eq:9Li}). They are determined by the variational equations obtained by the energy minimization of $E_c$.  We employ the shell model like basis wave function for $\phi_\alpha(A_c)$.

The wave function of the two valence neutrons $\chi(nn)$ in the hybrid-$VT$ model of two neutron halo nuclei ($N=2$) is expressed 
as the superposition of the COSM ($V$-type) and $T$-type wave functions as
\begin{eqnarray}
	\chi(nn)
&=&     \chi_{V}(\vc{\xi}_V)+\chi_{T}(\vc{\xi}_T).
\end{eqnarray}
Here, the coordinate sets $\vc{\xi}_V$ and $\vc{\xi}_T$ represent $V$-type and $T$-type ones, respectively, as shown in Fig.~\ref{fig:VT2}.
We take antisymmetrization between two neutrons, explicitly.
The radial components of the relative wave functions are expanded with a finite number of Gaussian functions centered at the origin with various length parameters \cite{Ka88}. 

In the hybrid-$VT$ model, the total wave function of the $A$-nucleon system and the corresponding Schr\"odinger equation are given as
\begin{eqnarray}
\Psi(A)&=&{\cal A} \left \{\sum_\alpha \phi_\alpha(A_c)\chi_\alpha(nn) \right \},
\label{eq:halo}
\\
H\Psi(A)&=&E\Psi(A)~,
\label{eq:halo2}
\end{eqnarray}
where the total Hamiltonian $H$ is given in Eq.~(\ref{eq:COSM}).
We omit the angular momentum coupling between the core nucleus and the valence neutrons for simplicity.  The operator ${\cal A}$ is the antisymmetrizer between core nucleons and valence neutrons.
The mixing amplitudes of the core configurations $\alpha$ are included in the wave functions $\chi_\alpha(nn)$.  Eq.~(\ref{eq:halo}) is useful to understand the asymptotic condition of the wave function, in which some of the valence neutrons are located far away from the core, such as the tail part of halo structure and the scattering states.
This will be discussed later in the numerical results of $^{11}$Li. 

In order to solve Eq.~(\ref{eq:halo2}), we employ the orthogonality condition model (OCM) \cite{Sa77,Ya86,Ao06} instead of the resonating group method (RGM) \cite{Ho77}.  In the OCM, the antisymmetrizer ${\cal A}$ between core nucleons and valence nucleons is replaced by introducing the projection operator to remove the Pauli forbidden states from the relative motion of the valence neutrons.  The projection is expressed by introducing the following one-body term in the original Hamiltonian in Eq.~(\ref{eq:COSM}).
\begin{eqnarray}
v^{\rm PF}_i &=& \lambda \sum_{k}^{N_i}|\phi^{{\rm PF }}_k\rangle \langle \phi^{{\rm PF }}_k|~,
\end{eqnarray}
where the indices $i$ and $k$ are the labels representing each valence neutron and each Pauli-forbidden state for one valence neutron.  $N_i$ is the number of Pauli-forbidden states for one valence neutron. We take a sufficiently large value for $\lambda$ in the numerical calculation.

In the calculation of the matrix element of the Hamiltonian in Eq.~(\ref{eq:COSM}), we fold the Hamiltonian by using the wave function of the core nucleus.  In the coupled channel OCM with the hybrid-$VT$ model, we obtain the following equation for the valence neutrons $\chi_\alpha(nn)$,
\begin{eqnarray}
&&\sum_{\beta} \Biggl[
H^c_{\alpha\beta}+ \sum_{i=1}^2 \Bigl\{\frac{\vec{p}_i^2}{2\mu}\delta_{\alpha\beta} + V^F_{i,\alpha\beta}
+v^{\rm PF}_i \delta_{\alpha\beta} \Bigr\} 
\nonumber\\
&&\hspace{2cm}+ \sum_{i<j}^2\Bigl\{ v_{ij} +\frac{\vec{p}_i\cdot\vec{p}_j}{(A_c+1)\mu}
\Bigr\}\delta_{\alpha\beta} \Biggr]\chi_\beta(nn) 
= E\chi_\alpha(nn),
\label{Eq-COSM}
\end{eqnarray}
where
\begin{eqnarray}
H^c_{\alpha\beta}&=& \langle \phi_\alpha(A_c)|H_c|\phi_\beta(A_c) \rangle,
\\
V^F_{i,\alpha\beta}&=& \langle \phi_\alpha(A_c)|V_i|\phi_\beta(A_c)\rangle
~=~\langle \phi_\alpha(A_c)|\sum_{j=1}^{A_c}v_{ij}|\phi_\beta(A_c)\rangle.
\end{eqnarray}

We explain here the Gaussian expansion method to describe the wave functions of valence neutrons $\chi_\alpha(nn)$ in the hybrid-$VT$ model. 
The spatial part of the basis functions for one relative motion $\vec{r}$ is given by the following Gaussian wave functions,
\begin{eqnarray}
    \psi^{b}_{l}(\vec{r})
&=& N_l(b)\ r^l\ \exp(-\frac{r^2}{2b^2}) \ Y_{l}(\hat{\vec{r}})~,
    \label{eq:Gauss}
    \\
    N_l(b)
&=&  \left[  \frac{2b^{-(2l+3)} }{ \Gamma(l+3/2)}\right]^{\frac12},
\end{eqnarray}
where $l$ is the orbital angular momentum. 
The set of the length parameter $b$ is usually chosen in geometric progression \cite{hiyama03}.  In $V$- and $T$-type basis functions, we commonly use these basis states.  We expand each relative motion of the hybrid-$VT$ model with a finite number of the above basis functions.  The Gaussian expansion method is able to describe the halo structure very nicely.

For the coupling with intrinsic spin of neutrons,
in the COSM ($V$-type), we adopt the $j$-$j$ coupling scheme in a sense of the shell model.  This representation is suitable to express the motion of each valence neutron in the mean field potential provided by the core nucleus.  In the $T$-type basis function, we take the $L$-$S$ coupling scheme.  This is because the di-neutron pair is considered to have the dominant component of the $0^+$ state with spin singlet state.  In the $T$-type basis, we directly take into account the $^1S_0$ component of the two neutrons.  Actually, the analysis of $^6$He provides more than $80\%$ in the spin singlet states.  In the COSM ($V$-type basis), its basis wave function corresponding the configuration of the core nucleus $\phi_\alpha$ is given as
\begin{eqnarray}
    \chi^J_{\alpha,V}(nn)
&=& \sum_{p} C^p_{\alpha,V}\ {\cal A}_{12}\left[[\psi^{b_1}_{l_1}(\eta_1),\chi^\sigma_{1/2}]_{j_1},[\psi^{b_2}_{l_2}(\eta_1),\chi^\sigma_{1/2}]_{j_2}\right]_J
\label{eq:COSM2}
    \\
    p
&=& \{b_1,b_2,l_1,l_2,j_1,j_2,J\}~,
\nonumber
\end{eqnarray}
where ${\cal A}_{12}$ is an antisymmetrizer between two valence neutrons, and $C^{p}_{\alpha,V}$ are variational coefficients for the basis set. 

The $T$-type basis function is similarly described as
\begin{eqnarray}
   \chi^J_{\alpha,T}(nn) 
&=&\sum_{q} C^q_{\alpha,T}\
   {\cal A}_{12} \left[ \left[\psi^{b_r}_{l}(\vec{r}),  \psi^{b_R}_{L}(\vec{R}) \right]_I \chi_S \right]_J,
   \label{eq:T-basis}
\\
   q&=& \{b_r,b_R,l,L,S,J\},~\qquad \chi_S=[\chi^\sigma_{1/2},\chi^\sigma_{1/2}]_S ~.
\nonumber
\end{eqnarray}
In the calculation of the hybrid-$VT$ model, we take various sets of orbital angular momenta and spins until we reach the convergence of the solutions.  The variation of the total energy $E$ with respect to the total wave function $\Psi(A)$ is given by
\begin{eqnarray}
\delta\frac{\bra\Psi|H|\Psi\ket}{\bra\Psi|\Psi\ket}&=&0\ ,
\end{eqnarray}
which leads to the following equations:
\begin{eqnarray}
    \frac{\del \bra\Psi| H - E |\Psi \ket} {\del C^p_{\alpha,V}}
&=& 0\ ,\qquad
    \frac{\del \bra\Psi| H - E |\Psi \ket} {\del C^q_{\alpha,T}}
=   0\ .
   \label{eq:vari}
\end{eqnarray}

\begin{figure}[b]
\centering
\includegraphics[width=8.5cm]{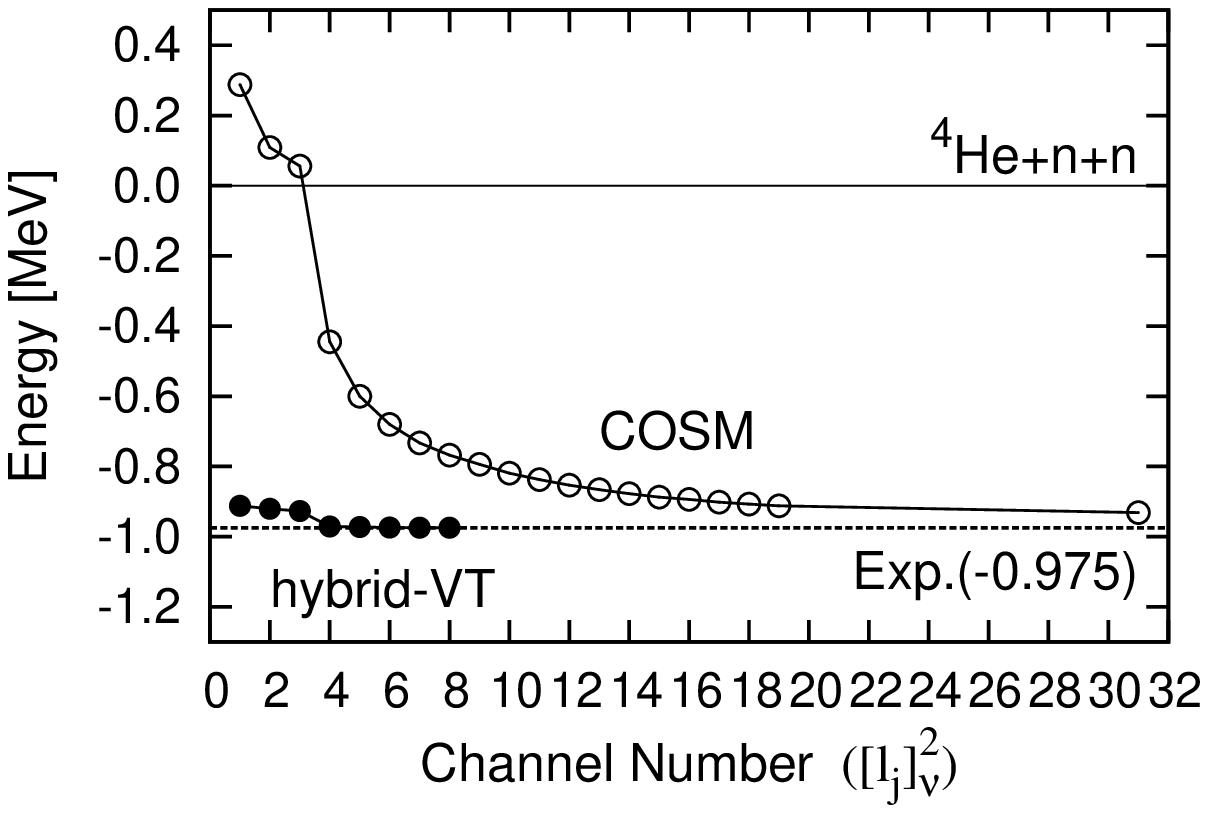}
\caption{The binding energy of the $^6$He ground state as a function of the channel number \cite{Ao95a}.  The open circles are the results of the COSM model and the solid circles are those of the hybrid-$VT$ model.}
\label{fig:conv_ene1}
\end{figure}

\subsection{Application of the hybrid-VT model to $^6$He}

We discuss the results of the hybrid-$VT$ model for the two-neutron halo nuclei $^6$He as $^4$He+$n$+$n$ and $^{11}$Li as $^9$Li+$n$+$n$.  An extension to a core plus many valence neutrons (e.g. $^7$He=$^4$He+$n$+$n$+$n$) is straightforward \cite{masui06,myo07c}. 
For $^6$He case, we take a single configuration for the $^4$He core.  The Hamiltonian in Eq.~(\ref{Eq-COSM}) can be written as
\begin{eqnarray}
H&=&H_c+
 \sum_{i=1}^2 \Bigl\{\frac{\vec{p}_i^2}{2\mu}+V^F_{cn}(\vc{\eta}_i)+v^{\rm PF}_i \Bigr\}
+  v_{nn}+\frac{\vec{p}_1\cdot\vec{p}_2}{(A_c+1)\mu}.
\label{Eq-Ham}
\end{eqnarray}
Here, $H_c$ can be replaced by the observed energy of $^4$He ($-28.3$ MeV) and we can discuss only the relative motion of valence two neutrons in $^6$He.  For the $^4$He-$n$ potential $V^F_{cn}$ we adopt the microscopic KKNN potential \cite{Ao95a,kanada79}, which reproduces the observed phase shift between $^4$He and $n$.  The Minnesota interaction \cite{Ta78} is used for $v_{nn}$ between two valence neutrons, where the exchange mixture $u$ is chosen to be 0.95.  
These choices are the same as those in Refs.~\cite{Mu98},~\cite{Ao97}~and~\cite{Ka99}.

It is important to understand the model performance for $^6$He as a simple system for the description of more complicated systems as $^{11}$Li.  The binding energy of the $^6$He ground state is shown as a function of the channel number in Fig.~\ref{fig:conv_ene1}. 
The calculated results in the COSM and the hybrid-$VT$ model are shown by the open and solid circles, respectively. 
In both cases, the convergence was achieved and it is found that the hybrid-$VT$ model converges much faster than the COSM.  
In the hybrid-$VT$ model, it is sufficient to take shell model states up to the $d_{5/2}$-shell orbit (Channel number is five) using the $V$-type coordinates 
and add the di-neutron channel of $l=L=0$ states in Eq.~(\ref{eq:T-basis}) using the $T$-type coordinates.  Thus, the di-neutron correlation is very important for the ground state of $^4$He.   The hybrid-$VT$ model is an efficient framework to treat the di-neutron correlation in the shell model basis.

\begin{figure}[t]
\centering
\includegraphics[width=7.0cm,clip]{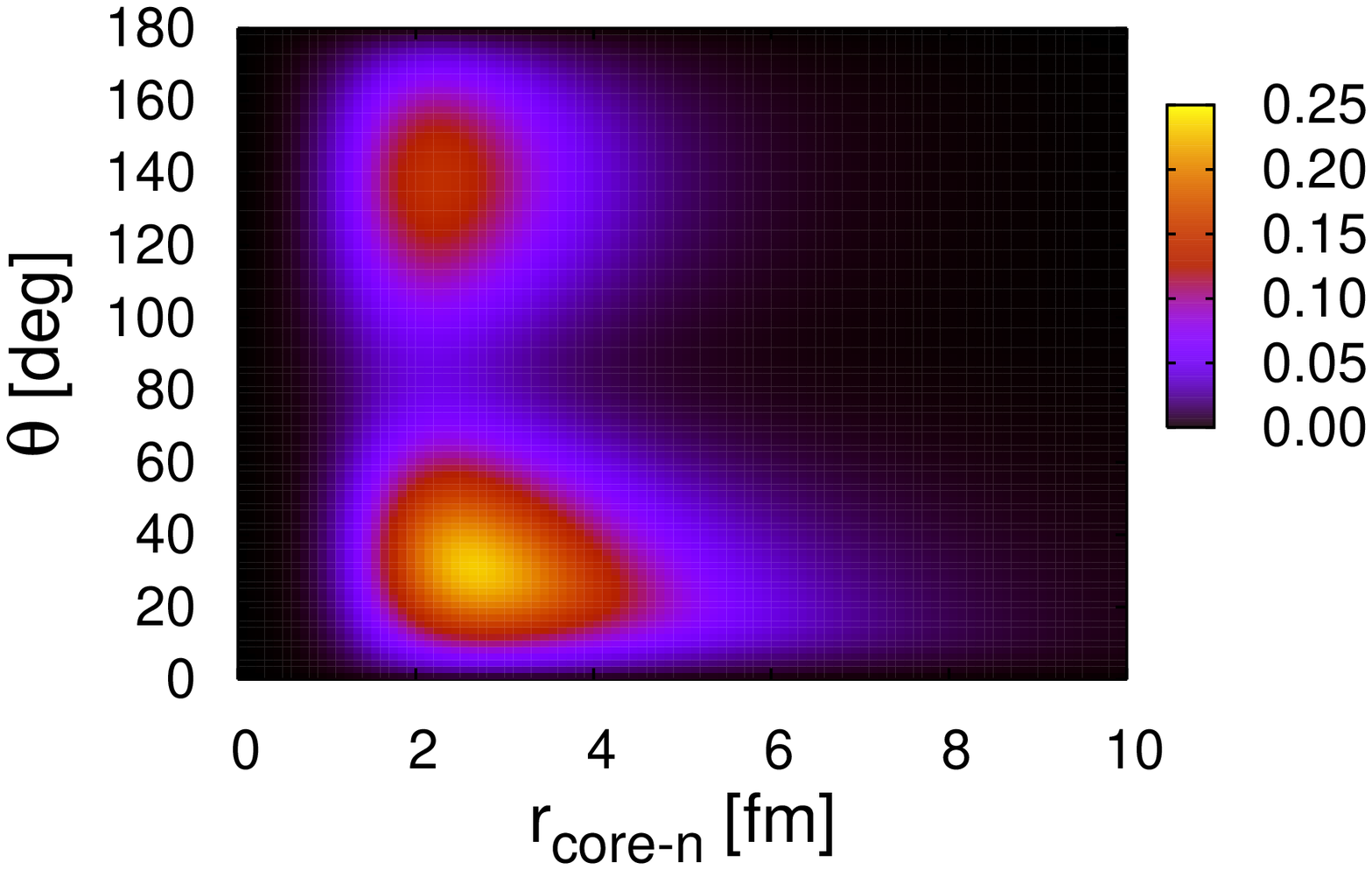}
\caption{The two neutron correlation density $\rho_{nn}(r_{{\rm core}-n},\theta)$ in $^{6}$He in the $r$ and $\theta$ plane \cite{kikuchi09}.  Here, $r_{core-n}$ denotes the relative distance between the core and one of the two neutrons and $\theta$ the opening angle between the two neutrons.}
\label{fig:density6}
\end{figure}

The radii of $^6$He are converged as 2.46 fm and 2.06 fm for matter and charge rms radii, respectively.  These values agree with the recent observations \cite{Ta88b,mueller07}.  The core-$n$ and $n$-$n$ mean distances are obtained as 3.42 fm and 4.90 fm, respectively.
In addition to the radius, the spatial correlations of the halo neutrons in $^6$He are interesting \cite{Es92,hagino,Zh93,Ni01} and the calculated results are shown in Fig.~\ref{fig:density6}.   We show the density distribution of halo neutrons $\rho_{nn}(r,\theta)$ in $^6$He as a contour in the plane of $^4$He-$n$ distance $r$ and the opening angle between two neutrons $\theta$ with the following definition \cite{kikuchi09,myo08}; 
\begin{eqnarray}
    \rho_{nn}(r,\theta)
&=& \int_0^\infty dr^\prime  \rho_{nn}(r,r^\prime,\theta)
    \\
    \rho_{nn}(r,r^\prime,\theta)
&=& 8\pi^2r^2 {r^\prime}^2 \sin\theta\
\bra \Psi^J(^{6}{\rm He},r,r',\theta) | \delta(r-r'')\delta(r'-r''')
    \nonumber
    \\
&\times& \delta(\theta-\theta')| \Psi^J(^{6}{\rm He},r'',r''',\theta')\ket,
    \label{density}
\end{eqnarray}
where the total wave function of $^{6}$He has three variables, the $^4$He-$n$ distances with $r$ and $r^\prime$ for each neutron and their opening angle $\theta$ in Eq.~(\ref{density}).
We integrate out only the variables of the ket part.  
It is confirmed that the di-neutron type configuration (a large $r$ and a small $\theta$) gives a maximum value of the density, 
although the density of neutrons is widely distributed.  
There is another component of the cigar type configuration (a small $r$ and a large $\theta$), which coexists with the di-neutron type configuration.  The characteristics of these two structures in the density distribution come from the $p_{3/2}^2$ configuration of two neutrons in $^6$He, which is the dominant component by 90.2 \% in the ground state wave function. 
The mixing of the higher orbital configurations makes an spatial extension of the distribution and enhances the di-neutron component,
such as $p_{1/2}^2$ with 4.3\%,  $1s_{1/2}^2$ with 1.2\%, $d_{5/2}^2$ with 2.6\% and  $d_{3/2}^2$ with 0.9\%. 
The results indicate that the $j$-$j$ coupling scheme is well established in \nuc{6}{He}.

\subsection{Hybrid-VT model on di-neutron clustering in $^{11}$Li}
\label{subsec:3-3}

We consider the three-body problem of \nuc{9}{Li}+$n$+$n$ using
the orthogonality condition model (OCM).  
The Hamiltonian consists of the similar form as given in Eq.~(\ref{Eq-Ham}) for $^6$He.
The difference from the $^6$He model is the configuration mixing for the $^9$Li core nucleus,
because of the small neutron separation energy of $^9$Li in comparison with $^4$He.
This means that we take into account the core excitation in $^{11}$Li.
In this section, we first take into account the neutron $0^+$ pairing correlation of the $^9$Li core 
and examine this effect on the structures of $^{11}$Li and $^{10}$Li.  
Later, we include the tensor correlation in $^9$Li.

Before showing the numerical results, we generally formulate the coupled \nuc{9}{Li}+$n$+$n$ model of $^{11}$Li, 
in which the configuration mixing is performed for the $^9$Li core. This framework is straightforward to apply when the tensor correlation is included in the core part, later.
In the coupled hybrid-$VT$ model of \nuc{9}{Li}+$n$+$n$, we consider the Pauli forbidden (PF) states in the \nuc{9}{Li}-$n$ relative motion \cite{Ku86}.  
In this model, PF states removed from the the relative motion depend on the configuration of \nuc{9}{Li}, namely the orbits occupied by neutrons in the \nuc{9}{Li} core.
The main configurations are given in Eq.~(\ref{eq:9Li}) and the amplitudes are written as $C_1$, $C_2$ and $C_3$. 
The PF states corresponding to those wave functions are given as
\begin{eqnarray}
	\phi_{PF}
&=&	\left\{
	\renewcommand{\arraystretch}{1.2}
	\begin{array}{ll}
	0s_{1/2},~0p_{3/2}                      &~~\mbox{for}\quad C_1\\
	0s_{1/2},~0p_{3/2},~(0p_{1/2})_{\nu\nu} &~~\mbox{for}\quad C_2\\
	0s_{1/2},~0p_{3/2},~(0p_{1/2})_{\pi\nu} &~~\mbox{for}\quad C_3~.
	\end{array}
	\right.
        \label{eq:PF}
\end{eqnarray}
In case of $C_2$ the PF $p_{1/2}$ orbit is used by the pairing state and indicated as $(p_{1/2})_{\nu\nu}$, while in case of $C_3$ the PF $p_{1/2}$ orbit is used by the deuteron-like tensor correlation and indicated as $(p_{1/2})_{\pi\nu}$.   From Eq.~(\ref{eq:halo}), the wave function of \nuc{11}{Li} is given as 
\begin{eqnarray}
	\Psi^J(^{11}{\rm Li})
&=&	\sum_\al^{N_\al}
        {\cal A}\left\{\, [\phi^{3/2^-}_\al, \chi^{j}_\al(nn)]^J\,\right\}.
	\label{eq:WF_11Li}
\end{eqnarray}
Here, $\chi^{j}_{\al}(nn)$ represents the wave functions of two valence neutrons, and $j$ and $J$ are the spin of two valence neutrons and the total spin of \nuc{11}{Li}, respectively. The three-body eigenstates are obtained by solving the eigenvalue problem for the coupled-channel Hamiltonian given in Eq.~(\ref{eq:COSM}).
\begin{eqnarray}
	H(^{11}{\rm Li})\Psi(^{11}{\rm Li})
&=&	E(^{11}{\rm Li})\Psi(^{11}{\rm Li})
\end{eqnarray}

We discuss the coupling between the \nuc{9}{Li} configurations $\phi^{3/2^-}_\al$ and the motion of valence neutrons.  
In \nuc{11}{Li}, the amplitudes $C_\al$ of each configuration of \nuc{9}{Li} in Eq.~(\ref{eq:corewf}) are determined variationally. 
Asymptotically, when the two valence neutrons are far away from \nuc{9}{Li}, the wave function of \nuc{11}{Li} becomes
\begin{eqnarray}
	\chi^j_\al(nn)
&~~	\mapleft{\eta_1,\eta_2\to\infty}~~&C_\alpha\cdot \chi^j(nn),
	\\
	\Phi^J(^{11}{\rm Li})
&~~	\mapleft{\eta_1,\eta_2\to\infty}~~&
	\left[
	\left(\sum_\al^{N_\al} C_\al \phi^{3/2^-}_\al \right),~
	\chi^j(nn)
	\right]^J.
	\label{eq:asympt}
\end{eqnarray}
The first equation implies that the asymptotic wave function of the two valence neutrons is decomposed into the internal amplitude $C_\alpha$ of the $^9$Li configuration and the relative wave function $\chi^j(nn)$, which is independent of the $^9$Li configuration.
This means that the coupling between the valence neutrons and \nuc{9}{Li} disappears.  As for the di-neutron wave function $\chi^j(nn)$, the correlation between the two neutrons disappears also at far distance, because the two neutrons do not form bound state in the free space.  The mixing amplitudes $\{C_\al\}$ of \nuc{9}{Li} in Eq.~(\ref{eq:asympt}) are the same as those of the isolated \nuc{9}{Li} in Eq.~(\ref{eq:corewf}).  
Contrastingly, when the two valence neutrons are close to the \nuc{9}{Li} core, 
the motions of the two valence neutrons dynamically couple to the configuration of \nuc{9}{Li} in order to satisfy the Pauli principle,
which changes the mixing amplitudes $\{C_\al\}$ in \nuc{9}{Li} from those of the isolated \nuc{9}{Li} core.

We now carry out the coupled-channel three-body calculation for \nuc{11}{Li}.   In the model, the $^9$Li-n interaction $V_{c n}$ is taken as a folding-type potential with the MHN interaction \cite{Ao06,Ka93,Fu80},
which is constructed from the $G$-matrix using the bare nucleon-nucleon interaction.  The folding potential for \nuc{9}{Li}-$n$ includes the coupling between intrinsic spins of the valence neutron and \nuc{9}{Li} ($3/2^-$).  This coupling produces splittings of the energy levels, for instance $1^+$--$2^+$ (for the $p_{1/2}$-neutron) and $1^-$--$2^-$ (for the $s_{1/2}$-neutron) in the \nuc{10}{Li} spectra.  

\begin{figure*}[b]
\centering
\includegraphics[width=8.5cm]{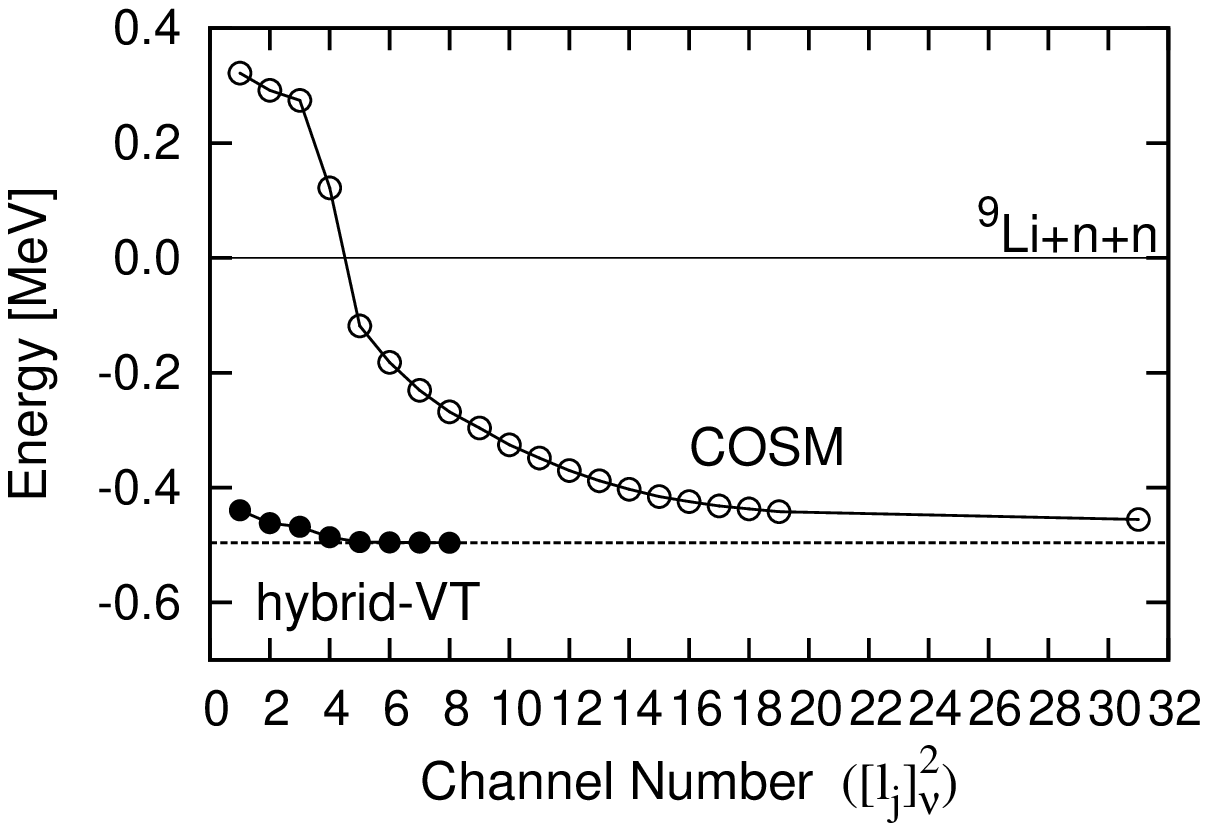}
\caption{Convergence of the \nuc{11}{Li} ground state energy with respect to the channel number in the COSM (open circles) and in the hybrid-$VT$ model (solid circles) \cite{Myo02}.  The dotted line represents energy to which the calculation converges ($-0.50$ MeV) in the present model setting.}
\label{fig:conv_ene}
\end{figure*}

The important points we wish to study in this calculation are whether the present model can solve the under-binding problem and describe the halo structure.  This is because the three-body model of $^{11}$Li by using only the inert core model of $^9$Li, which corresponds to the use of the $C_1$ term alone in Eq. (2), does not make a bound state \cite{Mu98}.  We also want to see how the pairing correlations act on the binding mechanism.  The results are shown in Fig.~\ref{fig:conv_ene}.  The binding energy of the \nuc{11}{Li} ground state measured from the three-body threshold is obtained as 0.5 MeV by considering the pairing correlation in $^9$Li.  The matter radius is obtained as 2.69 fm, which is smaller than the experimental value \cite{Ta88b,To97}.

We discuss the role of the pairing correlation between valence neutrons in $^{11}$Li.  
In Fig.~\ref{fig:conv_ene}, two kinds of the energy convergence of \nuc{11}{Li} are plotted as functions of the channel number of the 
$j^\pi=0^+$ pairing configuration for valence neutrons.  
One of them is the calculation employing only the COSM basis and the other is that with the hybrid-$VT$ basis.   In the calculation, we take the first channel as $(p_{1/2})^2$, and the order in which channels are added to the first one is $(s_{1/2})^2$,~$(p_{3/2})^2$,~$(d_{5/2})^2$,~$(d_{3/2})^2$, $\cdots$, $(l_j)^2$.  The maximum number of channel is 31, where the orbital angular momentum and the spin of one valence neutron are $l~=~15$ and $j~=~\frac{31}{2}$.  
We see rapid convergence of the energy in the hybrid-$VT$ model.  
This result indicates that the pairing correlation between valence neutrons is important to reproduce the weak binding state of \nuc{11}{Li}.  This result is similar to the $^6$He case shown in Fig.~(\ref{fig:conv_ene1}).  We comment here that the binding energy comes out to be $0.5$ MeV, which is larger than the experimental value of 0.3 MeV.  At the same time, the s-wave component is less than 10 \% as compared with the experimental value $\sim 50 \%$.  In the present analysis, we determine the $^{9}$Li-$n$ interaction to reproduce the $1^+$ state at 0.42 MeV and the virtual $s$-wave state just at the $^9$Li-n threshold energy in the $^{10}$Li spectrum.  In this case, two neutrons of $^{11}$Li are slightly overbound, which indicate that the pairing correlation of $^9$Li partially solves the problem of $^{10,11}$Li. This over-binding property together with other problems will be removed consistently by considering the deuteron-like tensor correlation, which pushes up the $(p_{1/2})^2$ state energetically close to the $(s_{1/2})^2$ state, as will be shown later.

\begin{figure}[t]
\centering
\includegraphics[width=10.0cm,clip]{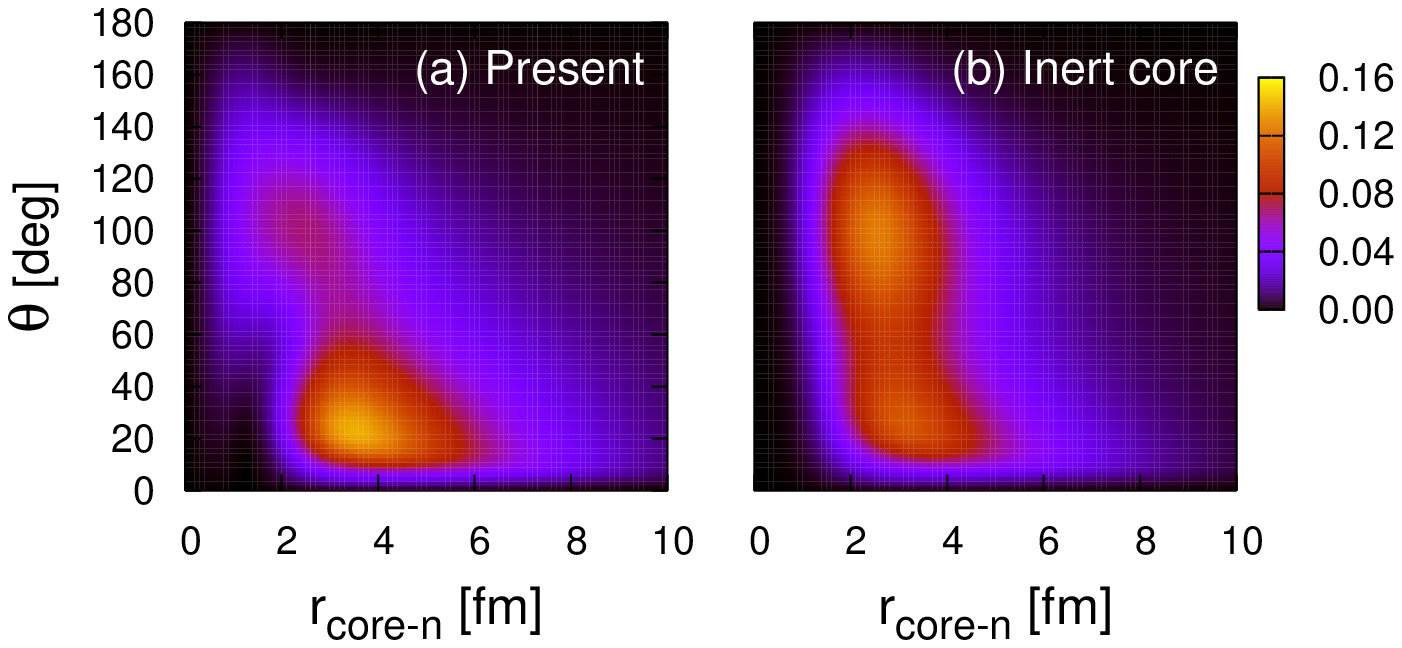}
\caption{Two neutron correlation density $\rho_{nn}(r_{{\rm core}-n},\theta)$ for $^{11}$Li \cite{myo08}.  The case (a) is the calculation with the TOSM of $^9$Li and (b) is the Inert Core case of $^9$Li, respectively.}
\label{fig:density}
\end{figure}

We also discuss the di-neutron correlation in $^{11}$Li.
The spatial correlations of the halo neutrons in $^{11}$Li are interesting \cite{Es92,hagino,Zh93,Ni01} and 
they are shown in Fig. \ref{fig:density}.  
We calculate the density distribution of halo neutrons $\rho_{nn}(r,\theta)$ in $^{11}$Li as functions of the $^9$Li-$n$ distance $r$ and the opening angle $\theta$ between two neutrons.  
In order to present the realistic case, which reproduce the large $s^2$ component of halo neutrons \cite{simon99},
we show here the result of calculations with the tensor optimized shell model (TOSM), which includes both the pairing and deuteron-like tensor correlations in $^9$Li \cite{myo08}.   This TOSM wave function contains the $s$-wave component in $^{11}$Li by a large amount 47\%.  
We will discuss the details of the TOSM in the following section.  In the TOSM case (a), it is confirmed that the di-neutron clustering configuration gives a maximum value of the density, although the density of neutrons is widely distributed.  Contrastingly, the Inert Core case (b) with a small $s^2$ component of 4\%, does not show much enhancement of the di-neutron clustering configuration and the cigar type configuration coexists with the di-neutron clustering configuration.  This feature of the case (b) is similar to $^6$He \cite{Zh93}.  These two results indicate the role of the $s^2$ component on the formation of the di-neutron clustering configuration as follows:   The $s^2$ component in $^{11}$Li increases the amplitude of the tail region of two neutrons far from $^9$Li, and these neutrons tend to come close to each other to gain the interaction energy between them.
As a result, the di-neutron clustering configuration is enhanced, although the spatial distribution of neutrons are still wide.
The spatial distribution of two neutrons also affects the opening angle $\theta$, where the TOSM case having large di-neutron component, 
shows a smaller $\theta$ value (65 deg.) than the Inert Core one (73 deg.).

\section{Continuum and resonance states in complex scaling method (CSM)}
\label{sec:3}

The halo nuclei have extremely small binding energies and it is very important to take into account the continuum and resonance states for quantitative account of the halo nuclei.  At the same time, neighboring nuclei have odd numbers of neutrons and often those nuclei do not have bound states.  Hence, in order to obtain precious informations on the neutron-core potentials, it is important to describe resonance states of the neighboring nuclei.  In this section, we would like to develop a powerful method of treating continuum and resonance states as if they are bound states.  This method is called the complex scaling method (CSM).  We can also apply the CSM for excitation functions of halo nuclei.  We would like to emphasize here that there has not been any method to treat three-body unbound systems.  In particular, there is a case where two-body system out of the three-body system may be in the resonance state.  We can treat these interesting possibilities in the CSM. 

\subsection{Formulation of CSM}
\label{subsec:3-1}

We explain the CSM, which describes resonances and non-resonant continuum states of a many-body system.  Hereafter, we refer to non-resonant continuum states as simply continuum states.  In the CSM, we transform every relative coordinates $\{\vec{r}_i\}$ of the system such as core+$n$+$n$ model, by the operator $U_\theta$ as
\begin{eqnarray}
	U_\theta~:~~\vec{r}_i
&\to&	\vec{r}_i\, e^{i\theta} 
	\qquad \mbox{for}~~i=1,\cdots,N \ ,
\end{eqnarray}
where $\theta$ is a scaling angle and $N$ the total number of particles in the system.  The Hamiltonian $H$ is transformed into the complex-scaled Hamiltonian $H_\theta=U_\theta H U_\theta^{-1}$, and the corresponding complex-scaled Schr\"odinger equation is given as
\begin{eqnarray}
	H_\theta\Psi^J_\theta
&=&     E\Psi^J_\theta,
	\label{eq:eigen}
	\\
        \Psi^J_\theta
&=&	e^{(3/2)i\theta X}\,
	\Psi^J(\{\vc{r}_i e^{i\theta}\}),
\end{eqnarray}
where $X$ stands for the number of degrees of freedom.  The phase factor $e^{(3/2)i\theta X}$ is attached here due to the phase freedom of wave function and originates from the Jacobian in the integral over the coordinates.  In the three-body model of $^{11}$Li, $X=2$.  The eigenstates $\Psi^J_\theta$ are obtained by solving the eigenvalue problem of $H_\theta$ in Eq.~(\ref{eq:eigen}).  In the CSM, we obtain all the energy eigenvalues $E$ of bound and unbound states on a complex energy plane, governed by the ABC theorem \cite{ABC}.  In this theorem, it is proved that the boundary condition of Gamow resonances is transformed to the damping behavior at the asymptotic region.  The Gamow resonance is a pole of $S$-matrix and has an complex energy eigenvalue of $E=E_r-i\Gamma/2$, where $E_r$ and $\Gamma$ are the resonance energies 
measured from the lowest threshold and the decay widths, respectively. 

For simple understanding of this theorem, we consider the asymptotic wave functions for Gamow states in the two-body case.  The Gamow states with complex wave number $k_p$ are described by the outgoing waves $\exp(ik_p r e^{i\theta})$.  It is easily understood that the bound state wave functions maintain the damping behavior for $\theta<\pi/2$. The wave functions of resonances, which had divergent behavior originally as
$e^{ik_R\cdot r}=e^{i(\kappa_r-i\gamma_r)r}=e^{\gamma_r r}\cdot e^{i\kappa_r r}$, behave as 
\begin{eqnarray}
e^{ik_R\cdot re^{i\theta}}&=&e^{i(\kappa_r-i\gamma_r)re^{i\theta}}=e^{ir(\kappa_r-i\gamma_r)
(\cos{\theta}+i\sin{\theta})} \nonumber \\
         &=&e^{(-\kappa_r\sin{\theta}+\gamma_r\cos{\theta})r}\cdot
e^{i(\kappa_r\cos{\theta}+\gamma_r\sin{\theta})r}~. 
\label{Eq2-10}
\end{eqnarray}
This equation shows that the divergent behavior of the resonant wave functions is regularized when we take the scaling angle $\theta$ to be larger than the angle $\theta_r=\tan^{-1}(\frac{\gamma_r}{\kappa_r})$ of the resonance position $\kappa_r-i\gamma_r$.   This damping condition enables us to use the same theoretical method to obtain many-body resonance states as those used for bound states.   For a finite value of $\theta$, every Riemann branch cut is commonly rotated down by $2\theta$.  We can identify the resonance poles of complex eigenvalues without any ambiguities.

\begin{figure}[th]
\sidecaption[t]
\includegraphics[width=7.0cm]{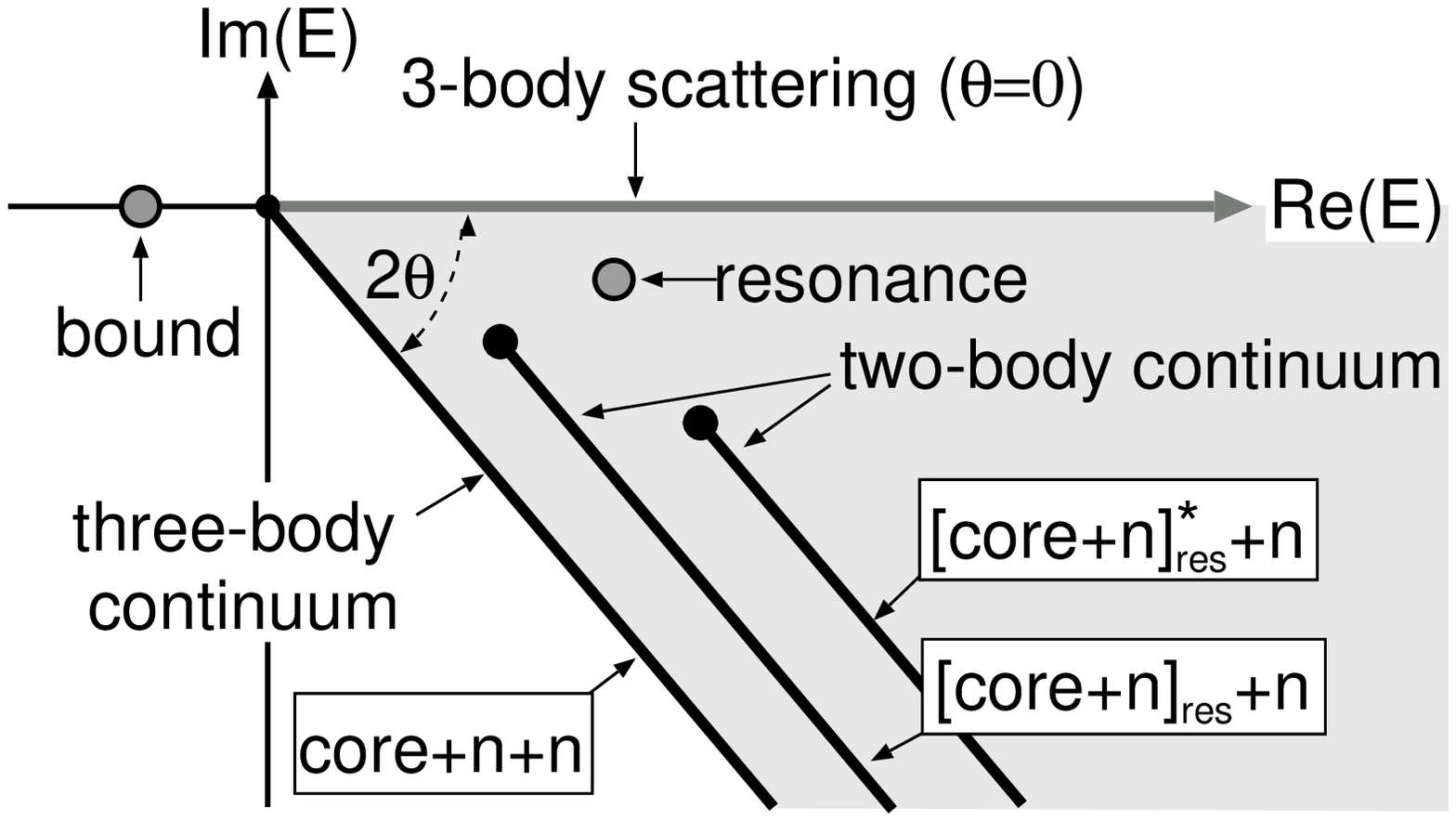}
\caption{A schematic distribution of energy eigenvalues of the Borromean core+$n$+$n$ system in the CSM, where the origin of energy is chosen as the three-body threshold energy \cite{Ao06}.}
\label{fig:3}       
\end{figure}

In the wave function, the $\theta$ dependence is included in the
variational coefficients in Eqs.~(\ref{eq:COSM2}) and (\ref{eq:T-basis}) as $\{C^{p,\theta}_{\alpha,V}\}$ and $\{C^{p,\theta}_{\alpha,T}\}$, respectively.  The wave functions are expanded with the finite number of the $L^2$ basis functions, so that all the obtained states are discretized on the complex energy plane.  The stability of the calculated matrix elements of resonant and continuum states using the CSM has been shown in many works \cite{Myo01,Ao06,My98}.  For continuum states, we adopt the discretized representation using the $L^2$ integrable basis functions.  This discretization has been checked to reproduce the genuine continuum states by using the CSM \cite{Kr07,Su08,Su05}.

In the study, we use the CSM not only to search for the resonance positions, but also to calculate the strength functions, such as $E1$ responses.  This is related to the continuum level density of the scattering states.  We have shown that the CSM provides us with the accurate continuum level density even if the states are discretized.
This fact means that the continuous strength function into scattering states can be obtained in the CSM, which is performable in the many-body case.  So far, we have succeeded to apply this characteristics of the CSM to calculate the electric responses, Gamow-Teller strengths, nucleon-removal strength and so on.  

In the calculation of the strength function, we need a complete set of the core+$n$+$n$ system including bound, resonant, and continuum states.  We express this complete set using the complex-scaled eigenstates $\Psi^J_\theta$ obtained in the core+$n$+$n$ model.  We briefly explain the extended completeness relation (ECR) using the CSM \cite{Myo01,My98,Be68}.  When we take a large $\theta$ like in Fig.~12, three-body scattering states are decomposed into three categories of discrete three-body resonances, three-body continuum states of core+$n$+$n$, and two-body continuum states of [core+$n$]$_{\rm res}$+$n$.  Here, the [core+$n$]$_{\rm res}$+$n$ two-body continuum states are obtained on the branch cuts, whose origins are resonance positions of the core+$n$ system, as shown in Fig.~12.  Using all the unbound states, we introduce the extended three-body completeness relation (ECR) of the complex-scaled Hamiltonian $H_\theta$ as
\begin{eqnarray}
        {\bf 1}
&=&     \sum_\nu\hspace*{-0.40cm}\int\kets{\Phi^\theta_\nu}\bras{\wtil{\Phi}^\theta_\nu}
        \nonumber
        \\
&=&     \{\mbox{three-body bound state}\}
        \nonumber
        \\
&+&     \{\mbox{three-body resonance}\}
        \nonumber
        \\
&+&    \{\mbox{three-body continuum states of core+$n$+$n$}\}
        \nonumber
        \\
&+&     \{\mbox{two-body continuum states of [core+$n$]$_{\rm res}$+$n$}\}\, ,
        \label{eq:3-ECR}
\end{eqnarray}
where $\{ \Phi_\nu^\theta,\wtil{\Phi}_\nu^\theta \}$ are the complex-scaled wave functions and form a set of biorthogonal bases.
This relation is an extension of the two-body ECR \cite{Ao06,My98}.
Because the detailed definition of the biorthogonal bases is written in the previous works \cite{Myo01,My98}, we only briefly explain it here.
When the wave number $k_\nu$ of $\Phi_\nu$ is for discrete bound and resonance states, the adjoint wave number $\wtil{k}_\nu$ of $\wtil{\Phi}_\nu$ is defined as $\wtil{k}_\nu=-k^*_\nu$,  which leads to the relation $\wtil{\Phi}_\nu$ = $(\Phi_\nu)^*$ \cite{My98,Be68,Mo78}. 
For continuum states, the same relation of the bi-orthogonal states of resonances is adopted, because we use a discretized representation.
In the core+$n$+$n$ model, the core+2$n$ channel is included in the three-body continuum components of the core+$n$+$n$ system, because two neutrons do not have any bound states or physical resonances. 

We explain how to calculate the strength function $S(E)$ using the ECR model.  The strength $S(E)$ is a function of the real energy of the whole system $E$.  We first introduce the Green's function ${\cal G}(E,\vc{\eta},\vc{\eta}')$, which is used in the derivation of the strength \cite{Myo01,Myo03}.
The coordinates, $\vc{\eta}$ and $\vc{\eta}'$, represent the set of $\vc{r}_i$ ($i=1,\cdots,X$) in Fig.~7.    Here, we introduce the complex-scaled Green's function ${\cal G}^\theta(E,\vc{\eta},\vc{\eta}')$ as
\begin{eqnarray}
        {\cal G}(E,\vc{\eta},\vc{\eta}')
&=&     \left\bra \vc{\eta} \left|
        \frac{ {\bf 1} }{ E-H }\right|\vc{\eta}' \right\ket
        \label{eq:green0}
        \\
~\to~   {\cal G}^\theta(E,\vc{\eta},\vc{\eta}')
&=&     \left\bra \vc{\eta} \left|
        \frac{ {\bf 1} }{ E-H_\theta }\right|\vc{\eta}' \right\ket
        \nonumber
        \\
&=&     \sum_\nu\hspace*{-0.4cm}\int\
        \frac{\Phi^\theta_\nu(\vc{\eta})\ [\wtil{\Phi}^*_\nu(\vc{\eta}')]^\theta}{E-E_\nu^\theta}
~=~     \sum_\nu\hspace*{-0.4cm}\int\
        {\cal G}^\theta_\nu(E,\vc{\eta},\vc{\eta}')\, .
        \label{eq:green1}
\end{eqnarray}
In the derivation from Eq.~(\ref{eq:green0}) to Eq.~(\ref{eq:green1}), we insert the ECR of the whole system given in Eq.~(\ref{eq:3-ECR}).   The total energy $E_\nu^\theta$ corresponds to the eigen wave function $\Phi^\theta_\nu$.  The $\theta$ dependence of $E_\nu^\theta$ appears only in the continuum spectra. 

The strength function $S(E)_\alpha$ for the arbitrary operator $O_\alpha$, in which $\alpha$ is the quantum number for the operator, is defined using the ordinary Green's function as
\begin{eqnarray}
        S_\alpha(E) 
&=&     \sum_\nu \hspace*{-0.4cm}\int\
        \bras{\wtil{\Psi}_0} O^\dagger_\alpha \kets{\Phi_\nu} \bras{\wtil{\Phi}_\nu} O_\alpha \kets{\Psi_0}\
        \delta(E-E_\nu)
        \label{eq:strength_org}
        \nonumber
        \\
&=&     -\frac1{\pi}\ {\rm Im}
        \left[
        \int d\vc{\eta} d\vc{\eta}'                    \:
        \wtil{\Psi}_0^*(\vc{\eta})\: O^\dagger_\alpha  \:
        {\cal G}(E,\vc{\eta},\vc{\eta}')               \:
        O_\alpha \Psi_0(\vc{\eta}')
        \right]\, .
        \label{eq:strength}
\end{eqnarray}
For simplicity, we omit the labels of the angular momenta and their $z$ components of the wave functions and of the operators.  The wave function $\Psi_0$ is the initial state.  We also consider the sum rule value of the strength $S_\alpha(E)$ in Eq.~(\ref{eq:strength}),
which is defined by the integration of $S_\alpha(E)$ over the real energy $E$.  Using the completeness relation of the final states of $^6$He, the sum rule value is given as
\begin{eqnarray}
        \int dE\ S_\alpha(E) 
&=&     \sum_{\nu}\hspace*{-0.4cm}\int\ \bras{\wtil{\Psi}_0}O^\dagger_\alpha \kets{\Phi_\nu}\bras{\wtil{\Phi}_\nu} O_\alpha \kets{\Psi_0}\
        \nonumber
        \\
&=&     \langle \widetilde{\Psi}_0|O^\dagger_{\alpha} O_{\alpha}| \Psi_0 \rangle~.
\end{eqnarray}
Thus, it is also confirmed that the energy integrated value of $S_\alpha(E)$ satisfies the expectation value of the operator $O_\alpha$ for the initial state.  In the case of the $E1$ transition of halo nuclei, the sum rule value corresponds to the relative distance between core and the center of mass of valence neutrons.  When $O_\alpha$ is an annihilation operator, for example, $^7$He into $^6$He+$n$, the integrated value satisfies the associated particle number of $^7$He, namely the number of three valence neutrons.

To calculate the strength function $S_\alpha(E)$ in Eq.~(\ref{eq:strength}),  we operate the complex scaling on $S_\alpha(E)$, 
and use the complex-scaled Green's function of Eq.~(\ref{eq:green1}) as
\begin{eqnarray}
        S_\alpha(E)
&=&     -\frac1{\pi}\ {\rm Im}
        \left[
        \int d\vc{\eta} d\vc{\eta}'                       \:
        [\wtil{\Psi}_0^*(\vc{\eta})]^\theta (O^\dagger_\alpha)^\theta\:
        {\cal G}^\theta(E,\vc{\eta},\vc{\eta}')            \:
        O_\alpha^\theta \Psi^\theta_0(\vc{\eta}')
        \right]
        \nonumber
        \\
&=&     \sum_\nu\hspace*{-0.4cm}\int\ S_{\alpha,\nu}(E)\, ,
        \label{eq:strength1}
        \\
        S_{\alpha,\nu}(E)
&=&     -\frac1{\pi}\ {\rm Im}
        \left[   \frac{
                \bras{\wtil{\Psi}_0^\theta}  (O^\dagger_\alpha)^\theta \kets{\Phi_\nu^\theta}
                \bras{\wtil{\Phi}_\nu^\theta} O_\alpha^\theta          \kets{\Psi_0^\theta}
          }{E-E_\nu^\theta}\right] .
        \label{eq:strength2}
\end{eqnarray}
In Eq.~(\ref{eq:strength2}), the strength function is calculated using the matrix elements $\bras{\wtil{\Phi}_\nu^\theta} a_\alpha^\theta\kets{\Psi^\theta_0}$.  It is noted that the function $S_{\alpha,\nu}(E)$ is independent of $\theta$ \cite{Myo01,myo07c,My98,Su05}.
This is because any matrix elements are obtained independently of $\theta$ in the complex scaling method, and also because the state $\nu$ is uniquely classified according to the ECR defined in Eq.~(\ref{eq:3-ECR}).  As a result, the decomposed strength $S_{\alpha,\nu}(E)$ is uniquely obtained.  Thus, the strength $S_\alpha(E)$ is calculated as a function of the real energy $E$ of the nucleus of interest.  When we discuss the structures appearing in $S_\alpha(E)$, it is useful to decompose $S_\alpha(E)$ into each component $S_{\alpha,\nu}(E)$ by using the complete set of the final state $\nu$ of the whole system.  We can categorize $\nu$ using the ECR in Eq.~(\ref{eq:3-ECR}).  Because of this decomposition of unbound states, we can unambiguously investigate how much each resonant and continuum state of the whole system exhausts the strength.  This can be performed in the Coulomb breakup strengths of $^{6}$He, $^{11}$Li and $^{11}$Be and many other three-body systems.

\subsection{Three body resonance and continuum states in $^6$He}
\label{subsec:3-2}

We discuss the results of the complex scaling method (CSM) for the halo system $^6$He.  We describe $^6$He as a $^4$He+$n$+$n$ system in the hybrid-VT model \cite{Ao95a,Myo01}.   Here, we briefly recapitulate the important properties of the CSM. The Hamiltonian of the model is the same as the one in Ref.~\cite{Ao95a} except for an introduction of a three-body interaction;
\begin{eqnarray}
	H
&=&	\sum_{i=1}^3{t_i} - T_G
+	\sum_{i=1}^2\left(V_{\alpha n,i}+v_i^{\rm PF}\right) + v_{nn} + V^3_{\alpha n n},
\end{eqnarray}
where $t_i$ and $T_G$ are kinetic energies of each particle and the center-of-mass of the three-body system, respectively.
The $^4$He core cluster is assumed to have the $(0s)^4$-closed configuration with the length parameter $b_c$=1.4 fm,
which reproduces the experimental charge radius of $^4$He.
The Pauli-forbidden {PF} state is the $0s$ orbit in the relative motion.
The two-body interactions $V_{\al n}$ and $v_{nn}$ are given by the microscopic KKNN potential \cite{kanada79} for $^4$He-$n$ and the Minnesota potential \cite{Ta78} for $n$-$n$, respectively. These potentials well reproduce the low-energy scattering data of each two-body system. 

A phenomenological three-body $^4$He-$n$-$n$ interaction $V^3_{\alpha n n}$ is introduced to fit the binding energy of the $^6$He ground state. This is introduced to overcome the small underbinding (few hundreds keV) of the $^6$He ground state with a frozen $^4$He core assumption. By taking into account the excitation or the dissociation of the $^4$He core, this underbinding problem in $^6$He is believed to be solved \cite{Cs93,Ar99}.  We include effectively the excitation of the $^4$He core inside $^6$He by this three-body interaction term. We introduce the three-body interaction assuming a single Gaussian function
\begin{eqnarray}
	V_{\alpha n n}^3 
&=&	V_3\ e^{-\nu({\bf r}_1^2+{\bf r}_2^2)}\, ,\\
	V_3
&=&	-0.218~{\rm MeV}\, ,\quad \nu~=~(0.1/b_c)^2~{\rm fm}^{-2}\, .
\end{eqnarray}
Using this Hamiltonian, the present hybrid-VT model reproduces the observed energies and decay widths of $^{5,6}$He, simultaneously \cite{Aj89}, namely, the threshold energies of the particle emissions for the He isotopes.

\begin{figure}[th]
\sidecaption[t]
\includegraphics[width=6.0cm]{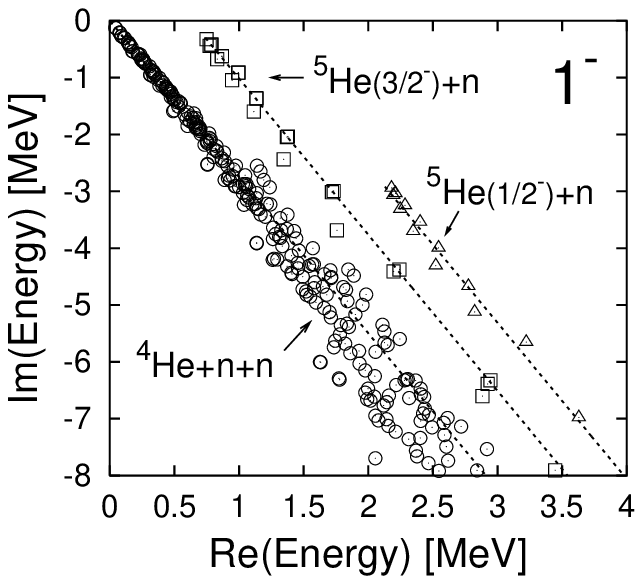}
\caption{
Energy eigenvalues of $1^-$ states calculated with the CSM where $\theta$ is 35 deg. \cite{Myo01}.  Squares and triangles indicate the two-body continuum states of \nuc{5}{He}($3/2^-$)+$n$ and \nuc{5}{He}($1/2^-$)+$n$, respectively.  Circles indicate the three-body continuum states of \nuc{4}{He}+$n$+$n$.} 
\label{fig:ene_1-}
\end{figure}

The three-body eigenstates are obtained by solving the eigenvalue problem of the complex-scaled Hamiltonian.  We use 30 Gaussian basis functions for one radial component in order to achieve stabilization of the calculated results for the position of resonances, distributions of continuum states and their transition matrix elements.  The maximum range of Gaussian basis functions is about 40 fm.  In Fig.~\ref{fig:ene_1-}, we show the eigenvalue distribution for $1^-$ states of $^6$He.  This result is obtained by diagonalization of the complex-scaled Hamiltonian of the \nuc{4}{He}+$n$+$n$ model with $\theta=35$ deg.  From Fig.~\ref{fig:ene_1-}, we see that all eigenvalues are obtained along three lines of rotated Riemann cuts corresponding to two two-body and one three-body continuum channels.   There is no $1^-$ resonance. Therefore, these results indicate that the $1^-$ unbound states above the \nuc{4}{He}+$n$+$n$ threshold are classified into two-body continuum states of \nuc{5}{He}($3/2^-$,$1/2^-$)+$n$ and three-body continuum states of \nuc{4}{He}+$n$+$n$.  
\begin{figure}[th]
\sidecaption[t]
\includegraphics[width=6.0cm]{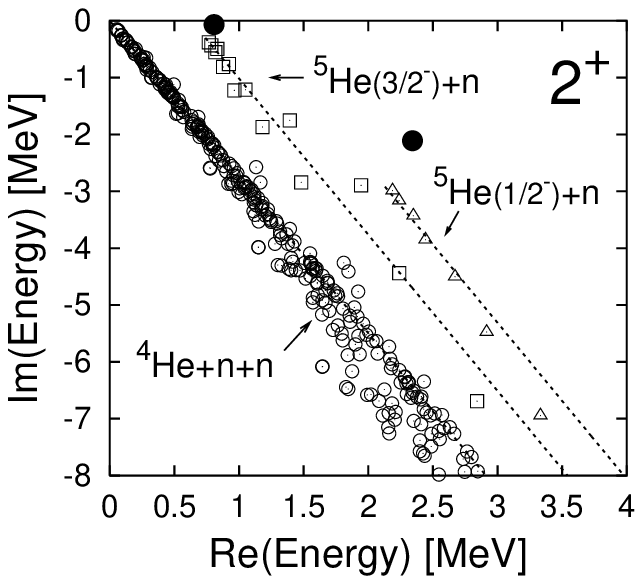}
\caption{Energy eigenvalues of $2^+$ states calculated with the CSM \cite{Myo01}, where two solid circles are $2^+_{1,2}$ resonances and other marks
indicate the same meanings as those in Fig.~\ref{fig:ene_1-}.}
\label{fig:ene_2+}
\end{figure}

\begin{figure}[th]
\sidecaption[t]
\includegraphics[width=6.0cm]{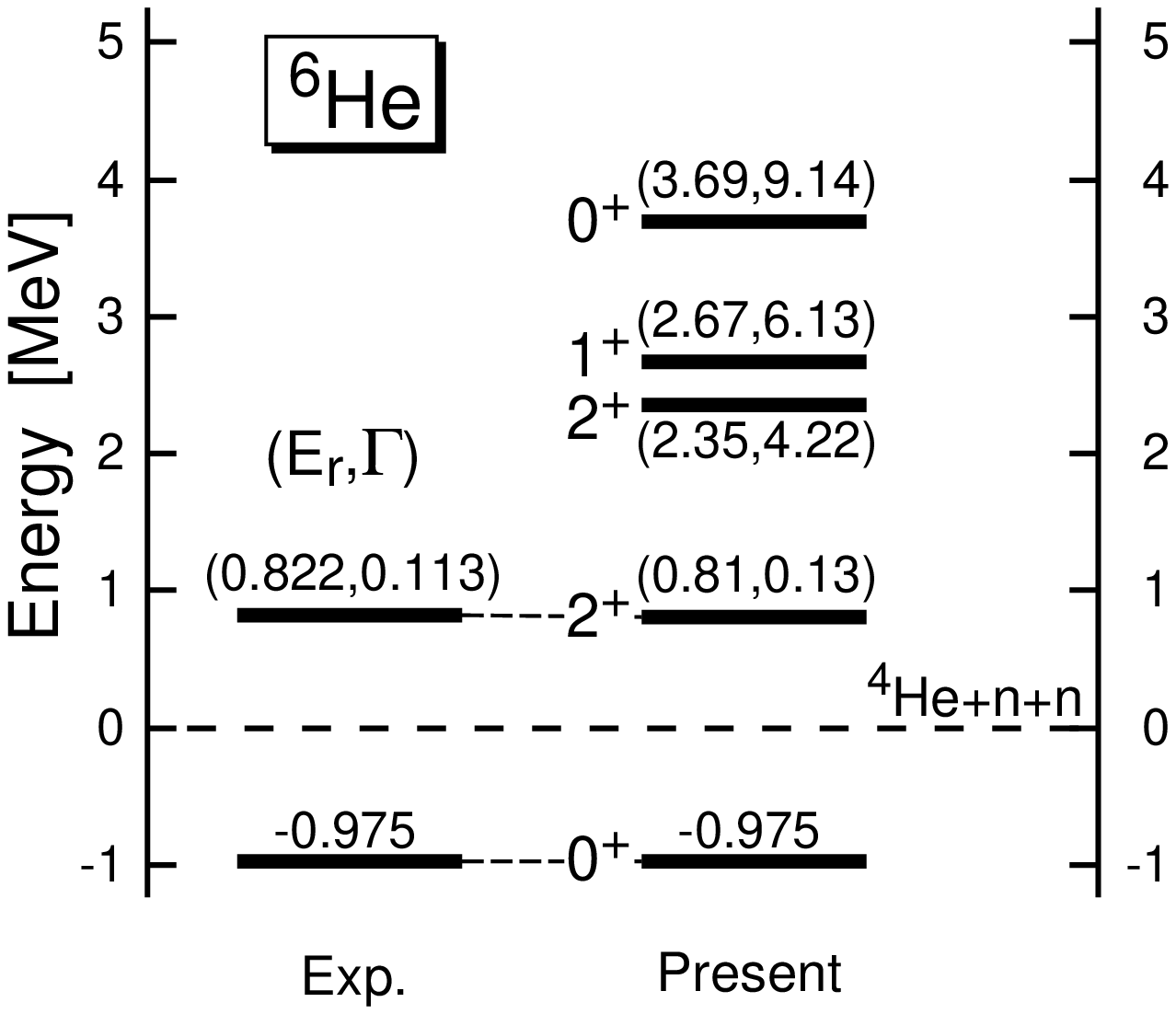}
\caption{Energy levels of $^6$He. Unit of energies and resonance widths is MeV \cite{Myo01}.}
\label{fig:6He_ene}       
\end{figure}

\begin{figure}[th]
\sidecaption[t]
\includegraphics[width=7.0cm]{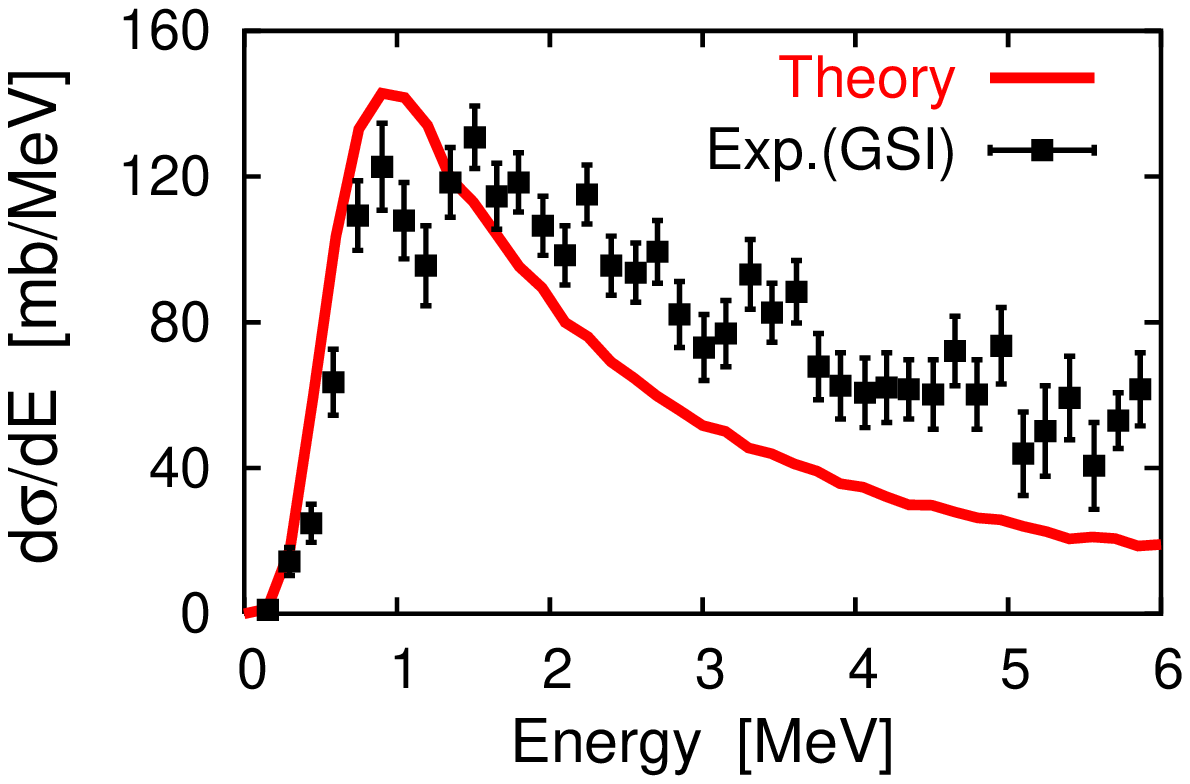}
\caption{Coulomb breakup cross section of $^6$He \cite{kikuchi09}.}
\label{fig:6He_E1}       
\end{figure}

In Fig.~\ref{fig:ene_2+}, we show the obtained $2^+_{1,2}$ resonances and continuum solutions which are decomposed into two- and three-body continuum states, similar to the $1^-$ spectra.
The whole energy levels of $^6$He are displayed in Fig.~\ref{fig:6He_ene}. We can see a good agreement between our calculation and experimental data.  The present calculation does not predict any $1^-$ resonance in the low excitation energy region. This result is consistent with the experimental results \cite{Aj89,Ja96}.

In Fig.~\ref{fig:6He_E1}, we show the Coulomb breakup cross section with respect to the excitation energy of $^6$He.  This cross section is calculated using the $E1$ strength with the equivalent photon method, considering the experimental resolution \cite{Au99}.  The target is Pb and the incident energy of the $^6$He projectile is 240 MeV/nucleon.  It is found that there is a low energy enhancement in the strength at around 1 MeV measured from the three-body threshold energy.
This energy is just above the two-body threshold (0.74 MeV) of the $^5$He($3/2^-$)+$n$ system \cite{Ao95a}, and the cross section gradually decreases with the excitation energy.  This structure of the strength indicates the sequential breakup process via the $^5$He($3/2^-$)+$n$ channel \cite{Myo01}. We also compare the strength with experiments \cite{Au99}.  The obtained result fairly reproduces the trend of the observed cross section, especially in the low excitation energy region below $E \sim 2$ MeV.  The height and position of the low-energy enhancement in the strength agree well with the experimental data.

In this section, we have discussed the complex scaling method (CSM) in order to treat resonance and continuum states in nuclear composite systems.  As an example of the usefulness of the CSM, we have discussed the case of $^6$He as a core+$n$+$n$ three-body system.  We can interpret the calculated results nicely and can distinguish any structure in the continuum spectrum.  As for the application of the CSM to the Li isotopes, we defer the detailed discussions after the introduction of the deuteron-like tensor correlation.  The interpretation of the experimental results further need the participation of the $s_{1/2}$ orbit in the wave functions of the Li isotopes.  For this, we have to introduce the deuteron-like tensor correlation, which is the subject of the next section.

\section{Deuteron-like tensor correlation and tensor optimized shell model (TOSM)}
\label{sec:4}

In this section, we would like to discuss the role of the deuteron-like tensor correlation on the nuclear structure in $^4$He before applying the developed method to the halo nucleus $^{11}$Li.  The tensor interaction is one of the most important ingredients of the bare nucleon-nucleon interaction and plays the central role for the formation of finite nuclei and nuclear matter.   In this lecture note, we have shown explicitly the case of the deuteron in Sec. 2, where the tensor interaction plays the central role to provide a strong binding energy through coupling of the $s$-wave component with the $d$-wave component.  Although the tensor interaction is known to be important, the standard approach of nuclear many-body problems is to obtain the effective interaction (G-matrix) by solving the Brueckner equation to include the high momentum components in the effective G-matrix interaction.  Hence, we are used to treat a well behaved central effective interaction and have not faced to treat the tensor interaction explicitly for shell model states.  This is the reason why we have not developed a method of treating the tensor interaction explicitly in nuclear physics.

Recently it became possible to calculate nuclei up to mass around $A\sim 12$ \cite{pieper01,pieper02,pudliner97} using the realistic nucleon-nucleon interaction.  The method used for the calculation is the Green's function Monte-Carlo method (GFMC) with the use of relative nucleon coordinates.   This method introduces various correlation functions with many variational parameters in the nuclear wave function.  In the GFMC, the nuclear structures and binding energies were successfully reproduced by including the three-body interaction. One big surprise is extremely a large contribution of the one pion exchange interaction, which is about 70 $\sim$ 80 \% of the whole nucleon-nucleon interaction.  In principle, they can extend this method to calculate heavier nuclei.  As for the tensor interaction, it contributes about 50 \% of the whole two-body matrix element.  It is however extremely time consuming even with the present computer power.  Hence, it is strongly desired to develop a new method to calculate nuclei with large nucleon numbers by using the nucleon-nucleon interaction.

The nucleon-nucleon interaction has distinctive features that there exists the strong tensor interaction at intermediate distance caused by pion exchange and strong short range repulsive interaction at short distance caused by quark dynamics.  The explicit form of the nucleon-nucleon interaction is presented in Sect. 2.  Although these two interactions have totally different characters, it is customary to adopt the Brueckner Hartree-Fock theory to integrate out the high momentum components on the same footing and use the resulting $G$-matrix as an effective interaction in the shell model.  In this way, we lose information of the tensor correlation and the short range correlation in the shell model wave function.  Hence, we search for a powerful method to treat explicitly both the tensor interaction and the short range interaction for the study of not only light nuclei but also medium and heavy nuclei.

There have been two important developments for this purpose.  One is to find out that the tensor interaction is of intermediate range and hence we are able to express the tensor correlation in a reasonable shell model space \cite{myo06,myo07}.  We name this method as the Tensor Optimized Shell Model (TOSM), where the nuclear wave function is written in terms of the standard shell model state and enough amount of two-particle two-hole ($2p2h$) states.  This TOSM formalism is based on the success of the parity and charge projection in the treatment of the pion exchange interaction \cite{sugimoto04,ogawa06}.  We have shown that the tensor interaction could be treated properly by taking a reasonable amount of multipoles ($l \leq 5$) in the $2p2h$ wave functions with the optimization of the radial parts of the particle states. The other is the Unitary Correlation Operator Method (UCOM) for the treatment of the short range correlation \cite{feldmeier98,neff03}.  The short range repulsive interaction is of very short range and it is suited to treat the short range correlation using a unitary transformation and take the approximation to use only up to the two-body operators.  This approximation is justified because the volume associated with the short range correlation is extremely small, where more than three nucleons rarely enter into the small volume.  This is not the case for the tensor correlation, since the tensor interaction is of intermediate and long range as discussed by Neff and Feldmeier \cite{neff03}.

Our idea is to combine these two methods, TOSM and UCOM, for our purpose to develop a theoretical framework to describe furthermore medium and heavy nuclei beyond the light nuclei using the realistic nucleon-nucleon interaction.  We can use the TOSM for the strong tensor interaction utilizing the intermediate nature caused by finite angular momentum of the relative wave function and the UCOM for the strong short range interaction utilizing the short range nature.  We use completely different methods for these two distinctive characters of the nucleon-nucleon interaction.  After demonstrating its power we hope to apply a newly developed method, which we name TOSMU, to many nuclei.  Using the TOSMU, we aim at understanding the roles of the tensor and short range correlations in nuclei using the bare nucleon-nucleon interaction.  As a good start, we would like to apply the TOSMU to $^4$He.  Hence, there are two purposes for this study.  One is to see how this method works for the treatment of the bare nucleon-nucleon interaction.  The other is to compare with rigorous calculations to check the accuracy of the results obtained in the TOSMU.  From this comparison, we can see how far we can describe the short range and tensor correlations and to find what we are supposed to do for further improvement of the TOSMU in order to solve the nucleus as precisely as possible.

On the other hand, we would like to develop a theoretical framework to describe wave functions in terms of single particle coordinates, which we call a V-coordinate method.  This V-coordinate method has ability to describe nuclei with many nucleons relatively easier than the T-coordinate method.  Furthermore, we are able to describe the wave function based on the shell model picture and hence it is easier to interpret the calculated results in the shell model sense.  The difficulty, on the other hand, is to express the correlations of the relative motion between two nucleons, which are caused by the short range repulsive interaction and the tensor interaction in the nucleon-nucleon interaction.  
We overcome this problem by developing the TOSMU to describe the short range and the tensor correlations simultaneously.

\subsection{Formulation of TOSM}
\label{subsec:4-1}

We explain the formulation of the Tensor optimized shell model (TOSM).   We shall begin with the many-body Hamiltonian with mass number $A$.
\begin{eqnarray}
H=\sum_i^A T_i - T_{\rm cm} + \sum_{i<j}^A V_{ij}
\label{eq:Ham}
\end{eqnarray}
with
\begin{eqnarray}
    V_{ij}
&=& v_{ij}^C + v_{ij}^{T} + v_{ij}^{LS} + v_{ij}^{Clmb} .
\end{eqnarray}
Here, $T_i$ is the kinetic energy of all the nucleons with $T_{cm}$ being the center of mass kinetic energy.  We take the bare nucleon-nucleon interaction for $V_{ij}$ such as the AV8$^\prime$ consisting of central ($v^C_{ij}$), tensor ($v^T_{ij}$) and spin-orbit ($v^{LS}_{ij}$) terms.  The $v_{ij}^{Clmb}$ is the Coulomb term. We describe the many-body system with many-body wave function, $\Psi$, by solving the Schr\"odinger equation $H \Psi=E \Psi$. In the TOSM, we take the $V$-coordinates to express $\Psi$.  

We have discussed the property of the deuteron, whose wave function is expressed in terms of the $s$-wave and $d$-wave components.  We ought to express the deuteron-like tensor correlation in finite nuclei.  We show the deuteron-like structure is expressed in terms of 2p-2h configurations in the shell model framework by taking the case of $^4$He.  The two nucleons in the $s$-orbit have two components.  One is the $^3 S_1$ pair and the other is the $^1 S_0$ pair.  The $^3 S_1$ pair can be expressed as
\beq
|[(0s_{1/2})^2]_{1M} \ket \sim \Psi_{L=0}(R) \psi_{l=0}(r) [Y_0(\hat r) \otimes \chi_1(\sigma_1 \sigma_2)]_{1M}~.
\eeq
Hence, the relative wave function is in the $s$-state.  On the other  hand, when two nucleons are in the $p$-orbit, we can write the wave function of the relative motion being in the triplet even channel as
\beq
|[(0p_{1/2})^2]_{1M} \ket \sim \Psi_{L=0}(R) \psi_{l=2}(r) [Y_2(\hat r) \otimes \chi_1(\sigma_1 \sigma_2)]_{1M}~.
\eeq
Hence, the relative wave function is in the $d$-state.  As discussed for the deuteron wave function being written in terms of the $s$ and $d$ wave components, we are able to express the deuteron-like wave function in the shell model framework by taking both the above two components.  Hence, we have to introduce 2p-2h wave functions to express the deuteron-like tensor correlation in the shell model basis.  There is the other $p$-wave component, $(p_{3/2})^2$.  In this case, the spin and the angular momentum have a stretched configuration and this state is not mixed by the tensor interaction.    This consideration naturally leads us to introduce the tensor optimized shell model (TOSM).  It remains for us to check if we can express the deuteron-like tensor correlation within reasonable amount of multipoles in the TOSM \cite{myo06}.   As for the short range correlation appearing in the $s$-wave component in the deuteron wave function, we ought to take a clever method.  For this problem, Feldmeier et al. have demonstrated that the unitary correlation operator method (UCOM) can be used to treat the short-range correlation \cite{feldmeier98,neff03}.

In the TOSM, the total wave function $\Psi$ is written in terms of a linear combination of 0p-0h and 2p-2h wave functions.
\begin{eqnarray}
\Psi=C_0 \kets 0 + \sum_p C_p \kets {2p2h}_p \ .
\end{eqnarray}
Here, the wave function $\kets {0}$ is a shell model wave function and $\kets {(0s)^4}$ for $^4$He.  $\kets {2p2h}$ represents a $2p2h$ state with various ranges for the radial wave functions of particle states. We can write $\kets {2p2h}$ as
\begin{eqnarray}
    \kets {2p2h}_p
&=& \kets {\left[ [\psi^{n_1}_{\alpha_1}(\vec x_1)\psi^{n_2}_{\alpha_2}(\vec x_2)]^J
    \otimes [\tilde\psi^{n_3}_{\alpha_3}(\vec x_1)\tilde\psi^{n_4}_{\alpha_4}(\vec x_2)]^J\right]^0}_A \ .
\label{eq:2p2h}
\end{eqnarray}
The suffix $A$ of the wave function indicates anti-symmetrization of the wave functions.  Here, $p$ denotes a set of quantum numbers of $2p2h$ states, which are expressed with particle (hole) wave functions $\psi^{n}_{\alpha}$ ($\tilde\psi^{n}_{\alpha}$).  The index $n$ is to distinguish the different radial components of the single-particle wave function $\psi$.  The index $\alpha$ is a set of three quantum numbers, $l$, $j$ and $t_z$, to distinguish the single-particle orbits,  where $l$ and $j$ are the orbital and total angular momenta of the single-particle states, respectively, and $t_z$ is the projection of the nucleon isospin.   The normalization factors of the two particle states are included in the wave functions given in Eq.~(\ref{eq:2p2h}).
For $^4$He, the coupled spin, $J$, of two nucleons is $J=0$ or $J=1$. We omit writing the coupled isospin, which should be either 0 or 1 depending on the value of $J$. We have used Gaussian functions for radial wave functions to express more effectively compressed radial wave functions \cite{myo06}.  The shell model technique is used to calculate all the necessary matrix elements, which are expressed explicitly in Ref.~\cite{myo09}.  In more heavier nuclei, such as $^{12}$C and $^{16}$O, the dominant configuration $\kets {0}$ can be extended to the superposed ones, 
which includes the few $\hbar \omega$ configurations such as $2\hbar \omega$. This part describes the low-momentum component of the wave function and the tensor force contribution is not so decisive to determine the nuclear structure.  The high momentum part is treated by considering the $2p2h$ excitations $\kets {2p2h}_p$ from the each low-momentum configurations.  In that case, the spatial shrinkage of particle states becomes important. 

We explain the Gaussian expansion technique for single-particle orbits \cite{Ao06,hiyama03}.  Each Gaussian basis function has the form of a nodeless harmonic oscillator wave function (HOWF), except for the $1s$ orbit.  When we superpose a sufficient number of Gaussian bases with appropriate length parameters, we can fully optimize the radial component of every orbit of every configuration with respect to the total Hamiltonian in Eq.~(\ref{eq:Ham}).  We construct the following ortho-normalized single-particle wave function $\psi^n_{\alpha}$ with a linear combination of Gaussian bases $\{\phi_\alpha\}$ with length parameters $b_{\alpha,m}$.
\begin{eqnarray}
        \psi^n_{\alpha}(\vc{r})
&=&     \sum_{m=1}^{N_\alpha} d^n_{\alpha,m}\ \phi_{\alpha}(\vc{r},b_{\alpha,m})
        \qquad
        {\rm for}~~n~=~1,\cdots,N_\alpha~.
        \label{eq:Gauss1}
\end{eqnarray}
Here, $N_\alpha$ is the number of basis functions for $\alpha$, and $m$ is an index that distinguishes the bases with different values of $b_{\alpha,m}$.
The explicit form of the Gaussian basis function is expressed as
\begin{eqnarray}
        \phi_{\alpha}(\vc{r},b_{\alpha,m})
&=&     N_l(b_{\alpha,m})\ r^l\ e^{-(r/b_{\alpha,m})^2/2}\ [Y_{l}(\hat{\bf r}),\chi^\sigma_{1/2}]_j \chi_{t_z} \ ,
        \label{eq:Gauss2}
        \\
        N_l(b_{\alpha,m})
&=&     \left[  \frac{2\ b_{\alpha,m}^{-(2l+3)} }{ \Gamma(l+3/2)}\right]^{\frac12}.
\end{eqnarray}
The coefficients $\{d^n_{\alpha,m}\}$ are determined by solving the eigenvalue problem for the norm matrix of the non orthogonal Gaussian basis set in Eq.~(\ref{eq:Gauss2}) with the dimension $N_\alpha$.
Following this procedure, we obtain new single-particle wave functions $\{\psi^n_{\alpha}\}$ using Eq.~(\ref{eq:Gauss1}).

We choose the Gaussian bases for the particle states to be orthogonal to the occupied single-particle states, which is $0s_{1/2}$ in the $^4$He case. For $0s_{1/2}$ states, we employ one Gaussian basis function, namely, the HOWF with the length parameter $b_{0s_{1/2},m=1}=b_{0s}$. For $1s_{1/2}$ states, we introduce an extended $1s$ basis function orthogonal to the $0s_{1/2}$ states 
and possessing a length parameter $b_{1s,m}$ that differs from $b_{0s}$ \cite{myo06}. 
In the extended $1s$ basis functions, we change the polynomial part from the usual $1s$ basis states to satisfy the conditions of the normalization and the orthogonality to the $0s$ state. 

Two-body matrix elements in the Hamiltonian are analytically calculated using the Gaussian bases \cite{Ao06,hiyama03}, 
whose explicit forms are given in Ref.~\cite{myo09}.
In the numerical calculation of following, we prepare 9 Gaussian functions at most with parameters of various ranges to obtain a convergence of the energy.  Furthermore, we have to take care of the center-of-mass excitations.  For this purpose, we use the well-tested method of introducing a center-of-mass term in the many-body Hamiltonian \cite{otsuka07,lawson}.  
\begin{eqnarray}
       H_{\rm cm}
&=&    \lambda \ \left( \frac{\vc{P}_{\rm cm}^2}{2A\, m}+ \frac12\, A\, m\, \omega^2\, \vc{R}_{\rm cm}^2-\frac{3}{2}\hbar\omega \right),
       \\
        \vec{P}_{\rm cm}
&=&     \sum_{i=1}^A \vec{p}_i\ ,\qquad
        \vec{R}_{\rm cm}
\,=\,   \frac1A\, \sum_{i=1}^A \vec{r}_i\ ,\qquad
        \omega
\,=\,   \frac{\hbar}{m\, b_{0s}^2}~.
\end{eqnarray}
Here, $m$ and $A$ are the nucleon mass and the mass number, respectively, and
$b_{0s}$ is the length parameter of the HOWF for the hole $0s$ state.
We take a sufficiently large coefficient, $\lambda$, to project out only the lowest HO state for the center-of-mass motion.
In the numerical calculation, the excitation of the spurious center-of-mass motion is suppressed to be less than 10 keV.

The variation of the energy expectation value with respect to 
the total wave function $\Psi(^{4}{\rm He})$ is given by
\begin{eqnarray}
\delta\frac{\bra\Psi|H|\Psi\ket}{\bra\Psi|\Psi\ket}&=&0\ ,
\end{eqnarray}
which leads to the following equations:
\begin{eqnarray}
    \frac{\del \bra\Psi| H - E |\Psi \ket} {\del b_{\alpha,m}}
&=& 0\ ,\quad
    \frac{\del \bra\Psi| H - E |\Psi \ket} {\del C_{p}}
=   0\ .
   \label{eq:vari}
\end{eqnarray}
Here, $E$ is a Lagrange multiplier corresponding to the total energy. The parameters $\{b_{\alpha,m}\}$ for the Gaussian bases appear in non linear forms in the energy expectation value. We solve two types of variational equations in the following steps. First, fixing all the length parameters $b_{\alpha,m}$, we solve the linear equation for $\{C_{p}\}$ as an eigenvalue problem for $H$ with partial waves up to $L_{\rm max}$. We thereby obtain the eigenvalue $E$, which is a function of $\{b_{\alpha,m}\}$. Next, we try to search various sets of the length parameters $\{b_{\alpha,m}\}$ to find the solution that minimizes the total energy. In this wave function, we can describe the spatial shrinkage with an appropriate radial form, which is important for the tensor correlation \cite{myo06}.

\subsection{Formulation of UCOM}
\label{subsec:4-2}

We employ the UCOM for the short-range correlation.  Feldmeier et al. worked out a unitary correlation operator in the form \cite{feldmeier98,neff03},
\begin{eqnarray}
C&=&\exp\left(-i\sum_{i<j} g_{ij}\right)~=~\prod_{i<j}c_{ij} 
\end{eqnarray}
with $c_{ij}= \exp(-i\ g_{ij})$. Here, $i$ and $j$ are the indices to distinguish particles.  
Here, the two-body operator $g_{ij}$ is a Hermite operator, and hence $C$ is a unitary operator.  We express the full wave function $\Psi$ in terms of less sophisticated wave function $\Phi$ as $\Psi=C\Phi$.  Hence, the Schr\"odinger equation, $H\Psi=E\Psi$ becomes $\hat{H}\Phi=E\Phi$, where $\hat{H}=C^\dagger H C$.  If we choose properly the unitary correlator $C$ we are able to solve more easily the Schr\"odinger equation.  Moreover, once we obtain $\Phi$, we can then obtain the full wave function, $\Psi$, by the unitary transformation $\Psi=C\Phi$.  Since $C$ is expressed with a two-body operator in the exponential, it is a many-body operator.  In the case of the short-range correlation, we are able to truncate modified operators at the level of two-body operators \cite{feldmeier98}.

In the actual calculation of the UCOM, we define the operator $g_{ij}$ as
\begin{equation}
g_{ij}=	\frac12 \left\{ p_{r,ij} s(r_{ij})+s(r_{ij})p_{r,ij}\right\},
\end{equation}
where the momentum $p_{r,ij}$ is the radial component of the relative momentum, which is conjugate to the relative coordinate $r_{ij}$.  $s(r_{ij})$ is the amount of the shift of the relative wave function at the relative coordinate $r_{ij}$ for each nucleon pair. 
Hereafter, we omit the indices $i$ and $j$ for simplicity. We also introduce $R_+(r)$ as
\begin{equation}
        \int_r^{R_+(r)}\frac{d\xi}{s(\xi)} =     1 ,
\end{equation}
which leads to the following relation,
\begin{equation}
	\frac{dR_+(r)}{dr}
=     \frac{s \left( R_+(r) \right)}{s(r)} .
\end{equation}
In the UCOM, we use $R_+(r)$ instead of $s(r)$ to use the UCOM prescription. $R_+(r)$ represents the correlation function to reduce the amplitude of the short-range part of the relative wave function in nuclei and can be determined for four spin-isospin channels independently.
The explicit form of the transformation of the operator for the relative motion is given as
\begin{eqnarray}
    	c^\dagger r c
&=&	R_+(r) , 
        \qquad
        c^\dagger p_r c
~=~     \frac{1}{\sqrt{R^\prime_+(r)}}  p_r \frac{1}{\sqrt{R^\prime_+(r)}} , 
        \qquad
        c^\dagger \vec{l} c
~=~     \vec{l} ,
        \\
        c^\dagger \vec{s} c
&=&     \vec{s} , 
        \qquad
        c^\dagger S_{12} c
~=~     S_{12} ,
        \qquad
        c^\dagger v(r)  c
~=~  v(R_+(r)) , 
\end{eqnarray}
where the operators $\vec{l}$, $\vec{s}$ and $S_{12}$ are the relative orbital angular momentum operator, the intrinsic spin operator and the tensor operator, respectively. $v(r)$ is the arbitrary function depending on $r$, such as potential.

In the calculation using the UCOM, we parametrize $R_+$(r) in the same manner as proposed by Neff-Feldmeier and Roth et al. 
\cite{feldmeier98,neff03,roth06}.
We assume the following forms for even and odd channels, respectively.
\begin{eqnarray}
	R_+^{\rm even}(r)
&=&	r + \alpha \left(\frac{r}{\beta}\right)^\gamma \exp[-\exp(r/\beta)],
	\\
	R_+^{\rm odd}(r)
&=&	r + \alpha \left( 1- \exp(-r/\gamma) \right) \exp[-\exp(r/\beta)]~.
\end{eqnarray}
Here, $\alpha$, $\beta$, $\gamma$ are the variational parameters to optimize the function $R_+(r)$ and minimize the energy of the system.
They are independently determined for four channels of the spin-isospin pair.
In the actual procedure of the variation, once we fix the parameters included in $R_+(r)$, we solve the eigenvalue problem 
of the Hamiltonian using Eq.~(\ref{eq:vari}) and determine the configuration mixing of the shell model-type bases.
Next, we try to search various sets of the $R_+(r)$ parameters to minimize the obtained energy.

In the present framework of the UCOM, we introduce the UCOM function $R_+(r)$ for each spin-isospin channel and ignore the partial wave dependence of $R_+(r)$.  It is generally possible to introduce the partial wave dependence in the UCOM and then $R_+(r)$ functions are determined in each relative partial wave in the two-body matrix elements.
Here, we consider the specific case of this extension of the UCOM by taking care of the characteristics of the short-range correlation.
One of the simplest cases of this extension is the UCOM for only the $s$-wave relative motion, since all the other partial waves $l$ except for $s$-wave ($l=0$) have $r^l$ behavior near the origin, where the short-range hard core is extremely large.  Hence, this $r^l$ behavior largely cuts down the effect of the short-range hard core.  However, only the $s$-wave function is finite at the origin, and the behavior in the origin is determined by the hard core dynamics.  In fact, the method used by Feldmeier et al. \cite{feldmeier98} is to determine the unitary operator to reproduce the short-range behavior of the $s$-wave relative wave function.

When we incorporate the $S$-wave UCOM ($S$-UCOM, hereafter) into the TOSM, we extract the relative $s$-wave component in all the two-body matrix elements in the TOSM using the $V$-type basis expanded by the Gaussian functions.  For numerical calculations, we prepare the completeness relation consisting of the $T$-type basis functions $\kets{T}$ as
\begin{eqnarray}
      1
&=&   \sum_i |T_i \ket \bra T_i|,
      \qquad
      |T_i\ket
~=~   |[[\psi^{\bf r}_{l}\psi^{\bf R}_L]_{L'}, \chi_{S}]_J\ \chi_T \ket ,
      \label{eq:T}
\end{eqnarray}
where the $T$-type basis is expanded by the two coordinates of the relative part $\vec{r}$ and the center of mass part
$\vec{R}$ of two nucleons, which are the set of Jacobi coordinates.
The orbital angular momenta of each coordinate, $\vec{r}$ and $\vec{R}$, are $l$ and $L$, respectively.
It is easy to prepare the $s$-wave relative part by considering $l$ as zero in the $T$-type basis.
We construct the above completeness relation of the $T$-type basis states by diagonalizing the norm matrix 
expanded by the finite number of Gaussian basis functions for two coordinates. 
In the actual calculation, we use 12 bases for each coordinate, with which convergence is achieved.

We calculate the matrix elements of the arbitrary two-body operator $\hat{O}$ including the $S$-UCOM correlator $C_s$ using the $V$-type basis with indices $\alpha$ and $\beta$.  Here, we insert the above $T$-type completeness relation in Eq.~(\ref{eq:T}) as
\begin{eqnarray}
      \bra V_\alpha | C_s^\dagger \hat{O} C_s | V_\beta \ket
&=&   \sum_{ij} \bra V_\alpha |T_i \ket \cdot 
      \bra T_i| C_s^\dagger \hat{O} C_s |T_j \ket \cdot  
      \bra T_j| V_\beta \ket.
\end{eqnarray}
The matrix element using the $T$-type base, $\bra T_i| C_s^\dagger \hat{O} C_s |T_j \ket$, is calculated for the two-body kinetic part and the central and tensor interactions.  For the kinetic part and the central interaction, the matrix elements conserve the relative angular momentum, and then we can easily calculate the matrix elements of the transformed operator $C_s^\dagger \hat{O} C_s$.
For the tensor interaction, the $sd$ coupling matrix elements are properly treated, in which $C_s$ is operated on only the $s$-wave part 
of the relative motion. In this case, the operator $C_s$ acts on the $s$-wave relative Gaussian basis function $\phi_{l=0}(r)$, which is transformed as
\begin{eqnarray}
      C_S\ \phi_{l=0}(r)
&=&  \frac{R_-(r)}{r}\ \sqrt{R^\prime_-(r)}\ \phi_{l=0}(R_-(r)),
\end{eqnarray}
where $R_-(r)$ is the inverse transformation of $R_+(r)$, namely, $R_-(R_+(r))=r$.
The matrix elements of the $T$-type basis function are calculated using the above transformed wave function. 

\subsection{Numerical results of TOSM for $^4$He}
\label{subsec:2-3}

\begin{figure}[th]
\centering
\includegraphics[width=5.2cm,clip]{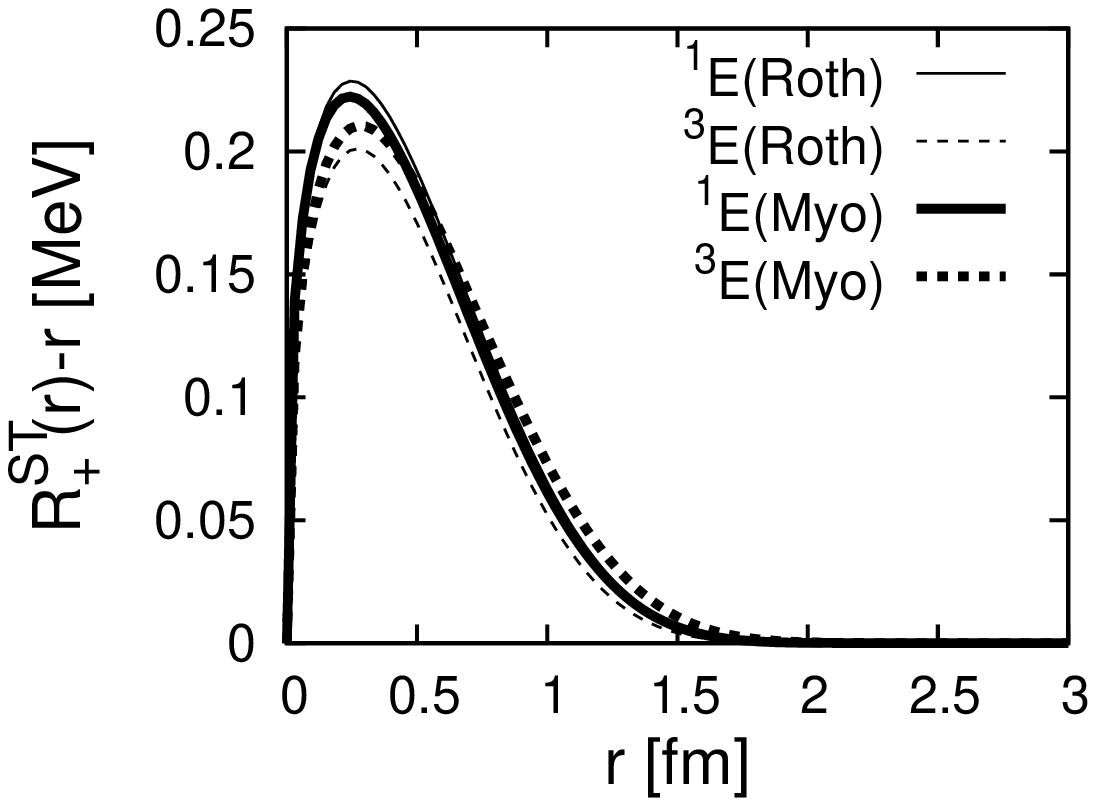}
\hspace*{0.5cm}
\includegraphics[width=5.2cm,clip]{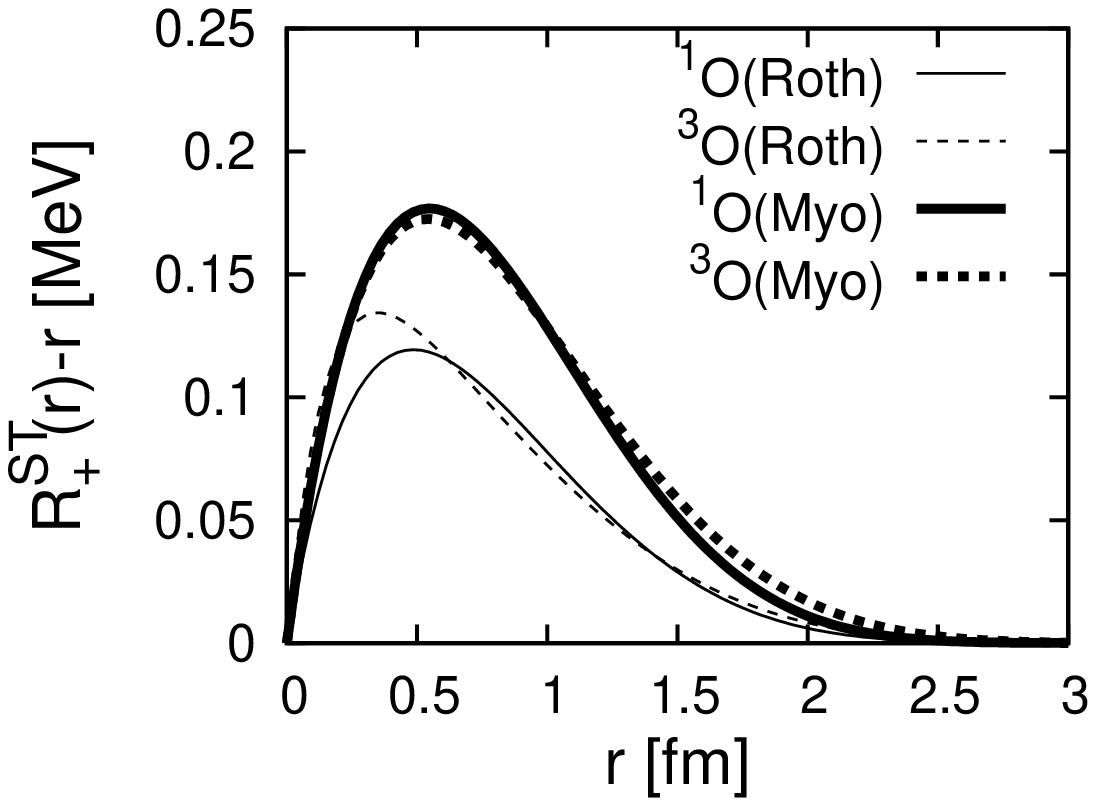}
\caption{The short range correlation functions, $R_+(r)$, for the UCOM in even and odd channels.  The thin curves are for the $R_+(r)$ function of Roth et al. and the thick curves are those of Myo et al.}
\label{R+}
\end{figure}

It is important to understand the origin of the large binding energy of $^4$He for the study of the Li isotopes.  Particularly, it is important to develop a method to describe the source of the large binding energy in the shell model language.  Hence, we describe in detail the structure of $^4$He in the TOSM and also the role of the UCOM.  First of all, we determine the UCOM functions $R_+(r)$ for the calculation of the TOSMU.  In the UCOM, we optimize the $R_+(r)$ function by changing the three parameters of $\alpha$, $\beta$ and $\gamma$ to search for the energy minimum in the TOSMU.  In Table \ref{tab:R+2}, the optimized three parameters in the $S$-UCOM are listed.  In Fig. \ref{R+}, $R_+(r)$ functions used in the present study are plotted in comparison with the case in Ref.~\cite{roth06}.  For the odd channel, in accordance with the discussion in Refs. \cite{neff03} and \cite{roth06}, we cannot find the optimum value of $R_+(r)$  in the two-body cluster approximation of the UCOM transformation for the Hamiltonian.  Hence, we decide to fix the range of $R_+(r)$, namely, $\beta$ as the same one adopted in Ref.~\cite{roth06} and optimize $\alpha$ and $\gamma$, while the variation of $R_+(r)$ for the odd channel does not have significant effects on the energy and other properties of $^4$He in comparison with the original case \cite{neff03,roth06}.  Essentially, two types of parameter set of $R_+(r)$ in the present study and Ref.~\cite{roth06} give the similar form of $R_+(r)$ for even channels, in which we omitted the correlation function for the even channels except for $s$-waves. This result indicates that the correlation functions for the short-range repulsion are uniquely determined for each channel.

\begin{table}[t]
\begin{center}
\caption{Optimized parameters in $R_+(r)$ in TOSM+UCOM for four channels in fm in the present work.}
\label{tab:R+2} 
\begin{tabular}{c|cccc}
\hline\noalign{\smallskip}
              &  $\alpha$ &  $\beta$  &  $\gamma$ \\
\noalign{\smallskip}\svhline\noalign{\smallskip}
singlet even  &  1.32  &  0.88  &  0.36 \\
triplet even  &  1.33  &  0.93  &  0.41 \\
singlet odd   &  1.57  &  1.26  &  0.73 \\
triplet odd   &  1.18  &  1.39  &  0.53 \\
\hline\noalign{\smallskip}
\end{tabular}
\end{center}
\end{table}

\begin{figure}[t]
\centering
\includegraphics[width=9cm,clip]{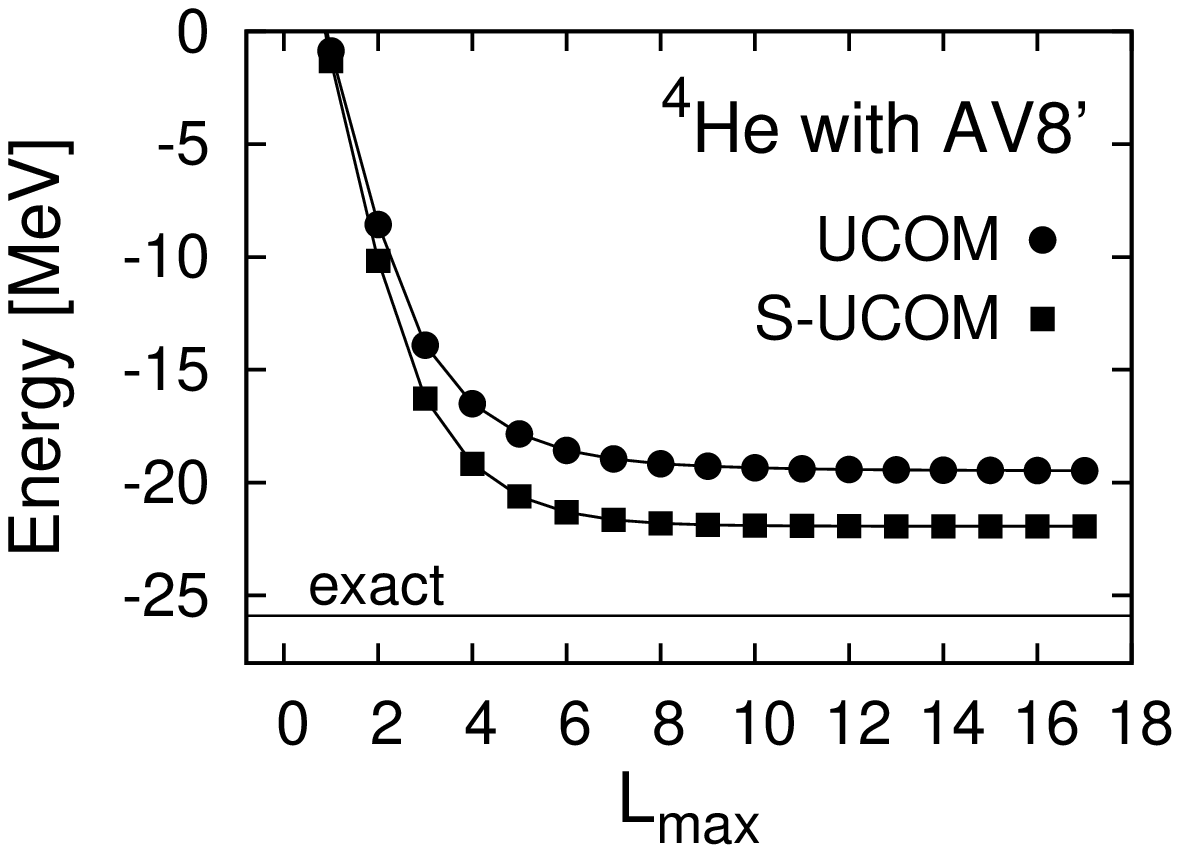}
\caption{The energy of $^4$He in the TOSMU as a function of the the maximum angular momentum $L_{\rm max}$.  
The circles are the results where the UCOM is used for all the partial waves. 
The squares are the results using $S$-UCOM in the tensor interaction matrix elements.}
\label{energy}
\end{figure}

Next, we show the calculated results of the energy of $^4$He as a function of $L_{\rm max}$ in Fig.~\ref{energy}.  We shall then compare the obtained results with the benchmark calculation given in Ref.~\cite{kamada01}.  To start with, we show the ordinary UCOM case where the UCOM is used for all the partial waves.  The calculated results of the energy are indicated in Fig. \ref{energy} by circles as a function of the maximum angular momentum $L_{\rm max}$.  The results show good convergence to reach $-19$ MeV, while the exact value of the few-body calculations is approximately $-26$ MeV as indicated in Fig. \ref{energy}.
Although the binding energy is small, we point out here that we can calculate the binding energy directly using the nucleon-nucleon interaction in the TOSMU.  The tensor interaction matrix element is approximately $-50$ MeV.   On the other hand, in the previous study\cite{myo06}, we obtained approximately $-60$ MeV for the tensor interaction matrix element to check the validity of the TOSM, when we used the $G$-matrix for the central interaction to renormalize the short-range repulsion and retained the bare tensor interaction of AV8$^\prime$ in our previous calculation.  This fact indicates that the treatment of the short-range repulsive interaction is interfering with the contribution of the tensor interaction.   This is due to a large removal of the short-range part of the relative wave functions in the UCOM, in particular, in the $d$-wave part of the $sd$ coupling of the tensor interaction matrix element, where the tensor interaction possesses some amount of strength.  We have also calculated the contributions beyond the $2p2h$ configurations in the TOSM such as $3p3h$ and $4p4h$ configurations.  When we include the $4p4h$ configurations within the $p$-shell,  their contribution to the binding energy is approximately 50 keV.
This fact denotes that these more complicated wave functions contribute very little in the total $^4$He wave function. 

We have decided to restrict the use of the UCOM to the relative $s$-wave only for the even channel ($S$-UCOM), where the treatment of the short-range repulsion is absolutely necessary.  In other partial waves, we have the centrifugal potential that cuts out the short-range part from the wave functions of the higher partial waves.   In this case, we can use the modified interaction and the kinetic energy only for the relative $s$-wave component in the even channels.   Since the use of the UCOM for the odd partial wave is slightly better, we use the UCOM for all odd partial waves.   As a starting calculation, we have neglected the $S$-UCOM correlation in the calculation of the tensor interaction matrix elements.  The energy converges to $-24$ MeV, which is now very close to the exact one as shown in Fig. \ref{energy}.   In this case, the tensor interaction matrix element is $-61$ MeV, which becomes close to the exact value of $-68$ MeV.  This improvement mainly comes from the increase in the $sd$ coupling of the tensor interaction matrix element, however, this calculation is still not yet perfect.
We have to treat the effects of the short-range repulsion on the tensor interaction matrix element.  Hence, we have worked out the formulation to treat the rigorous $s$-wave function with the effect of the short-range repulsion for the calculation of the tensor interaction matrix elements as explained in the previous section.

\begin{figure}[t]
\centering
\includegraphics[width=9cm,clip]{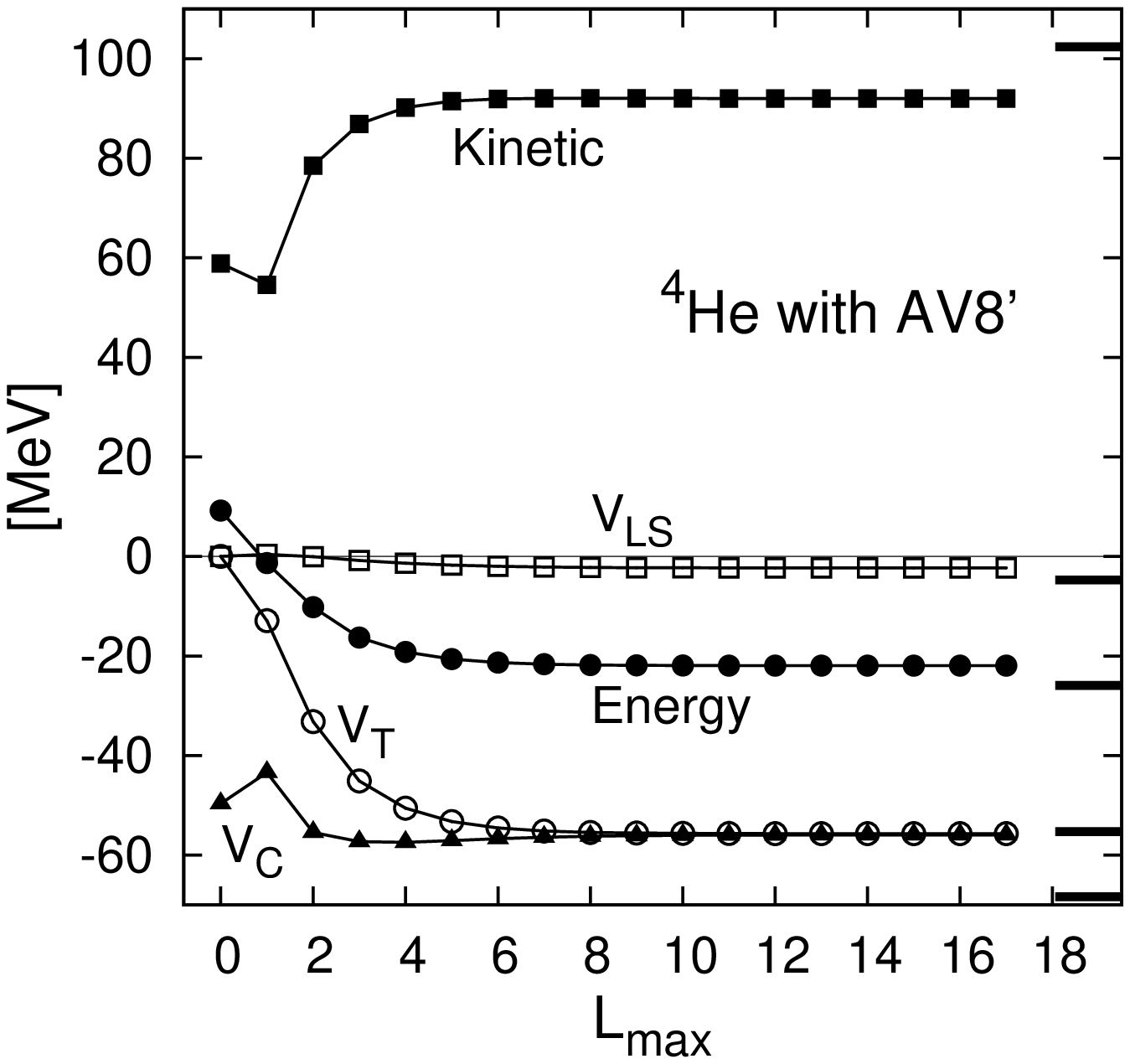}
\caption{Matrix elements of the central interaction ($V_C$), tensor interaction ($V_T$) and the spin-orbit interaction ($V_{LS}$) together with the kinetic energy (Kinetic) and total energy (Energy) in the Hamiltonian for $^4$He as function of $L_{\rm max}$. We observe good convergence for all the matrix elements.  These values are compared with the benchmark results of Ref.~\cite{kamada01}, which are indicated by the thick short solid lines on the right-hand side of the figure.}
\label{H_comp}
\end{figure}

The numerical calculation is quite involved in the $S$-UCOM case, since the $s$-wave relative wave function with the effect of short-range repulsion should be used for the tensor interaction matrix element.  
We show the calculated results for the total energy by the squares in Fig. \ref{energy}.  
We see quite a satisfactory result for the total energy, which is approximately $-22$ MeV.  We show now all the components of the energy for $^4$He in Fig. \ref{H_comp}.  All the energy components show the saturation behavior as function of $L_{\rm max}$.  In the tensor component, the saturation is obtained at around $L_{\rm max}$ being $8$.  For the other components, their saturation points are seen at the similar $L_{\rm max}$.  A very interesting feature is the kinetic energy, which goes up to a large value as the tensor interaction matrix element becomes large. As for the comparison with the rigorous calculation, we see that $V_c$ satisfies the rigorous value, which is approximately $-55$ MeV.  On the other hand, the tensor interaction matrix element, $V_T$ converges to $-55$ MeV, while the rigorous one is $-68$ MeV.  
The kinetic energy is approximately $90$ MeV, while the rigorous one is $102$ MeV.  The LS matrix element is also smaller than the rigorous value. As the net value, the total energy, $E$, is $-22$ MeV and the rigorous value is $-26$ MeV.   
A detailed comparison is performed in Table \ref{tab:benchmark}, in which the converged energies in the TOSMU are shown with the rigorous calculations.  One of the possibilities for the lack of the energy in the TOSMU is due to the separate treatment of the short-range and tensor correlations.  Although the dominant part of the tensor interaction is of intermediate and long range, there may remain some small strength in the short-range part of the tensor interaction, which can couple with the short-range correlations. This effect can be included by extending the truncation of the UCOM transformation in the Hamiltonian to more than the two-body level.  Three-body term of the UCOM transformation is one of the possibilities to overcome the lack of energy in the TOSMU \cite{feldmeier98}.
\begin{table}[t]
\caption{Total energy, matrix elements of the Hamiltonian and radius of $^4$He are compared with the benchmark results denoted by FY.  Units are in MeV for the total energy and the matrix elements, and fm for the radius of $^4$He.}
\label{tab:benchmark}
\begin{center}
\begin{tabular}{c|cccccc}
                           & Energy     &  Kinetic &  Central   &
Tensor   &
LS     & Radius \\
\hline\noalign{\smallskip}
Present(UCOM)              &  $-$19.46  &   88.64  &  $-56.81$ &
$-50.05$ &
$-1.24$ & 1.555 \\
Present($S$-UCOM)          &  $-$22.30  &   90.50  &  $-55.71$ &
$-54.55$ &
$-2.53$ & 1.546 \\
FY  &  $-$25.94  &  102.39  &  $-55.26$ &
$-68.35$ &
$-4.72$ & 1.485 \\
\hline\noalign{\smallskip}
\end{tabular}
\end{center}
\end{table}

In Table~\ref{tab:mixing}, we list the mixing probabilities of the dominant configurations in $^4$He.  The subscripts 00 and 10 represent $J$ and $T$, the spin and isospin quantum numbers, respectively.  It is found that the $2p2h$ configurations with ($J$, $T$)=$(1,0)$ for the particle pair state are significantly mixed.  These spin and isospin are the same as those for the deuteron, and thus, this two-nucleon coupling can be understood as a deuteron-like correlation \cite{myo06}.
\begin{table}[t]
\caption{Mixing probabilities in the $^4$He ground state in \%.}
\label{tab:mixing} 
\begin{center}
\begin{tabular}{c|c}
\hline\noalign{\smallskip}
$(0s)_{00}^4$                                  & 82.48  \\
$(0s)_{10}^{-2}(0p_{1/2})_{10}^2$              &  2.54  \\
$(0s)_{10}^{-2}[(1s_{1/2})(0d_{3/2})]_{10}$    &  2.34  \\
$(0s)_{10}^{-2}[(0p_{3/2})(0f_{5/2})]_{10}$    &  1.90  \\
$(0s)_{10}^{-2}[(0p_{1/2})(0p_{3/2})]_{10}$    &  1.55  \\
$(0s)_{10}^{-2}[(0d_{5/2})(0g_{7/2})]_{10}$    &  0.79  \\
$(0s)_{10}^{-2}(0d_{3/2})_{10}^2$              &  0.44  \\
\mbox{remaining part}                          &  7.96  \\
\hline\noalign{\smallskip}
\end{tabular}
\end{center}
\end{table}

We have been describing $^4$He as an example of the TOSMU for the strong tensor correlation with the use of the UCOM for the short range correlation.  This method is able to describe the $^4$He system almost precisely with the use of the bare nucleon-nucleon interaction.  We are working out further the small difference present between the TOSM and the few body many body methods.  We believe now that the difference comes from the competition of the short range repulsion and the tensor attraction in the very short range part of the relative wave function.  However, we believe that the TOSM is able to describe the deuteron-like tensor correlation and we shall apply this method for the description of the Li isotopes.

\section{Di-neutron clustering and deuteron-like tensor correlation in Li isotopes}
\label{sec:5}

The biggest puzzle from the theory side is the large $s$-wave component for the halo neutrons in $^{11}$Li.  If we interpret this fact in the shell model, the shell gap at $N=8$ has to disappear.   However, the mean field treatment of a central force is not able to provide the disappearance of the $N=8$ shell gap.  So far, there were many theoretical studies for $^{11}$Li~\cite{To94,Es92,Va02,To90b,Myo02,Mu98,De97,Ba97,Mau96,Ga02a,Ga02b,blanchona07} and essentially all the theoretical works of $^{11}$Li had to accept that the $1s_{1/2}$ single particle state is brought down to the $0p_{1/2}$ state without knowing its reason~\cite{To94,Ga02a}.   It is therefore the real challenge for theoretician to understand this disappearance of the $N=8$ shell gap, to be called $s$-$p$ shell gap problem, which is discussed in this section by developing a framework to treat the deuteron-like tensor correlation explicitly using the nucleon-nucleon interaction.  The halo structure of $^{11}$Li is also related with the $1s$-state and the $0p$-state in $^{10}$Li.  Several experiments suggest the dual states of the $s_{1/2}$-state coupled to the $3/2^-$ proton state appears close to the threshold energy of $^9$Li+$n$ together with the dual states of the $p_{1/2}$-state \cite{Th99,Cha01}.

The tensor interaction, on the other hand, plays an important role in the nuclear structure.  For example, the contribution of the tensor interaction in the binding of $^4$He is comparable to that of the central force~\cite{kamada01,Ak86}.  The tensor correlation induced by the tensor interaction was demonstrated important for the $^4$He+n system \cite{Te60,Na59,My05}.  We treated there the tensor interaction in the shell model basis by $2p$-$2h$ excitations, and found that the $(0s_{1/2})^{-2}(0p_{1/2})^2$ excitation of the proton-neutron pair has a special importance in describing the tensor correlation in $^4$He \cite{sugimoto04,My05}.  This $2p$-$2h$ excitation causes the strong Pauli-blocking in the $^4$He+$n$ system for the $p_{1/2}$-orbit of the last neutron, which contributes to the $p$-wave doublet splitting in $^5$He \cite{My05}.  The same effect of the deuteron-like tensor correlation with the Pauli-blocking of the additional two neutrons in the $p_{1/2}$ orbit is expected in $^{11}$Li.  The occupation of the two neutrons in the $p_{1/2}$ orbit interferes with the deuteron-like tensor correlation, which is used to provide a large binding effect of $^4$He in $^9$Li.
 
Hence, it is important to study the effect of the deuteron-like tensor correlation together with the pairing correlation for the $s$-$p$ shell gap problem in $^{11}$Li.  This is the purpose of this section.  We shall perform the configuration mixing based on the shell model framework for $^9$Li to describe the tensor and pairing correlations explicitly. In particular, we pay attention to the special features of the tensor correlation.  For $^{11}$Li, we shall solve the coupled $^9$Li+$n$+$n$ problem which treats both correlations and investigate further the Coulomb breakup strength of $^{11}$Li and other observables to see the effect of these correlations.

\subsection{Model of Li isotopes}

We shall begin with the introduction of the model for $^9$Li, whose Hamiltonian is given as
\begin{eqnarray}
    H(\mbox{$^9$Li})
&=& \sum_{i=1}^9{t_i} - t_G  + \sum_{i<j}^9 v_{ij}\ .
    \label{H9}
\end{eqnarray}
Here, $t_i$, $t_G$, and $v_{ij}$ are the kinetic energy of each nucleon, the center-of-mass (c.m.) term and the two-body $NN$ interaction consisting of central, spin-orbit, tensor and Coulomb terms, respectively. The wave function of $^9$Li($3/2^-$) is described in the tensor-optimized shell model~\cite{myo06,My05}. We express $^9$Li by a multi-configuration,
\begin{eqnarray}
  \Psi(^{9}\mbox{Li})=\sum_i^N C_i\, \Phi^{3/2^-}_i,
  \label{WF9}
\end{eqnarray}
where we consider up to the $2p$-$2h$ excitations within the $0p$ shell for $\Phi^{3/2^-}_i$ in a shell model type wave function, and
$N$ is the configuration number. Based on the previous study of the tensor-optimized shell model~\cite{myo06,My05}, we adopt the spatially modified harmonic oscillator wave function (Gaussian function) as a single particle orbit and treat the length parameters $b_\alpha$ of every orbit $\alpha$ of $0s$, $0p_{1/2}$ and $0p_{3/2}$ as variational parameters. This variation is shown to be important to optimize the tensor correlation~\cite{myo06,sugimoto04,ogawa06,My05}. 

Following the procedure of the tensor-optimized shell model, we solve the variational equation for the Hamiltonian of $^9$Li and determine $\{C_i\}$ in Eq.~(\ref{WF9}) and the length parameters $\{b_{\alpha}\}$ of three orbits. The variation of the energy expectation value with respect to the total wave function $\Psi(^9{\rm Li})$ is given by
\begin{eqnarray}
\delta\frac{\bra\Psi|H(\mbox{$^9$Li})|\Psi\ket}{\bra\Psi|\Psi\ket}&=&0\ ,
\end{eqnarray}
which leads to the following equations:
\begin{eqnarray}
    \frac{\del \bra\Psi| H(\mbox{$^9$Li}) - E |\Psi \ket} {\del b_{\alpha}}
=   0,~~
    \frac{\del \bra\Psi| H(\mbox{$^9$Li}) - E |\Psi \ket} {\del C_{i}}
=   0.
\end{eqnarray}
Here, $E$ is the total energy of $^9$Li. The parameters $\{b_{\alpha}\}$ for the Gaussian bases appear in non-linear forms in the total energy $E$. We solve two kinds of variational equations in the following steps. First, fixing all the length parameters $b_{\alpha}$, we solve the linear equation for $\{C_{i}\}$ as an eigenvalue problem for $H$($^9$Li). We thereby obtain the eigenvalue $E$, which is a function of $\{b_{\alpha}\}$. Next, we try various sets of the length parameters $\{b_{\alpha}\}$ to find the solution which minimizes the energy of $^9$Li. In this wave function, we can optimize the radial form of single particle orbit appropriately so as to describe the spatial shrinkage of the particle state, which is important for the tensor correlation \cite{myo06,sugimoto04,ogawa06,My05}.

For $^{11}$Li and $^{10}$Li, their Hamiltonians are written in terms of $^9$Li+$n$+$n$ and $^9$Li+$n$, respectively, and are given as
\begin{eqnarray}
  H(\mbox{$^{11}$Li})
&=&H(\mbox{$^9$Li})
+   \sum_{k=0}^2{T_k} - T^{(3)}_G
+   \sum_{k=1}^2V_{cn,k}
+ V_{nn}, 
    \label{H11}
    \\
  H(\mbox{$^{10}$Li})
&=&H(\mbox{$^9$Li})
+   \sum_{k=0}^1{T_k} - T^{(2)}_G + V_{cn},
    \label{H10}
\end{eqnarray}
where $H(\mbox{$^9$Li})$, $T_k$, $T^{(3)}_G$ and $T^{(2)}_G$  are the internal Hamiltonian of $^9$Li given by Eq.~(\ref{H9}), the kinetic energies of each cluster ($k=0$ for $^9$Li) and the c.m. terms of three or two cluster systems, respectively. ${V}_{cn,k}$ are the $^9$Li core-$n$ interaction ($k=1,2$) and $V_{nn}$ is the interaction between last two neutrons. The wave functions of $^{11}$Li and $^{10}$Li with the spin $J$ and $J^\prime$, respectively, are given as
\begin{eqnarray}
    \Psi^J(^{11}{\rm Li})
&=& \sum_i^N{\cal A}\left\{ [\Phi^{3/2^-}_i, \chi^{J_0}_i(nn)]^J \right\},
    \label{WF11}
    \\
    \Psi^{J^\prime}(^{10}{\rm Li})
&=& \sum_i^N{\cal A}\left\{ [\Phi^{3/2^-}_i, \chi^{J^\prime_0}_i(n)]^{J^\prime} \right\}.
    \label{WF10}
\end{eqnarray}
We obtain the coupled differential equations for the neutron wave functions $\chi^{J_0}(nn)$ and $\chi^{J^\prime_0}(n)$, where $J_0$ and $J_0^\prime$ are the spins of the additional neutron part of $^{11}$Li and $^{10}$Li, respectively. To obtain the total wave function $\Psi^J(^{11}{\rm Li})$ and $\Psi^{J^\prime}(^{10}{\rm Li})$, we actually use the orthogonality condition model (OCM) \cite{To90b,Myo02,Ao06} to treat the antisymmetrization between last neutrons and neutrons in $^9$Li. In OCM, the neutron wave functions $\chi$ are imposed to be orthogonal to the occupied orbits by neutrons in $^9$Li, which depend on the configuration $\Phi^{3/2^-}_i$ in Eq.~(\ref{WF9}). We obtain the following coupled Schr\"odinger equations with OCM for the set of the wave functions $\{\chi_i^{J_0}(nn)\}$ for $^{11}$Li and $\{\chi_i^{J^\prime_0}(n)\}$ for $^{10}$Li, where $i=1,\cdots,N$:
\begin{eqnarray}
\sum_{j=1}^N \left[ \left(T_{\rm rel}^{(3)} + \sum_{k=1}^2V_{cn,k}+V_{nn}+ \Lambda_i \right) \delta_{ij} + h_{ij}(^9{\rm Li})\right]
\chi_j^{J_0}(nn)&=&E\ \chi_i^{J_0}(nn),
\label{OCM11}
\\
\sum_{j=1}^N \left[ \left(T_{\rm rel}^{(2)} + V_{cn} + \Lambda_i \right) \delta_{ij} + h_{ij}(^9{\rm Li})\right]
\chi_j^{J_0^\prime}(n)&=&E\ \chi_i^{J_0^\prime}(n),
\label{OCM10}
\end{eqnarray}
\begin{eqnarray}
\Lambda_i
&=& \lambda \sum_{\alpha\in \Phi_i(^{9}{\rm Li})} |\phi_\alpha \ket \bra \phi_\alpha|,
\end{eqnarray}
where $h_{ij}(^9{\rm Li})=\bra \Phi_i^{3/2^-} | H(^{9}{\rm Li}) | \Phi_j^{3/2^-} \ket$. $T^{(3)}_{\rm rel}$ and $T^{(2)}_{\rm rel}$ are the total kinetic energies consisting of the relative motions for $^{11}$Li and $^{10}$Li, respectively. $\Lambda_i$ is the projection operator to remove the Pauli forbidden states $\phi_\alpha$ from the relative wave functions \cite{Ka99,Ku86}, where $\phi_\alpha$ is the occupied single particle wave function of the orbit $\alpha$ in $^9$Li defined in Eq.~(\ref{eq:PF}). This $\Lambda_i$ depends on the neutron occupied orbits in the configuration $\Phi^{3/2^-}_i$ of $^{9}$Li and plays an essential role to produce the Pauli-blocking in $^{11}$Li and $^{10}$Li, explained later.  This Pauli blocking term $\Lambda_i$ reduces the pairing and the deuteron-like tensor correlations depending on the occupation of the additional neutrons in shell model single particle orbits.  In $^{10}$Li and $^{11}$Li, the term is particularly effective when the neutron or neutrons occupy the $p_{1/2}$ single particle orbit.  The value of $\lambda$ is taken large as $10^6$~MeV in the present calculation in order to project out the components of the Pauli forbidden states into an unphysical energy region.  Here, we keep the length parameters \{$b_\alpha$\} of the single particle wave functions as those obtained for $^9$Li. 

We explain the method of treating the orthogonality condition including the particle-hole excitations of $^9$Li in more detail \cite{Myo02,My05}. When the neutron orbit in $^9$Li is fully occupied, the orthogonality condition for the last neutrons to this orbit is given by $\Lambda_i$ in Eqs.~(\ref{OCM11}) and (\ref{OCM10}). When neutron orbits in $^9$Li are partially occupied, such as in the $2p$-2$h$ states, the last neutrons can occupy these orbits with particular probabilities, which are determined by the fractional parentage coefficients of the total wave functions of $^{10,11}$Li consisting of $^9$Li and the last neutrons.

We describe the two neutron wave functions $\chi(nn)$ in Eq.~(\ref{OCM11}) for $^{11}$Li precisely in a few-body approach of the hybrid-$VT$ model \cite{To90b,Ao95a,Myo02};
\begin{eqnarray}
      \chi^{J_0}_i(nn)
&=&   \chi^{J_0}_i(\vc{\xi}_V)+\chi^{J_0}_{i}(\vc{\xi}_T),
      \label{TV}
\end{eqnarray}
where $\vc{\xi}_V$ and $\vc{\xi}_T$ are V-type and T-type coordinate sets of the three-body system, respectively,
as shown in Fig. \ref{fig:VT2}.

Here, we discuss the coupling between $^9$Li and the last neutrons, whose details were already explained in the pairing-blocking case \cite{Myo02,Myo03,Ka99,Sag93}. We consider the case of $^{11}$Li. In the present three-body model, the Pauli forbidden states for the relative motion provides the Pauli-blocking effect caused by the last two neutrons \cite{Myo02,Ka99}. This blocking depends on the relative distance between $^9${Li} and the two neutrons, and change the structure of $^9$Li inside $^{11}$Li, which is determined variationally to minimize the energy of the $^{11}$Li ground state. 
Asymptotically, when the last two neutrons are far away from $^9${Li} ($\vc{\xi}_{V,T}\to\infty$), the effects of antisymmetrization and the interaction between $^9$Li and two neutrons vanish in Eq.~(\ref{OCM11}). Therefore, any coupling between $^9$Li and two neutrons disappears and $^9$Li becomes its ground state. Namely, the mixing coefficients $\{C_i\}$ are the same as those obtained in Eq.~(\ref{WF9}). Contrastingly, when the two neutrons are close to $^9$Li, the two neutrons dynamically couple to the configuration $\Phi^{3/2^-}_i$ of $^9$Li satisfying the Pauli principle. This coupling changes $\{C_i\}$ of $^9$Li from those of the $^9${Li} ground state, and makes the tensor and pairing correlations to be different from those in the isolated case.  For $^{10}$Li, the similar coupling scheme is considered. The dynamical effect of the coupling arising from the Pauli-blocking is explained in the results in detail.

\subsection{Effective interactions}

We explain here the interactions employed in Hamiltonians in Eqs.~(\ref{H9}), (\ref{H11}) and (\ref{H10}). Before explaining the present interactions, we give a brief review of the situation of the treatment of the effective interactions for the study of $^{9,10,11}$Li. As was mentioned, most theoretical studies based on the three-body model of $^{11}$Li employ the state-dependent $^9$Li-$n$ potential where only the $s$-wave potential is made deeper than other partial waves \cite{To94}, while the $^9$Li core is described as inert.  This state-dependence in the $^9$Li-$n$ potential is phenomenologically determined in order to satisfy the experimental observations of a large $s^2$ component and a two-neutron-separation energy of $^{11}$Li, and a virtual $s$-state in $^{10}$Li, simultaneously. On the other hand, for the $nn$ part, the interaction having a mild short-range repulsion \cite{Mu98,Ba97} or the density-dependent one are often used \cite{Es92}. However, even in the microscopic cluster models using an unique effective $NN$ interaction consisting of the central and $LS$ forces \cite{Va02,De97}, the $s$-$p$ shell gap problem in $^{11}$Li and $^{10}$Li cannot be solved simultaneously.
From these results, we consider that the usual approach based on the effective central and $LS$ interactions may be insufficient to explain the exotic structures of $^{10,11}$Li. For this problem, even the so-called ab-initio calculations using the realistic $NN$ interactions, such as Green's function Monte Calro \cite{Pi04}, do not provide good results for $^{11}$Li. 

In this study, we focus on the tensor correlation, which is newly considered to figure out the $s$-$p$ shell gap problem. To do this, we extend the three-body model of $^{11}$Li to incorporate the tensor correlation fully, in particular, for the $^9$Li part. In the present study, our policy for the study of $^{11}$Li is to use the experimental informations and the corresponding theoretical knowledge for $^9$Li and $^{10}$Li as much as possible. Following this policy, we explain our interactions in three terms; $v_{ij}$ of $H(^9{\rm Li})$ in Eq.~(\ref{H9}), core-$n$ $V_{cn}$ and $n$-$n$ $V_{nn}$ of the Hamiltonians in Eqs.~(\ref{H11}) and (\ref{H10}). 

For the potential $V_{nn}$ between the last two neutrons, we take a realistic interaction AV8$^\prime$ in Eq.~(\ref{H11}). Our interest is to see the $n$-$n$ correlation in the two-neutron halo structure, and therefore it is necessary to solve two-neutron relative motion without any assumption. For this purpose, our model space of two neutrons using the hybrid-$VT$ model shown in Eq.~(\ref{TV}) has no restriction and wide enough to describe the short range correlation under the realistic nuclear interaction AV8$^\prime$. Therefore, there is no parameter in the potential $V_{nn}$.  

The $^9$Li-$n$ potential, $V_{c n}$, in Eqs.~(\ref{H11}) and (\ref{H10}) is given by folding an effective interaction, the MHN interaction \cite{Fu80,Ha71}, which is obtained by the $G$-matrix calculation and frequently used in the cluster study of light nuclei \cite{To90b,Ao02,Ao06,Ka99,Fu80}. In the $^9$Li+$n$ system, the folding potential for the $^9$Li density calculated by using H.O. wave function has been discussed to reproduce the proper energies of the $^{10}$Li spectra \cite{To90b,Myo02,Ka99}. Furthermore, considering the small one-neutron-separation energy of $^9$Li and a long-range exponential tail of the density, we improve the tail behavior of the folding potential to have a Yukawa type form \cite{Myo02,Myo03}. Any state-dependence is not used in the present $^9$Li-$n$ potential, such as a deeper potential for the $s$-wave.  This is possible because the Pauli blocking effect of the single particle state is in action and the state with the $p_{1/2}$ orbit is pushed up in energy and becomes close to the state with the $s_{1/2}$ state \cite{Myo02,Ka99}. 
We will discuss the results on $^{10}$Li after the discussion on $^{11}$Li.  We introduce one parameter $\delta$, which is the second-range strength of the MHN $G$-matrix interaction in the calculation of the $^9$Li-$n$ potential as shown in Table~\ref{tab:energy_diff}.  The paramter $\delta$ is to describe the starting energy dependence dominantly coming from the tensor interaction in the $G$-matrix calculation \cite{Ao06,Fu80}. 
In the present calculation, we chose this $\delta$ parameter to reproduce the two-neutron-separation energy of $^{11}$Li as 0.31 MeV after working out the tensor and the pairing correlation effects as explained later.  It is found that this folding potential also reproduces the positions of the $p$-wave resonances in $^{10}$Li, just above the $^9$Li+$n$ threshold energy \cite{Th99}, as shown in the results.

Now we discuss the choice of the interaction between nucleons in the $^9$Li core; $v_{ij}$ in $H(^9{\rm Li})$, where we use the limited shell model wave functions up to the $p$ shell for the $^9$Li core in Eq.~(\ref{WF9}). Since our main interest in this work is to investigate the role of the tensor interaction and the Pauli blocking effect on the two-neutron halo formation, we describe the tensor correlation in addition to the pairing correlation in the $^9$Li core.  Along this line, recently we have many interesting works \cite{sugimoto04,ogawa06,To02,Ak04,Ik04}. We have also studied the role of the tensor interaction in the shell model framework, and proposed the tensor-optimized shell model \cite{myo06,My05}.   As a reliable effective interaction considered from those studies, in this calculation, we use GA proposed by Akaishi \cite{My05,Ak04,Ik04} for $v_{ij}$ in Eqs.~(\ref{H9}), (\ref{H11}) and (\ref{H10}). This effective interaction GA has a term of the tensor interaction obtained from the $
 G$-matrix calculation using the AV8$^\prime$ realistic potential keeping the large momentum space \cite{Ak04,Ik04}. Since we limit the shell model space, we increase the tensor strength by 50\%.  This increase of the tensor strength has been studied in the full TOSM to provide a quantitative account of the tensor correlation in the limited shell model space.  In GA, the obtained $^9$Li wave function in Eq.~(\ref{WF9}) shows smaller matter radius than the observed one due to the high momentum component produced by the tensor correlation \cite{sugimoto04,ogawa06,My05}.  Hence, we have to adjust the central force, which is done by changing the second range of the central force by reducing the strength by $21.5\%$ and increasing the range by 0.185 fm to reproduce the observed binding energy and the matter radius of $^9$Li in the same manner as done for $^4$He \cite{myo06,My05}.

\subsection{$^9$Li}

\begin{figure}[t]
\centering
\includegraphics[width=8.2cm,clip]{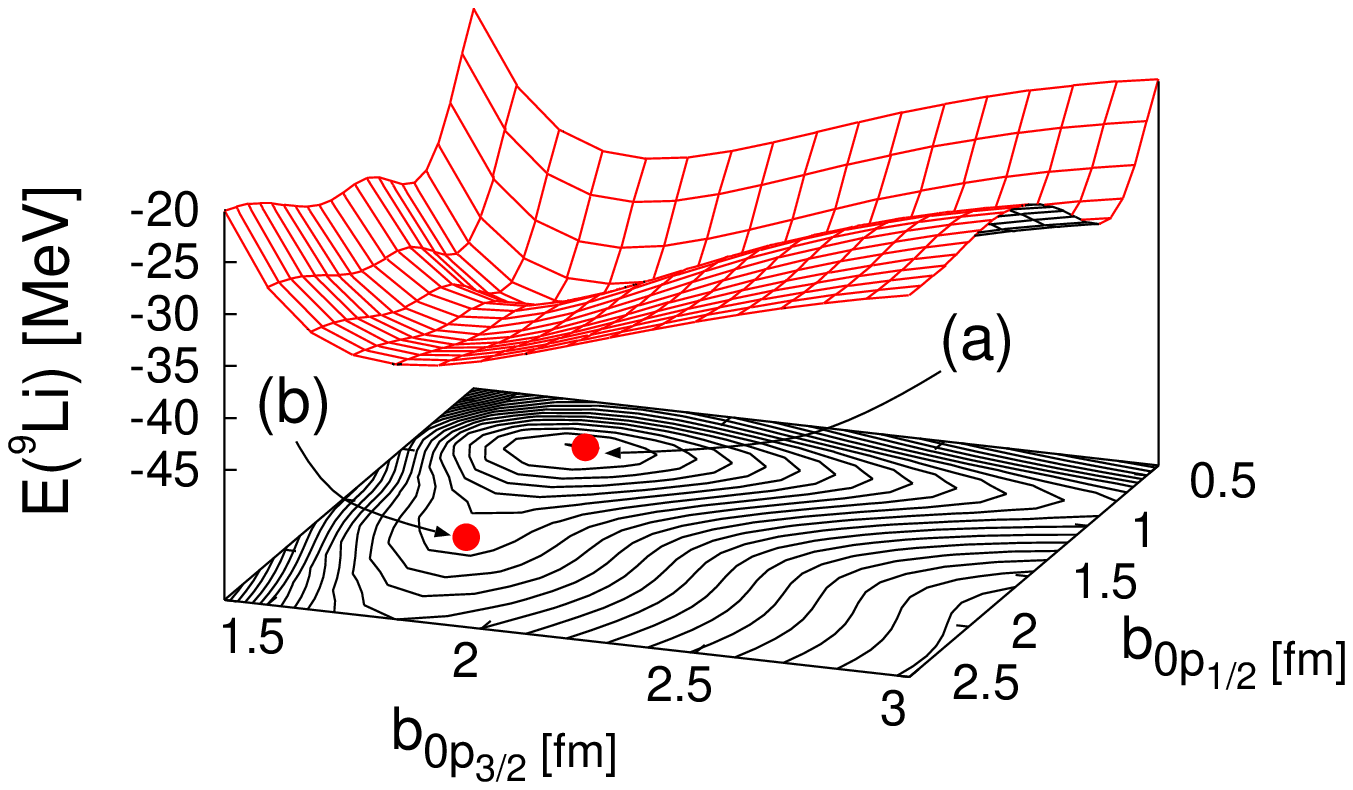}
\caption{(Color online) Energy surface of $^9$Li with respect to the length parameters {$b_\alpha$} of $0p$ orbits \cite{myo07}.  The two minima indicated by (a) and (b) in the contour map correspond to the states due to the tensor correlation and the paring correlation, respectively.
}
\label{9Li_ene}
\end{figure}

We first show the results of the $^9$Li properties, which give a dynamical influence on the motion of last neutrons above the $^9$Li core in $^{11,10}$Li.  In Fig.~\ref{9Li_ene}, we display the energy surface of $^9$Li as functions of the length parameters of two $0p$ orbits, where $b_{0s}$ is already optimized as 1.45 fm. There are two energy minima, (a) and (b), which have almost a common $b_{0p_{3/2}}$ value of 1.7-1.8 fm, and a small (0.85 fm) and a large (1.8 fm) $b_{0p_{1/2}}$ values, respectively. The properties of two minima are listed in Table \ref{tab:9Li} with the dominant $2p$-$2h$ configurations and their probabilities. It is found that the minimum (a) shows a large tensor contribution, while the minimum (b) does not.  Among the $2p$-$2h$ configurations, the largest probabilities are given by $(0s)^{-2}_{10}(0p_{1/2})^2_{10}$ for (a), similar to the results in Ref.~\cite{My05}, and $(0p_{3/2})^{-2}_{01}(0p_{1/2})^{2}_{01}$, namely the $0p$ shell pairing correlation for (b). These results indicate that the minima (a) and (b) represent the different correlations of the tensor and pairing characters, respectively.  The spatial properties are also different from each other; the tensor correlation is optimized with spatially shrunk excited nucleons for (a) and the pairing correlation is optimized when two $0p$ orbits make a large spatial overlap for (b). In Table~\ref{tab:9Li}, we show the results of the superposition of minima (a) and (b), named as (c), to obtain a $^9$Li wave function including the tensor and pairing correlations, simultaneously. For (c), the favored two configurations in each minimum (a) and (b) are still mixed with the $0p$-$0h$ one, and the property of the tensor correlation is kept in (c). The superposed $^9$Li wave function possesses both the tensor and pairing correlations.

\begin{table}[t]
\caption{Properties of $^9$Li with configuration mixing. The states (a) and (b) correspond to the each energy minimum shown in Fig.~\ref{9Li_ene}, respectively. The states (c) is obtained by superposing (a) and (b).}
\label{tab:9Li}
\renc{\baselinestretch}{1.15}
\begin{center}
\begin{tabular}{c|ccccc}
                                                   & \multicolumn{3}{c}{Present}  & Expt. \\
                                                   &   (a)   &  (b)     &  (c)    &       \\
\noalign{\hrule height 0.5pt}
 E [MeV]                                           & $-43.8$ & $-37.3$  & $-45.3$ &  $-45.3$\\
 $\langle V_T\rangle $ [MeV]                       & $-22.6$ &~~$-1.8$  & $-20.7$ &  ---  \\
\noalign{\hrule height 0.5pt}
 $R_m$                 [fm]                        &  2.30   & 2.32     & $2.31$  &   2.32$\pm$0.02\cite{Ta88b} \\
\noalign{\hrule height 0.5pt}
 $0p$-$0h$                                         & 91.2  &  60.1 & $82.9$  & --- \\
 $(0p_{3/2})^{-2}_{01}(0p_{1/2})^2_{01}$           & 0.03  &  37.1 & ~$9.0$  & --- \\
 $(0s_{1/2})^{-2}_{10}(0p_{1/2})^2_{10}$           & 8.2   &   1.8 & ~$7.2$  & --- \\
\end{tabular}
\end{center}
\end{table}

\begin{figure}[t]
\centering
\includegraphics[width=8.0cm,clip]{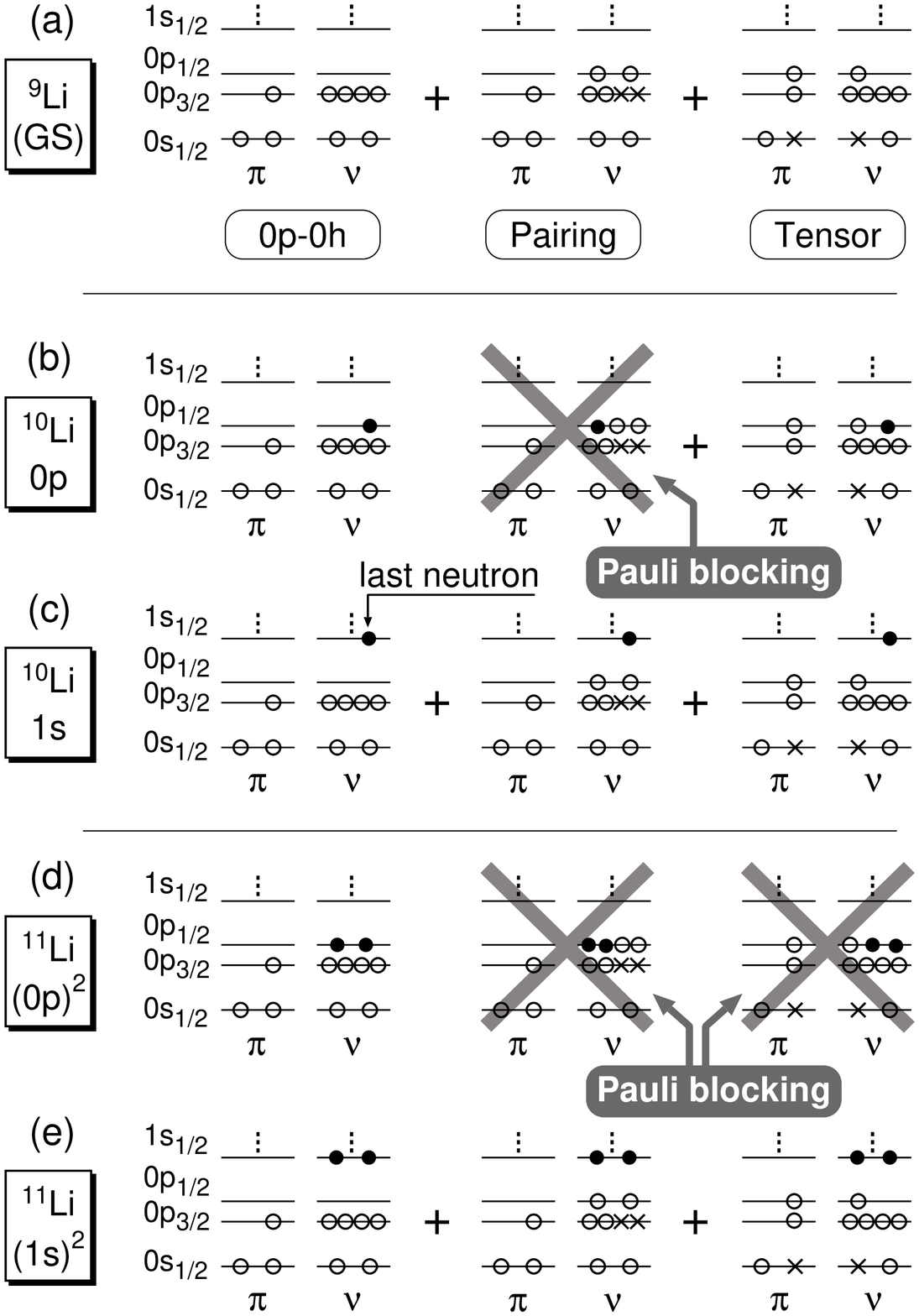}
\caption{Schematic illustration for the Pauli-blocking in $^{11}$Li.
Details are described in the text.}
\label{fig:Pauli}
\end{figure}

\subsection{Pauli-blocking effect in $^{11}$Li}

We discuss here the Pauli-blocking effect in $^{11}$Li and $^{10}$Li. 
Considering the properties of the configuration mixing of $^9$Li,
we discuss the Pauli-blocking effects in $^{10}$Li and $^{11}$Li and their difference as shown in Fig.~\ref{fig:Pauli}.  For (a) in Fig.~\ref{fig:Pauli}, the $^9$Li ground state (GS) is strongly mixed, in addition to the $0p$-$0h$ state, with the $2p$-$2h$ states caused by the tensor and pairing correlations.

Let us add one neutron to $^9$Li for $^{10}$Li. 
For (b) in Fig.~\ref{fig:Pauli}, when a last neutron occupies the $0p_{1/2}$ orbit for the $p$-state of $^{10}$Li, the $2p$-$2h$ excitation of the pairing correlation in $^9$Li are Pauli-blocked.  The tensor correlation is also blocked partially, but not fully by the Pauli principle because the $0p_{1/2}$ orbit is not fully occupied by a last neutron.   Accordingly, the correlation energy of $^9$Li is partially lost inside $^{10}$Li.  Contrastingly, for (c) in Fig.~\ref{fig:Pauli}, the $1s$ state of $^{10}$Li, the Pauli-blocking does not occur and $^9$Li gains its correlation energy fully by the configuration mixing with the $2p$-$2h$ excitations.  Hence, the energy difference between $p$ and $s$ states of $^{10}$Li becomes small to explain the inversion phenomenon \cite{Myo02,Ka99}.

For $^{11}$Li, let us add two neutrons to $^9$Li.
The similar blocking effect is expected for $^{11}$Li, whose important properties were given in the previous paper \cite{myo07}.
For (d) in Fig.~\ref{fig:Pauli}, when two neutrons occupy the $0p_{1/2}$-orbit,
the $2p$-$2h$ excitations of the tensor and pairing correlations in $^9$Li are Pauli-blocked.
In particular, the blocking of the tensor correlation in $^{11}$Li is expected to work stronger than the $^{10}$Li case,  due to the presence of the last two neutrons in the $p_{1/2}$ orbit.  Accordingly, the correlation energy of $^9$Li is lost inside $^{11}$Li stronger than the $^{10}$Li case.  For (e) in Fig.~\ref{fig:Pauli}, $(1s)^2$ of two neutrons, the Pauli-blocking does not occur, similar to the $1s$ state of $^{10}$Li.  Hence, the relative energy distance between $(0p)^2$ and $(1s)^2$ configurations of $^{11}$Li becomes small to break the magicity in $^{11}$Li.  It is found that there is a difference in the blocking effects between $^{11}$Li and $^{10}$Li.  It is interesting to examine how this difference affects the $s$-$p$ shell gap problem in these nuclei.  For the pairing correlation, we already pointed out the different blocking effects between $^{10}$Li and $^{11}$Li in the previous study \cite{Myo02}.  We further consider the blocking effect in the dipole excited states of $^{11}$Li later, which is also different from the ground state case.  In the previous paper \cite{myo07}, we examined that the configuration mixing of the $sd$-shell for $^9$Li would give a small influence on the blocking effect on the $(1s)^2$ configuration of $^{11}$Li.

\subsection{$^{10}$Li}

We solve $^{10}$Li in a coupled $^9$Li+$n$ model, in which 
the $^9$Li-$n$ folding potential is determined to reproduce the two-neutron separation energy $S_{2n}$ of $^{11}$Li as 0.31 MeV.
The resonant states are described using the complex scaling method.
On this condition, we investigate the spectroscopic properties of $^{10}$Li.  In Table \ref{tab:10}, it is shown that using the TOSM for $^9$Li, 
the dual $p$-state resonances are obtained near the $^9$Li+$n$ threshold energy.  The dual states come from the coupling of $[(0p_{3/2})_\pi(0p_{1/2})_\nu]_{1^+/2^+}$, while the experimental uncertainty is still remaining including the spin assignment \cite{Je06}.
The $1^+$ state is predicted at a lower energy than the $2^+$ state due to the attractive effect of the triplet-even $^3E$ channel in the $pn$ interaction.

For the $s$-wave states, their scattering lengths $a_s$ of the $^9$Li+$n$ system show negative values.  In particular, the $2^-$ state shows a large negative value of $a_s$, which is comparable to that of the  $nn$ system ($-18.$5 fm) with the $^1S_0$ channel\cite{Te87}, and indicates the existence of the virtual $s$-state near the $^9$Li+$n$ threshold energy.  Therefore the inversion phenomenon in $^{10}$Li is reasonably explained in the present model.  The order of $2^-$ and $1^-$ also comes from the attractive $^3E$ component in the $pn$ interaction.

For comparison, we calculate $^{10}$Li without the core excitations of $^9$Li (``Inert Core''), namely, we adopt only the single $0p$-$0h$ configuration for $^9$Li without the Pauli blocking effect explained in Fig.~\ref{fig:Pauli} (d).  In this case, we adjust the $\delta$ parameter  of the potential strength $(1+\delta)V_{cn}$ to be 0.066.  In Table \ref{tab:10}, the $p$-wave resonances are obtained just above the $^9$Li+$n$ threshold energy, and $a_s$ values show small positive values for both $1^-$ and $2^-$ states, which means that the virtual $s$- states are not located near the $^9$Li+$n$ threshold, and the $s$-$p$ shell gap is large.  These results mean that the Pauli-blocking nicely describes the spectroscopic properties of $^{10}$Li.

The $d$-wave resonance states of $^{10}$Li are also predicted using the TOSM case of $^9$Li, as shown in Table~\ref{tab:10Li_d}, 
whose excitation energies are higher than those of the $p$-states.
The whole spectra of $^{10}$Li is summarized in Fig.~\ref{fig:10Li} in comparison with the experimental data \cite{bohlen93}.

\begin{figure}[t]
\begin{minipage}[t]{6.0cm}
{\makeatletter\def\@captype{table} 
\caption{The resonance energies $E_r$ and the decay widths $\Gamma$ of the $p$-wave resonance states (1$^+$ and 2$^+$ states) of $^{10}$Li
in unit of MeV, measured from the $^9$Li+$n$ threshold. 
The scattering lengths $a_s$ of the $s$-states (1$^-$ and 2$^-$ states) are shown in unit of fm.  
We show here the two kinds of the results using TOSM and Inert Core for $^9$Li.}
\label{tab:10} 
\begin{tabular}{c|cc}
\hline\noalign{\smallskip}
                              &    TOSM      &  Inert Core    \\
\hline\noalign{\smallskip}
$(E_r,\Gamma)(1^+)$ [MeV]~    &~ (0.22,~ 0.09)~ &  (0.03,~ 0.005) \\
$(E_r,\Gamma)(2^+)$ [MeV]~    &~ (0.64,~ 0.45)~ &  (0.33,~ 0.20)  \\
$a_s(1^-)$ [fm]               &  $ -5.6$     &    1.4         \\
$a_s(2^-)$ [fm]               &  $-17.4$     &    0.8         \\
\hline\noalign{\smallskip}
\end{tabular}
\makeatother}
\end{minipage}
\hspace*{0.5cm}
\begin{minipage}[t]{4.5cm}
{\makeatletter\def\@captype{table} 
\caption{The resonance energies $E_r$ and the decay widths $\Gamma$ of the $d$-wave resonance states in $^{10}$Li in unit of MeV.}
\label{tab:10Li_d} 
\begin{tabular}{c|c}
\hline\noalign{\smallskip}
                        &    TOSM        \\
\hline\noalign{\smallskip}
$(E_r,\Gamma)(1^-)$~    &~ (5.84,~ 5.16) \\
$(E_r,\Gamma)(2^-)$~    &~ (5.81,~ 5.20) \\
$(E_r,\Gamma)(3^-)$~    &~ (6.57,~ 6.31) \\
$(E_r,\Gamma)(4^-)$~    &~ (5.30,~ 3.84) \\
\hline\noalign{\smallskip}
\end{tabular}
\makeatother}
\end{minipage}
\end{figure}

\begin{figure}[t]
\sidecaption[t]
\includegraphics[width=6.0cm,clip]{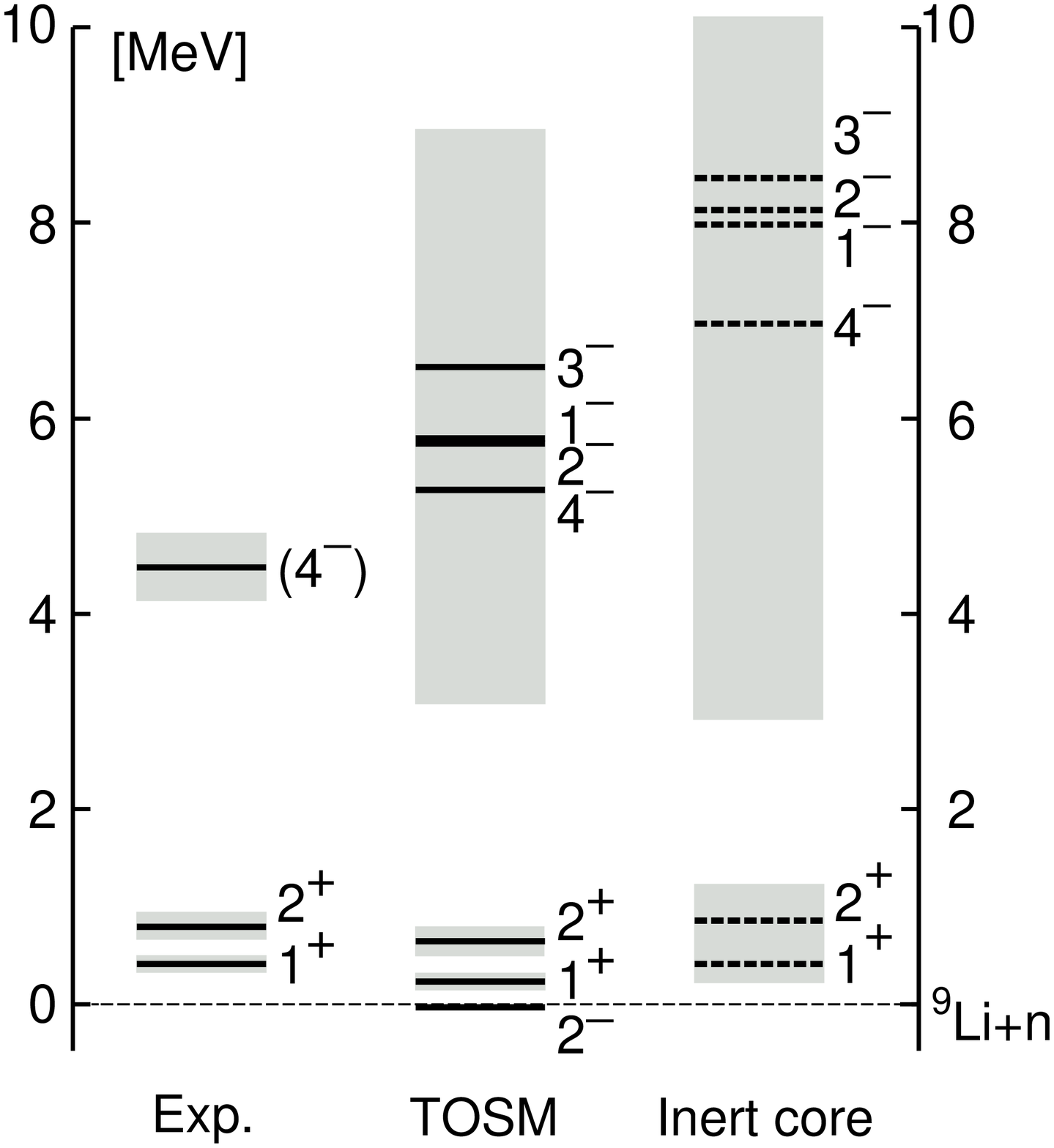}
\caption{$^{10}$Li spectrum using TOSM and Inert core for $^9$Li description \cite{myo08}.
Experimental data is taken from Ref.~\cite{bohlen93}.}
\label{fig:10Li}
\end{figure}

\subsection{$^{11}$Li}
\begin{table}[t]
\caption{
$\delta$ used in the $^9$Li-$n$ potential, and the energy differences $\Delta E$ between the $(1s)^2$ and $(0p)^2$ configurations of $^{11}$Li in MeV.}
\label{tab:energy_diff} 
\begin{center}
\begin{tabular}{c|cccc}
                    &~~Inert core~~  &~~Pairing~~& ~~Tensor~~  &  Present~~ \\
\hline\noalign{\smallskip}
$\delta$        ~~  &  $0.066$ &  $0.143$  & $0.1502$   &  $0.1745$~~  \\
$\Delta E$      ~~  &    2.1   &  1.4      &  0.5       &  $-0.1$~~    \\
\end{tabular}
\end{center}
\end{table}

\begin{figure}[b]
\centering
\includegraphics[width=9.0cm,clip]{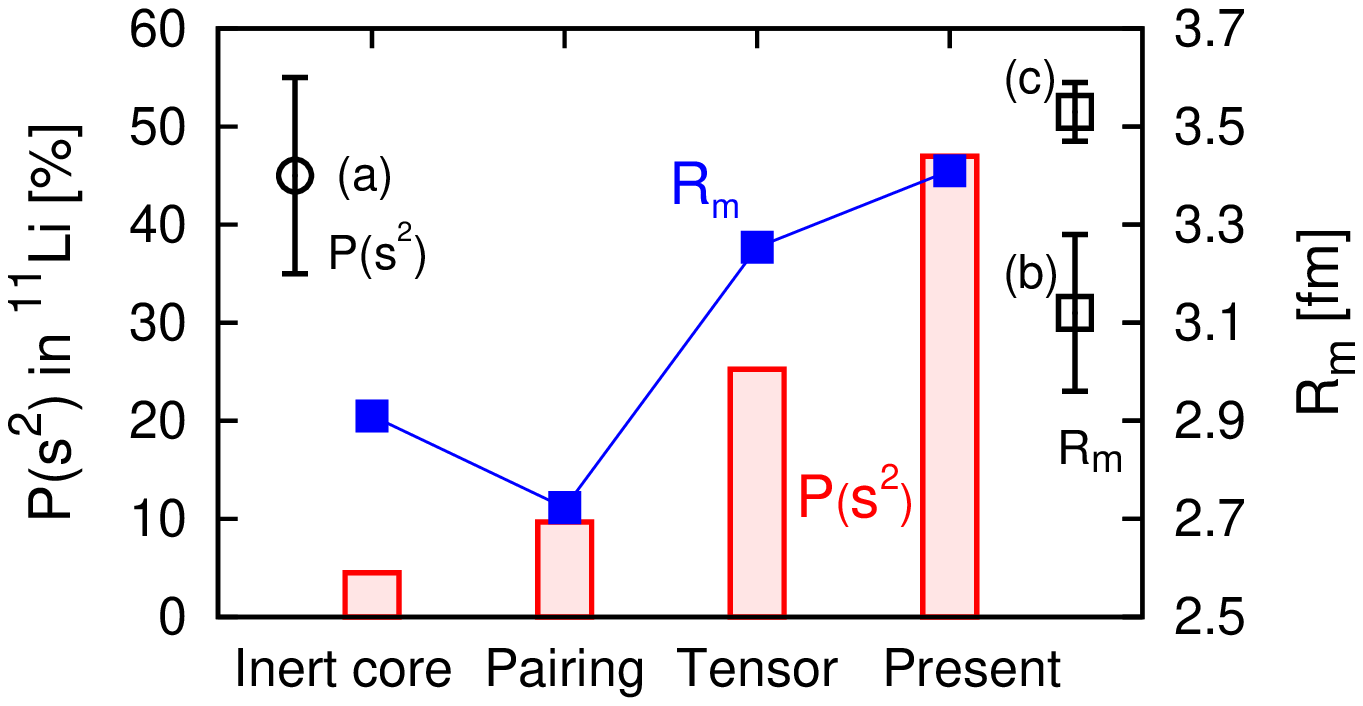}
\caption{(Color online) $(1s)^2$ probability $P(s^2)$ and matter radius $R_m$ of $^{11}$Li with four models in comparison with the experiments
((a)~\cite{simon99}, (b)~\cite{Ta88b} and (c)~\cite{To97}). The scale of $P(s^2)~$($R_m$) is right (left) hand side.}
\label{11Li}
\end{figure}

We solve $^{11}$Li in a coupled $^9$Li+$n$+$n$ model
and show the detailed properties of $^{11}$Li in Table~\ref{tab:11}, 
the partial wave components $P((nlj)^2)$ for halo neutrons, the various r.m.s. radius, the relative distance between halo neutrons ($R_{n\mbox{-}n}$) and the core-$2n$ distance ($R_{c\mbox{-}2n}$), and the expectation value of the opening angle between two neutrons $\theta$
measured from the $^9$Li core, respectively.  The case using TOSM for $^9$Li gives a large $P((1s)^2)$ value, comparable to $P((p_{1/2})^2)$ for halo neutrons and a large matter radius $R_m$ for $^{11}$Li, which are enough to explain the observations.  The case of ``Inert Core'' gives small $P((1s)^2)$ and small $R_m$ values, which disagree with the experiments.  From the difference between two models, it is found that the tensor and pairing correlations in $^9$Li play important roles to break the magicity and make the halo structure of $^{11}$Li, in addition to the $s$-$p$ inversion phenomenon in $^{10}$Li.  As was shown in the previous study \cite{myo07}, the Pauli blocking effect from the tensor correlation is stronger than the pairing one to break the magicity of $^{11}$Li.

In Fig.~\ref{11Li}, ``Present'' is found to give a large amount of the $(1s)^2$ probability $P(s^2)$, 46.9\% for the last two neutrons and a large matter radius $R_m$, 3.41 fm for $^{11}$Li, which are enough to explain the observations. The probabilities of $(p_{1/2})^2$, $(p_{3/2})^2$, $(d_{5/2})^2$ and $(d_{3/2})^2$ for the last two neutrons are obtained as $42.7\%$, $2.5\%$, $4.1\%$ and $1.9\%$, respectively. In Fig.~\ref{11Li}, when we individually consider the tensor and pairing correlations for $^9$Li, $P(s^2)$ is larger for the tensor case than for the pairing case.  Finally, both blocking effects furthermore enhance $P(s^2)$ and provide almost equal amount of $(1s)^2$ and $(0p)^2$ configurations. Hence, two correlations play important roles to break the magicity and make the halo structure for $^{11}$Li.

In Table \ref{tab:energy_diff}, we also estimate the relative energy difference $\Delta E$ between $(1s)^2$ and $(0p)^2$ configurations for $^{11}$Li using the mixing probabilities of these configurations and the coupling matrix element between them as 0.5 MeV obtained in Ref.~\cite{Myo02}. The present model is found to give the degenerated energies enough to cause a large coupling between the $(0p)^2$ and $(1s)^2$ configurations by the pairing interaction between the last neutrons.

\begin{table}[t]
\caption{Ground state properties of $^{11}$Li with $S_{2n}=0.31$ MeV, where two kinds of the $^9$Li descriptions of TOSM and Inert Core are shown, respectively. 
Details are described in the text.}
\label{tab:11} 
\begin{center}
\renc{\baselinestretch}{1.15}
\begin{tabular}{c|cccc|c}
                           &~~TOSM~~~& Inert Core & Expt. \\
\hline\noalign{\smallskip}
$P((p_{1/2})^2))$ [\%]     &  $42.7$  &  90.6      & ---  \\
$(1s_{1/2})^2 $            &  $46.9$  &   4.3      & 45$\pm$10\cite{simon99} \\
$(p_{3/2})^2  $            &  $2.5$   &  $0.8$     & ---  \\
$(d_{3/2})^2  $            &  $1.9$   &  $1.3$     & ---  \\
$(d_{5/2})^2  $            &  $4.1$   &  $2.1$     & ---  \\
$(f_{5/2})^2  $            &  $0.5$   &  $0.2$     & ---  \\
$(f_{7/2})^2  $            &  $0.6$   &  $0.3$     & ---  \\
\hline\noalign{\smallskip}
$R_m $ [fm]                &  3.41    &   2.99     & 3.12$\pm$0.16\cite{Ta88b} \\
                           &          &            & 3.53$\pm$0.06\cite{To97} \\
                           &          &            & 3.71$\pm$0.20\cite{Do06} \\
$R_p $                     &  2.34    &   2.24     & 2.88$\pm$0.11\cite{Ta88b} \\
$R_n $                     &  3.73    &   3.23     & 3.21$\pm$0.17\cite{Ta88b} \\
$R_{ch}   $                &  2.44    &   2.34     & 2.467$\pm$0.037\cite{Sa06} \\
                           &          &            & 2.423$\pm$0.034\cite{Pu06} \\
$R_{n\mbox{-}n}  $         &  7.33    &   6.43     & \\
$R_{c\mbox{-}2n} $         &  5.69    &   4.26     & \\
\hline\noalign{\smallskip}
$\theta   $ [deg.]         & 65.3     &  73.1      & \\
\hline\noalign{\smallskip}
\end{tabular}
\end{center}
\end{table}

In addition to the matter radius, the halo structure also affects the proton radius of $^{11}$Li, because of the recoil effect of the c.m. motion. 
In the three-body model of $^{11}$Li, its proton radius ($R_p$) consisting of the proton radius of $^9$Li and the relative distance between $^9$Li and the c.m. of two neutrons ($R_{\rm c-2n}$) with the following relation
\begin{eqnarray}
  \bra R^2_p(^{11}{\rm Li})\ket=\bra R^2_p(^9{\rm Li})\ket + \left(\frac{2}{11}\right)^2 \bra R_{\rm c-2n}^2\ket,
\end{eqnarray}
where the second term represents the recoil effect.
When the halo structure develops, $\bra R_{\rm c-2n}^2\ket$ is expected to be large. Experimentally, considering the nucleon radius,
the charge radius of $^{11}$Li was measured recently and its value is 2.467$\pm$0.037 fm, which is enhanced from the one of $^9$Li, 2.217$\pm$0.035 fm \cite{Sa06}.  The improved calculation for the isotope shift determination\cite{Pu06} shows that 2.423$\pm$0.037 fm and 2.185$\pm$0.033 fm for $^{11}$Li and $^9$Li, respectively.
The present wave functions provide 2.44 fm and 2.23 fm for $^{11}$Li and $^9$Li, respectively, which are in a good agreement with the experimental values. This enhancement is mainly caused by the large value of $\sqrt{\bra R^2_{\rm c-2n}\ket}$ obtained as 5.69 fm.
For comparison, the distance between last two neutrons is 7.33 fm, which is larger than the core-$2n$ case.

\subsection{Electromagnetic properties of Li isotopes}

We show the $Q$ and $\mu$ moments of $^9$Li and $^{11}$Li in Tables \ref{tab:Q} and \ref{tab:mu}, where only the absolute value of the $Q$ moment of $^{11}$Li is reported in the experiments \cite{Ar94}.  The present model describes reasonably those values for $^9$Li and $^{11}$Li.  For $Q$ moments, the values of $^9$Li and $^{11}$Li do not differ so much to each other.  This result is similar to that of the anti-symmetrized molecular dynamics (AMD) model \cite{En04} and different from the stochastic variational method (SVM) based on the multi-cluster model \cite{Va02}.  Here, similar to the charge radius, 
we discuss the recoil effect in the $Q$ moment of $^{11}$Li by expanding its operator $Q(^{11}{\rm Li})$ into the core part ($Q(^{9}{\rm Li})$), the last neutron part and their coupling part as
\begin{eqnarray}
    Q(^{11}{\rm Li})
&=& Q(^{9}{\rm Li})+ \sqrt{\frac{16\pi}{5}} 3 e \left(\frac{2}{11}\right)^2 {\cal Y}_{20}({\bf R}_{c\mbox{-}2n})
    \nonumber\\
&-& 8\pi \sqrt{\frac{2}{3}} [O_1(^9{\rm Li}),{\cal Y}_1({\bf R}_{c\mbox{-}2n})]_{20},
    \\
    O_{1m}(^9{\rm Li})
&=& e \sum_{i\in {\rm proton}} {\cal Y}_{1m}({\bf a}_i),
\end{eqnarray}
where ${\cal Y}_{lm}({\bf r})\equiv r^l Y_{lm}(\hat{\bf r})$ and $\{{\bf a}_i\}$ are the internal coordinates of protons in $^9$Li.
In our wave function of $^{11}$Li, last two neutrons almost form the $0^+$ state with the probability of around 99\%.  In that case, the $Q$ moment for the relative motion part of $^9$Li-$2n$($c$-$2n$) having the relative coordinate ${\bf R}_{c\mbox{-}2n}$ almost vanishes because of the non-zero rank properties of the $Q$ moment operator. 
This means that the recoil effect from the clusterization is negligible. 
Therefore, the $Q$ moment of $^{11}$Li is caused mainly by the $^9$Li core part inside $^{11}$Li.  In our wave function, the spatial properties of the proton part of $^9$Li inside $^{11}$Li do not change so much.
Hence, the $Q$ moment of $^{11}$Li is similar to the value of the isolated $^9$Li.  Small enhancement from $^9$Li to $^{11}$Li mainly comes from the lacking of the high momentum component of the tensor correlation due to the Pauli-blocking in $^{11}$Li, which extends the radial wave function of $^{11}$Li.  The experimental information of the $Q$ moment is important to understand the structure of $^{11}$Li.
It is highly desired that further experimental data are available for the $Q$ moment of $^{11}$Li.

\begin{figure}[b]
\begin{minipage}[l]{6.0cm}
{\makeatletter\def\@captype{table} 
\caption{$Q$ moments of $^9$Li and $^{11}$Li in units of $e$ fm$^2$.}
\label{tab:Q}
\renc{\baselinestretch}{1.15}
\begin{center}
\begin{tabular}{c|cc}
\hline\noalign{\smallskip}
                      & $^9$Li  & $^{11}$Li \\
\hline\noalign{\smallskip}
TOSM                  & $-2.65$ &  $-2.80$  \\
AMD\cite{En04}        & $-2.66$ &  $-2.94$  \\ 
SVM\cite{Va02}        & $-2.74$ &  $-3.71$  \\
Expt.\cite{Ar94}      & $-2.74\pm 0.10$~~~& $ 3.12\pm 0.45$ ($|Q|$) \\
Expt.\cite{Bo05}      & $-3.06\pm 0.02$ &  --- \\
\hline
\end{tabular}
\end{center}
\makeatother}
\end{minipage}
\hspace*{0.7cm}
\begin{minipage}[r]{4.5cm}
{\makeatletter\def\@captype{table} 
\caption{$\mu$ moments of $^9$Li and $^{11}$Li in units of $\mu_N$.}
\label{tab:mu}
\renc{\baselinestretch}{1.15}
\begin{center}
\begin{tabular}{c|cc}
\hline\noalign{\smallskip}
                      &~~~$^9$Li~~~&~~~$^{11}$Li~~~\\
\hline\noalign{\smallskip}
TOSM                  &   3.69   &   3.77    \\
AMD\cite{En04}        &   3.42   &   3.76    \\
SVM\cite{Va02}        &   3.43   &   3.23    \\
Expt.\cite{Ar94}      &   3.44   &   3.67    \\
\hline
\end{tabular}
\end{center}
\makeatother}
\end{minipage}
\end{figure}

For the $\mu$ moment, the value observed in $^{11}$Li is almost the Schmidt value of 3.79 $\mu_N$ of the $0p_{3/2}$ proton.  In $^9$Li, the $p_{1/2}$ proton is slightly mixed, which decreases the $\mu$ moment.  In $^9$Li, this $p_{1/2}$ proton is excited from the $0p_{3/2}$-orbit in a pair with the $p_{1/2}$ neutron and the $2p$-$2h$ excitation is Pauli-blocked in $^{11}$Li due to the additional neutrons.  As a result, the excitation of $p_{1/2}$ is suppressed, which makes the $\mu$ moment of $^{11}$Li close to the Schmidt value of the $p_{3/2}$ orbit.  The tendency of the increase of the $\mu$ moment from $^9$Li to $^{11}$Li can be obtained also in the shell model analysis using various effective interactions \cite{Su03}.

\begin{figure}[t]
\centering
\includegraphics[width=8.2cm,clip]{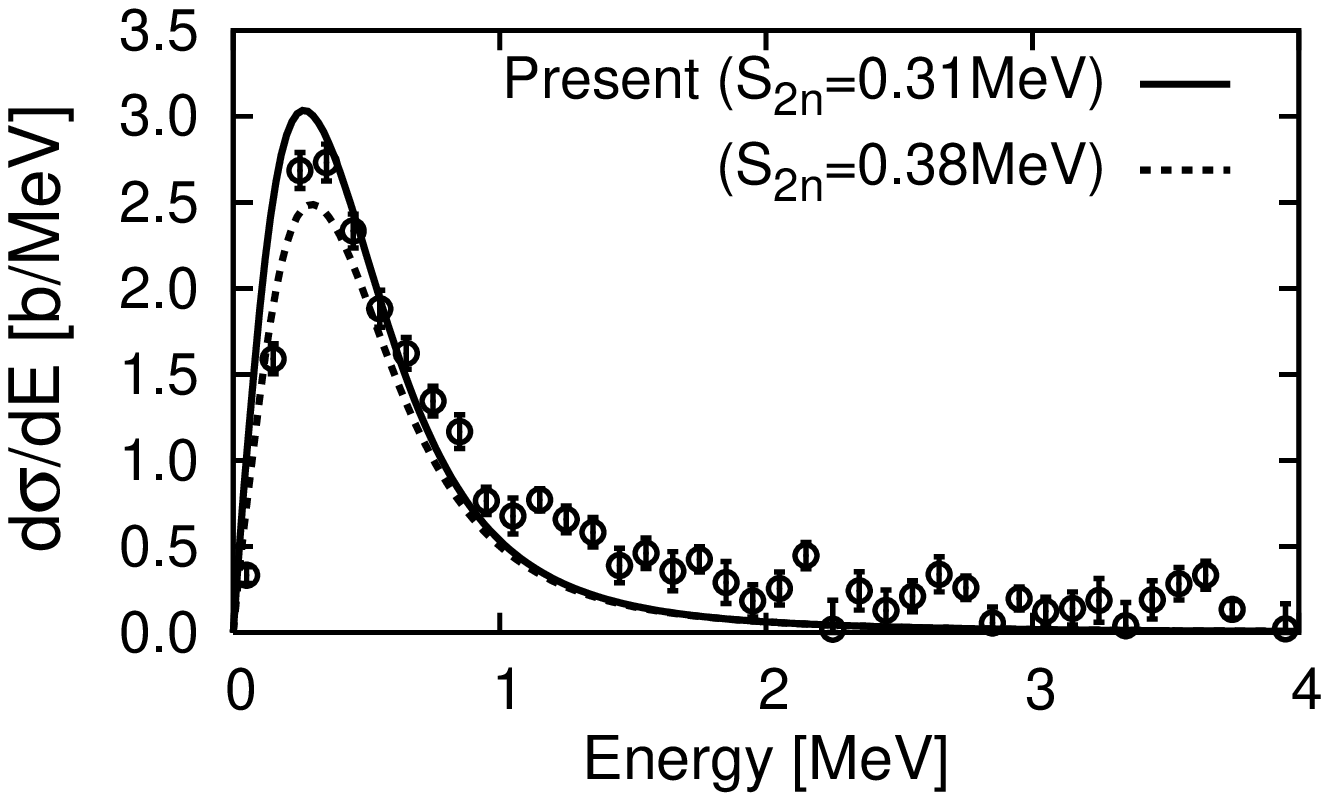}
\caption{Calculated Coulomb breakup cross section of $^{11}$Li into $^9$Li+$n$+$n$ measured from the $^9$Li+$n$+$n$ threshold energy. Data are taken from Ref. \cite{Na06}.}
\label{Li_cross}
\end{figure}

We further calculate the three-body Coulomb breakup strength of $^{11}$Li into the $^9$Li+$n$+$n$ system to investigate the properties of the dipole excited states and compare the strength with the new data from the RIKEN group \cite{Na06}. We use the Green's function method combined with the complex scaling method \cite{Ao06} to calculate the three-body breakup strength \cite{Myo03} using the dipole strength and the equivalent photon method, where the experimental energy resolution is taken into account \cite{Na06}. We do not find any resonances with a sharp decay width enough to make a resonance structure. In Fig.~\ref{Li_cross}, it is found that the present model well reproduces the experiment, in particular, for low energy enhancement and its magnitude.  Seeing more closely, however, our results seem to underestimate the cross section at $E>1$ MeV, while overestimate at low energy peak region slightly.  As a result, the integrated dipole strength for $E\leq 3$ MeV gives 1.35 $e^2 {\rm fm}^2$, which agrees with the experimental value of $1.42\pm0.18$ $e^2 {\rm fm}^2$\cite{Na06}.  We also slightly change the two neutron separation energy of the ground state of $^{11}$Li, 
which is close to the recent observation \cite{bachelet04,bachelet08}.

Summarizing this section, we have considered newly the tensor correlation in $^{11}$Li based on the extended three-body model.  We have found that the tensor and pairing correlations play important roles in $^9$Li with different spatial characteristics, where the tensor correlation prefers a shrunk spatial extension. The tensor and pairing correlations in $^9$Li inside $^{11}$Li are then Pauli-blocked by additional two neutrons, which makes the $(1s)^2$ and $(0p)^2$ configurations close to each other and hence activates the pairing interaction to mix about equal amount of two configurations. As a result we naturally explain the breaking of magicity and the halo formation for $^{11}$Li.  We also reproduce the recent results of the Coulomb breakup strength and the charge radius of $^{11}$Li.  For $^{10}$Li, the inversion phenomenon is explained from the Pauli-blocking effect.

\section{Conclusion}
\label{sec:6}

We have presented the physics of the di-neutron clustering and the deuteron-like tensor correlations by focusing on the halo structure of $^{11}$Li.  The halo structure provides an ideal platform for the di-neutron clustering to play an important role.  This di-neutron clustering phenomenon is strongly related with the central interaction in the $^1$S$_0$ channel, where the large scattering length in the nucleon-nucleon scattering suggests the appearance of the bound state by changing slightly the environment.  On the other hand, in order to produce the halo structure, there should be an active participation of the $s_{1/2}$ configuration in the neutron wave function.   This participation of the $s_{1/2}$ orbit was very difficult in the standard shell model framework.  We had to invoke the deuteron-like tensor correlation in $^{11}$Li, which blocks the two neutrons to enter the $p_{1/2}$ orbit and hence provides a mechanism to put neutrons in the $s_{1/2}$ orbit.

The theoretical description of these two new correlations to realize in $^{11}$Li was very difficult.  In fact, we had to develop the cluster orbit shell model (COSM) to handle an extended object and further the hybrid-$VT$ model (Hybrid-$VT$) to treat the di-neutron clustering correlation in finite nuclei.  We had to deal with unbound and resonance states quantitatively, since the halo structure appears when the binding energy of the last neutrons is very small.  The excitation function of several MeV forces us to treat the spectral function in the continuum region.  At the same time,  the neighboring nuclei are also unbound.  Hence, we had to develop the complex scaling method (CSM) to treat the continuum and resonance states.  All these theoretically involved methods have been developed for the quantitative description of the halo nucleus $^{11}$Li.  The most essential mechanism to bring down the $s_{1/2}$ configuration for the halo formation came from the deuteron-like tensor correlation in nuclei.  The strong binding energy of $^4$He comes from the strong tensor interaction.  We had to develop the tensor optimized shell model (TOSM) to describe the deuteron-like tensor correlation in the shell model basis.  The TOSM provided a clear key to push up the configuration involving the $p_{1/2}$ orbit, since the strong binding of the $^4$He nucleus requires the use of the $p_{1/2}$ orbit.  Hence, naturally the $s_{1/2}$ configuration is energetically favored for the halo structure formation.

We are used to treat the central interaction with the spin-orbit interaction in the shell model basis to treat many body systems.  In the theoretical challenge to describe the halo structure of $^{11}$Li, we had to face to treat the strong tensor interaction for the important role of the deuteron-like tensor correlation in many body systems.  We have seen that the deuteron-like correlation worked out in the shell model basis is able to treat the Pauli blocking effect due to the $(p_{1/2})^2$ neutrons to wash away the N=8 shell gap.  This effect allows the participation of the $(s_{1/2})^2$ configuration to provide a platform to develop the di-neutron clustering correlation in the halo structure of $^{11}$Li.  Hence, we expect many interesting many-body phenomena in unstable nuclei to be found in near future, where the deuteron-like tensor and/or di-neutron clustering correlations play important roles.

In this lecture note, we have gone through all these theoretical materials in details.  We have tried to make the motivation of the development of the theoretical tools in each step to describe the halo nucleus quantitatively.  These theoretical frameworks are not only for use of the halo structure, but should play a very important role for the description of finite nuclei.  In fact, the TOSM is essential to describe the deuteron-like tensor correlation, which should be the most important ingredient to provide large binding energies for all the nuclei.

\begin{acknowledgement}
We are grateful to all the collaborators for fruitful collaborations and continuous discussions.  We would like to thank Prof. Hisashi Horiuchi for continuous support and encouragement.  This work is supported by the JSPS grant: No. 18540269, 21540267 and 21740194.
\end{acknowledgement}

%
%
%
\def\JL#1#2#3#4{ {{\rm #1}}\ {\bf #2}, #4 (#3)}  
\nc{\PRC}[3]    {\JL{Phys. Rev. C}{#1}{#2}{#3}}
\nc{\PRA}[3]    {\JL{Phys. Rev. A}{#1}{#2}{#3}}
\nc{\PRL}[3]    {\JL{Phys. Rev. Lett.}{#1}{#2}{#3}}
\nc{\NP}[3]     {\JL{Nucl. Phys.}{#1}{#2}{#3}}
\nc{\PL}[3]     {\JL{Phys. Lett.}{#1}{#2}{#3}}
\nc{\PTP}[3]    {\JL{Prog. Theor. Phys.}{#1}{#2}{#3}}
\nc{\PTPS}[3]   {\JL{Prog. Theor. Phys. Suppl.}{#1}{#2}{#3}}
\nc{\SPTP}[3]   {\JL{Prog. Theor. Phys. Suppl.}{#1}{#2}{#3}}
\nc{\PRP}[3]    {\JL{Phys. Rep.}{#1}{#2}{#3}}
\nc{\ZP}[3]     {\JL{Z. Phys.}{#1}{#2}{#3}}
\nc{\JP}[3]     {\JL{J. of Phys.}{#1}{#2}{#3}}
\nc{\andvol}[3] {{\it ibid.}\JL{}{#1}{#2}{#3}}
\nc{\PPNP}[3]   {\JL{Prog. Part. Nucl. Phys.}{#1}{#2}{#3}}
\nc{\SJNP}[3]   {\JL{Sov. J. Nucl. Phys.}{#1}{#2}{#3}}
\nc{\CMP}[3]    {\JL{Commun. Math. Phys.}{#1}{#2}{#3}}

\end{document}